\documentclass[11pt]{article}
\usepackage[a4paper,margin=2.6cm]{geometry}
\usepackage{amsmath,amssymb,mathtools,bm}
\numberwithin{equation}{section}
\usepackage{authblk}

\usepackage{slashed}
\usepackage{booktabs}
\usepackage{hyperref}
\usepackage{xcolor}
\hypersetup{colorlinks=true,linkcolor=blue!20!black,citecolor=red!30!black,urlcolor=blue!20!black}

\usepackage{mathrsfs}

\newcommand{\g}{\mathfrak g}
\newcommand{\h}{\mathfrak h}
\newcommand{\m}{\mathfrak m}
\newcommand{\su}{\mathfrak{su}}
\newcommand{\so}{\mathfrak{so}}
\newcommand{\uu}{\mathfrak{u}}
\newcommand{\Tr}{\operatorname{Tr}}
\newcommand{\ad}{\operatorname{ad}}
\newcommand{\vol}{\operatorname{vol}}
\newcommand{\dd}{\mathrm d}
\newcommand{\ii}{\mathrm i}
\newcommand{\cP}{\mathcal P}
\newcommand{\cG}{\mathcal G}
\newcommand{\cC}{\mathcal C}
\newcommand{\cH}{\mathcal H}
\newcommand{\cL}{\mathcal L}

\newcommand{\cJ}{\mathcal J}
\newcommand{\cM}{\mathcal M}
\newcommand{\vev}[1]{\left\langle #1\right\rangle}

\newcommand{\ussvaldivia}{Facultad de Ingenier\'ia, Universidad San Sebasti\'an, General Lagos 1163, Valdivia, Chile}
\newcommand{\cecs}{Centro de Estudios Cient\'ificos, Universidad San Sebasti\'an, Avenida Arturo Prat 514, Valdivia, Chile}

\begin{document}

\title{Projected Yang-Mills Theory and the MacDowell-Mansouri Analogy}
% \author{Draft}
\date{}
\author[1,2]{P. D. Alvarez \thanks{E-mail \href{mailto:pedro.alvarez@uss.cl}{\nolinkurl{pedro.alvarez@uss.cl}}}}

\affil[1]{\ussvaldivia}
\affil[2]{\cecs}

\maketitle

\begin{abstract}
We construct Yang-Mills theories with projected curvature kinetic terms, developing their analogy with MacDowell-Mansouri gravity and their connection to nonlinear sigma models. We begin with models exhibiting explicit symmetry breaking and use them to motivate gauge-covariant Higgs extensions. For $SO(2n)$, an orthogonal complex structure defines a degenerate kinetic operator with image $\mathfrak{su}(n)$, while the adjoint Higgs vacuum has stabilizer $U(n)$. We compare the fixed-projector theory with gauge-covariant formulations involving either unrestricted adjoint Higgs fields or Higgs fields constrained exactly to the orbit $SO(2n)/U(n)$. We express the Stelle-West Higgs-gravity action through a single Clifford involution whose complementary projectors select the Lorentz curvature, isolate the radial Higgs derivative, and construct the induced spacetime metric. This formulation makes the common projector structure explicit and clarifies the distinct roles of curvature projection, the internal dual, and the coframe. In the orbit-constrained Yang-Mills theory with a positive Higgs kinetic coefficient the coset connection is auxiliary and the central connection drops out of the action. A nonlinear Dirac analysis establishes a regular weak-curvature branch with $2(n^2-1)$ physical degrees of freedom in four dimensions, locally equivalent to $SU(n)$ Yang-Mills theory. In the unrestricted theory, normal Higgs displacements can change the kinetic rank, so this degree-of-freedom count does not automatically extend beyond the constrained field space. We also characterize possible nontrivial algebraic branches and introduce projected Pontryagin theories and a full-rank equivariant completion.
\end{abstract}

\tableofcontents

%%%%%%%%%%%%%%%%%%%%%%%%%%%%%%%%%%%%%%%%%%%%%%%
\section{Introduction}
%%%%%%%%%%%%%%%%%%%%%%%%%%%%%%%%%%%%%%%%%%%%%%%

Ordinary Yang-Mills theory equips the gauge algebra with a nondegenerate
invariant quadratic form, giving a kinetic term to the curvature in every
algebra direction~\cite{Yang:1954ek}. Spontaneous symmetry breaking changes
the vector mass matrix but does not normally change the rank of the gauge
kinetic operator. Here we instead insert a projector into the curvature
kinetic term. The internal structure defining this projector selects which
curvature components enter the action. Determining the propagating fields
then requires a constraint analysis, since the connection components in
the kernel of the kinetic operator may still enter the action algebraically.

A motivation for this construction comes from MacDowell-Mansouri
gravity~\cite{MacDowell:1977jt}, whose action selects the Lorentz
components of an (A)dS curvature~\cite{Wise:2009fu}. In the Stelle-West
formulation, a compensator makes this selection covariant under the
full (A)dS group~\cite{Stelle:1979aj}. The compensator selects a Lorentz
stabilizer, while the component of its covariant derivative tangent
to the compensator orbit supplies the coframe. The associated
gravitational Higgs sector was reviewed by Leclerc~\cite{Leclerc:2005qc},
and Westman and Zlosnik developed Cartan-gravity models in which the
gauge-invariant Higgs norm is itself dynamical~\cite{Westman:2013mf}.
These constructions provide a gravitational counterpart to the
distinction between a Higgs field restricted to a fixed orbit and an
unrestricted Higgs field with normal fluctuations. The gravitational
construction also uses an internal dual and a nondegenerate coframe,
which are essential to its torsion mechanism. Comparing it with a
projection in an ordinary spacetime-Hodge Yang-Mills action helps
distinguish the consequences of the algebraic splitting from those of
these additional geometric structures.

Our starting point is an $SO(2n)$ connection and an orthogonal complex
structure $J$, which selects the splitting
$\so(2n)=\uu(n)\oplus\m$. Removing the central direction generated by
$J$ gives a projector onto $\su(n)$. Promoting $vJ$ to an adjoint Higgs
field makes this projector equivariant on the orbit $SO(2n)/U(n)$.
The stabilizer of the Higgs vacuum is still $U(n)$: selecting $\su(n)$
as the image of the kinetic projector does not further break the gauge
symmetry to $SU(n)$. The central connection instead drops out of the projected action.

The exact orbit restriction also connects the construction to nonlinear
sigma models. The space $SO(2n)/U(n)$ is a compact Hermitian symmetric
space of type DIII and is a standard target in supersymmetric and gauged
sigma-model constructions~\cite{Higashijima:1999ki,GrootNibbelink:2000hu}.
It also appears as the target of effective world-sheet sigma models
describing the orientational modes of non-Abelian
vortices~\cite{Gudnason:2010rm}. In our orbit-constrained theory, the
Higgs kinetic term built from $D_A\Phi$ is precisely the corresponding
gauged nonlinear sigma-model term, with the target-space metric induced
by the invariant trace pairing. The full $SO(2n)$ symmetry is gauged,
and this scalar sector is coupled to a degenerate curvature kinetic
operator. The resulting propagating content must therefore be
determined from the complete action, rather than inferred from the
sigma-model target alone.

Several ingredients have established precedents. Adjoint Higgs fields
and the breaking $SO(2n)\rightarrow U(n)$ occur in monopole and
grand-unified constructions~\cite{Auzzi:2004if,Bertolini:2009es}.
Decompositions based on a local internal direction are central to the
Cho-Faddeev-Niemi program~\cite{Faddeev:1999cj,Kondo:2005eq}, while
nonlinear realizations provide a general language for broken gauge
theories~\cite{Goon:2014ika}. Scalar-dependent gauge structures also
appear in curved Yang-Mills-Higgs theories~\cite{Kotov:2015nuz}.
Auxiliary gauge fields and the possible quantum generation of their
kinetic terms are familiar from hidden-local-symmetry
formulations~\cite{Bando:1987br}. Our focus here is the use of the adjoint
complex structure to define a degenerate gauge kinetic operator whose
kernel includes both the coset connection and the central direction,
and the resulting classical constraint structure.

We distinguish three theories. The first uses a fixed projector,
which explicitly reduces the connected gauge symmetry for $n\ge3$.
The exceptional case $n=2$, where the projection selects an ideal
of $\so(4)$, is discussed separately in Sec.~\ref{sec:fixed}.
The second promotes the defining structure to an unrestricted adjoint Higgs with a finite potential.
Its kinetic map remains equivariant away from the vacuum orbit, but
generally ceases to be a projector and changes rank. The third restricts
the Higgs field exactly to the orbit, so that the projector identities
hold throughout its field space. We formulate this constraint both
intrinsically and with a matrix Lagrange multiplier, and establish their
local classical equivalence.

The Hamiltonian analysis follows the Dirac-Bergmann
method~\cite{Dirac:1950pj,Henneaux:1992ig}. For $n\ge 3$ in the fixed-projector
theory, the coset constraint system is irregular at the trivial vacuum.
In the exactly constrained theory with $Z_\Phi>0$, the Higgs kinetic
term supplies a nondegenerate algebraic quadratic form for the coset
connection. Its constraints are then regularly second class near the
broken vacuum. In four spacetime dimensions, the regular weak-curvature
branch locally reduces to $SU(n)$ Yang-Mills theory with
$2(n^2-1)$ physical polarizations. This nonlinear result applies within
the constrained field space. It does not extend directly to the
unrestricted theory, where normal Higgs displacements can activate
vector kinetic terms absent from the quadratic vacuum action. We
therefore distinguish the quadratic spectrum of that theory from a
nonlinear degree-of-freedom count.

Section~\ref{sec:algebra} develops the algebraic splitting and its
projectors. Sections~\ref{sec:fixed}, \ref{sec:dynamical}, and
\ref{sec:constrained} treat the fixed, unrestricted, and exactly
constrained theories, respectively. Section~\ref{sec:MM} compares their
constraint mechanisms with MacDowell-Mansouri and Stelle-West gravity
and examines the projected Pontryagin action.
Section~\ref{sec:fullrank} constructs a full-rank equivariant kinetic
operator. We conclude by discussing quantum consistency, global
questions, and the scope of the classical results.

%%%%%%%%%%%%%%%%%%%%%%%%%%%%%%%%%%%%%%%%%%%%%%%
\section{Algebraic decomposition selected by a complex structure}
\label{sec:algebra}
%%%%%%%%%%%%%%%%%%%%%%%%%%%%%%%%%%%%%%%%%%%%%%%

In this section we fix our notation and define the algebraic objects that
will be important for the subsequent developments.  We first explain how an
orthogonal complex structure splits the real gauge algebra, then establish
the central charge convention and the representation content of the
resulting sectors.  We also distinguish the transformations of an adjoint
scalar from those of a connection.  Finally, we construct the projectors
that implement this splitting and verify the properties needed when they
are inserted into the gauge kinetic term.

%%%%%%%%%%%%%%%%%%%%%%%%%%%%%%%%%%%%%%%%%%%%%%%
\subsection{The complex structure and the real algebraic splitting}
\label{sec:real-algebra-splitting}
%%%%%%%%%%%%%%%%%%%%%%%%%%%%%%%%%%%%%%%%%%%%%%%

Let $V\simeq\mathbb R^{2n}$ carry its Euclidean metric
$\langle\,\cdot\,,\,\cdot\,\rangle_V$, and let $J$ be a real matrix satisfying
\begin{equation}
 J^T=-J,
 \qquad
 J^2=-\mathbf1_{2n}.
 \label{eq:Jdef}
\end{equation}
These conditions imply $J^TJ=\mathbf1_{2n}$, so $J$ is an orthogonal
complex structure on the real vector space $V\simeq\mathbb R^{2n}$.
This means that $J$ gives the real-linear realization of multiplication
by $\ii$ after identifying $(V,J)$ with $\mathbb C^n$.
%; it does not mean that the real matrix $J$ is equal to $\ii\mathbf1_{2n}$.
 To make this identification explicit, consider the canonical form
\begin{equation}
 J_0=
 \begin{pmatrix}
  \mathbf0_n & \mathbf1_n\\
  -\mathbf1_n & \mathbf0_n
 \end{pmatrix}.
 \label{eq:J-canonical-real-form}
\end{equation}
Writing a real vector as $v=(x,y)\in\mathbb R^n\oplus\mathbb R^n$, one
has
\begin{equation}
 J_0(x,y)=(y,-x).
\end{equation}
Now define the real-linear identification
\begin{equation}
 \iota:\mathbb R^{2n}\longrightarrow\mathbb C^n,
 \qquad
 \iota(x,y)=x-\ii y.
\end{equation}
Then
\begin{equation}
 \iota\!\left(J_0(x,y)\right)
 =\iota(y,-x)
 =y+\ii x
 =\ii(x-\ii y)
 =\ii\,\iota(x,y).
 \label{eq:J-realizes-complex-multiplication}
\end{equation}
Thus $J_0$ is precisely the real matrix representing multiplication by
$\ii$ under the identification $\iota$.  If instead one uses the
convention $z=x+\ii y$, the same matrix represents multiplication by
$-\ii$ or, equivalently, multiplication by $+\ii$ is represented by
$-J_0$.  The two choices differ only by the convention $J\mapsto-J$.
Thus
\begin{equation}
 \operatorname{Stab}_{SO(2n)}(J) := \left\{
 g\in SO(2n)
 \,\middle|\,
 gJg^{-1}=J
 \right\} =U(n).
 \label{eq:stabilizer-J}
\end{equation}
Indeed, since $g$ is invertible, the stabilizer condition is equivalent to
\begin{equation}
 gJg^{-1}=J
 \quad\Longleftrightarrow\quad
 gJ=Jg.
\end{equation}
Multiplication by $\ii$ on the complex vector space $(V,J)$ is defined by
$\ii v:=Jv$.  Therefore,
\begin{equation}
 g(\ii v)=g(Jv)=Jg(v)=\ii g(v),
\end{equation}
so the condition $gJ=Jg$ states precisely that $g$ is complex-linear.

The real Euclidean metric and the complex structure $J$ together define
the Hermitian form
\begin{equation}
 h(u,v)
 =
 \langle u,v\rangle_V
 -\ii\langle Ju,v\rangle_V.
 \label{eq:J-Hermitian-form}
\end{equation}
If $g\in SO(2n)$ also satisfies $gJ=Jg$, then
\begin{align}
 h(gu,gv)
 &=
 \langle gu,gv\rangle_V
 -\ii\langle Jgu,gv\rangle_V
 \nonumber\\
 &=
 \langle u,v\rangle_V
 -\ii\langle gJu,gv\rangle_V
 \nonumber\\
 &=
 \langle u,v\rangle_V
 -\ii\langle Ju,v\rangle_V
 =h(u,v).
\end{align}
Thus $g$ is complex-linear and preserves the Hermitian form $h$, so
$g\in U(n)$.

Conversely, a unitary transformation is complex-linear and therefore
commutes with $J$, while preservation of $h$ implies preservation of its
real part,
\begin{equation}
 \langle u,v\rangle_V=\operatorname{Re}h(u,v).
\end{equation}
Its determinant as a real transformation is
\begin{equation}
 \det_{\mathbb R}g
 =
 \left|\det_{\mathbb C}g\right|^2
 =1,
\end{equation}
so its underlying real transformation is orientation preserving and therefore belongs to $SO(2n)$.  

At the Lie-algebra level, the generators commuting with $J$ are
complex-linear, while those anticommuting with $J$ are complex-antilinear.
They define the decomposition
\begin{equation}
 \so(2n)=\uu(n)\oplus\m,
 \label{eq:sodecomp}
\end{equation}
where
\begin{equation}
 \uu(n)=\{X\in\so(2n)\mid[J,X]=0\},
 \qquad
 \m=\{X\in\so(2n)\mid\{J,X\}=0\}.
 \label{eq:real-sector-definitions}
\end{equation}
To see why these are the appropriate conditions, identify
$V$ with $\mathbb C^n$, so that $Jz=\ii z$.  Every real-linear map has a
unique decomposition
\[
 Xz=H_Xz+K_X\bar z.
\]
Here $H_X$ and $K_X$ are complex matrices describing the complex-linear
and complex-antilinear parts, respectively.  Skew-adjointness with respect
to the real Euclidean metric
$\langle z,w\rangle_V=\operatorname{Re}(z^\dagger w)$ is equivalent to
\[
 H_X^\dagger=-H_X,
 \qquad
 K_X^T=-K_X.
\]
The action of $J$ then gives
\begin{align*}
 [X,J]z&=X(\ii z)-\ii Xz=-2\ii K_X\bar z,\\
 \{X,J\}z&=X(\ii z)+\ii Xz=2\ii H_Xz.
\end{align*}
Thus $[X,J]=0$ selects $K_X=0$, leaving precisely the anti-Hermitian
complex-linear generators of $\uu(n)$.  Conversely, $\{X,J\}=0$ selects
$H_X=0$, leaving the skew-adjoint complex-antilinear generators of $\m$.
The anticommutator here is a matrix condition in $\operatorname{End}(V)$,
not a second Lie bracket on $\so(2n)$.

Thus $\m$ is the canonical complement to $\uu(n)$ selected by the metric
and the complex structure $J$.  The orthogonality of these sectors with
respect to the trace inner product, together with the construction of the
corresponding projectors, is established in
Subsection~\ref{sec:fixed-projector-properties}.

Let us remark that equation~\eqref{eq:real-sector-definitions} characterizes two
pure sectors, not an either-or test for an arbitrary element.
Every $X$ decomposes as $X=X_{\uu}+X_{\m}$, and a mixed element can satisfy
neither condition.  In particular, $[X,J]\ne0$ only implies
$X_{\m}\ne0$, whereas $\{X,J\}=0$ requires $X_{\uu}=0$.
For example, for $n\ge2$, a nonzero $Y\in\m$ gives
$X=J+Y$ with $[X,J]\ne0$ but $\{X,J\}=-2 \, \mathbf1\ne0$.
A basis can be chosen adapted to the splitting, so that each basis
generator either commutes or anticommutes with $J$; arbitrary linear
combinations need not do either.

The complementary space $\m$ is not generally a Lie subalgebra.  Instead,
the two spaces form a symmetric pair:
\begin{equation}
 [\uu(n),\uu(n)]\subset\uu(n),\qquad
 [\uu(n),\m]\subset\m,\qquad
 [\m,\m]\subset\uu(n).
 \label{eq:symmetricpair}
\end{equation}
These relations control how the separate connection components contribute
to the curvature.

On the compact real algebra we use the positive, invariant inner product
\begin{equation}
 \langle X,Y\rangle=-\Tr(XY),
 \label{eq:compact-inner-product}
\end{equation}
where $\Tr$ denotes the trace in the defining real $2n$-dimensional
representation.  The central $\uu(1)$ direction of $\uu(n)$ is generated
by $J$.  Removing this direction leaves the traceless unitary algebra
$\su(n)$, where tracelessness refers to the complex $n$-dimensional defining
representation, and the orthogonal splitting becomes
\begin{equation}
 \so(2n)=\su(n)\oplus\mathbb RJ\oplus\m.
 \label{eq:fullsplit}
\end{equation}
We denote the orthogonal projections onto $\uu(n)$, $\m$, $\mathbb RJ$,
and $\su(n)$ by $\cP_{\uu}$, $\cP_{\m}$, $\cP_J$, and $\cP_{\su}$,
respectively.  Their explicit formulas and proofs are given in
Subsection~\ref{sec:fixed-projector-properties}.

The distinction between $U(n)$ and $SU(n)$ is important.  Let
$V^{1,0}\subset V_{\mathbb C}$ denote the $+\ii$ eigenspace of $J$.  A
compatible unit complex volume form is a nonzero tensor
\begin{equation}
 \Omega\in\Lambda^n(V^{1,0})^*
          =\Lambda^{n,0}V_{\mathbb C}^*
 \label{eq:complex-volume-form}
\end{equation}
The Hermitian metric $h$ induces a norm on
$\Lambda^{n,0}V_{\mathbb C}^*$.  In components, this norm is
\begin{equation}
 \|\Omega\|_h^2
 =
 \frac1{n!}
 h^{i_1\bar j_1}\cdots h^{i_n\bar j_n}
 \Omega_{i_1\cdots i_n}
 \overline{\Omega}_{\bar j_1\cdots\bar j_n}.
 \label{eq:complex-volume-norm}
\end{equation}
In a unitary complex coframe,
$\Omega=c\,z^1\wedge\cdots\wedge z^n$, so that
$\|\Omega\|_h^2=|c|^2$.  The unit-normalization condition
$\|\Omega\|_h=1$ therefore fixes the magnitude of the complex volume
form, while its phase remains part of the choice of $SU(n)$ structure.

The form $\Omega$
is an internal algebraic tensor on $V$, not a spacetime differential
form.  Under $g\in U(n)$ it transforms according to the complex
determinant,
\begin{equation}
 g^*\Omega=\det_{\mathbb C}(g)\,\Omega.
\end{equation}
Consequently,
\begin{equation}
 \operatorname{Stab}_{SO(2n)}(J,\Omega)
 =
 \{g\in U(n)\mid g^*\Omega=\Omega\}
 \simeq SU(n).
 \label{eq:stabilizer-J-Omega}
\end{equation}
Thus $J$ alone defines a $U(n)$ structure, whereas the pair $(J,\Omega)$
defines an $SU(n)$ structure.

In the models below, projecting the kinetic term onto $\su(n)$ makes only the $SU(n)$
vectors propagate, but an adjoint $J$-type Higgs alone still leaves a local
$U(n)$ invariance.  A further Higgs datum carrying central $U(1)$ charge is
required if one wants the vacuum stabilizer itself to be exactly $SU(n)$.

We can see that the grading defined by $J$ is analogous to the chirality grading in gravitational gauge theories. Indeed, the commuting-anticommuting splitting above shares a precise algebraic
mechanism with the Lorentz-coset grading used in MacDowell-Mansouri
gravity and 4d unconventional supersymmetry models \cite{MacDowell:1977jt,Alvarez:2021zsw}.
More generally, an invertible matrix $S$ whose square is a nonzero scalar
multiple of the identity induces the involution
\[
 \vartheta_S(X)=SXS^{-1},
 \qquad S^2=c\mathbf1\quad(c\ne0),
 \qquad \vartheta_S^2=\mathrm{Id}.
\]
Provided conjugation preserves the Lie algebra under consideration, its
$+1$ and $-1$ eigenspaces are respectively the generators commuting and
anticommuting with $S$.  Moreover,
$\vartheta_S([X,Y])=[\vartheta_S(X),\vartheta_S(Y)]$, so their brackets
obey the same even-even, even-odd, and odd-odd rules as
Eq.~\eqref{eq:symmetricpair}.  This is an ordinary Lie-algebra
$\mathbb Z_2$ grading: ``odd'' refers to the involution, not to fermionic
statistics, and all these Lie brackets remain commutators.

For example, in a four-dimensional Clifford representation with
$\gamma_5^2=\mathbf1$ and $\{\gamma_5,\gamma_a\}=0$, chirality implements
the involution $\vartheta_5(X)=\gamma_5X\gamma_5^{-1}$.  The conformal
generators may be represented, up to normalization and convention-dependent
signs, by
\[
 M_{ab}\propto[\gamma_a,\gamma_b],\qquad
 D\propto\gamma_5,\qquad
 P_a\propto\gamma_a(1+\gamma_5),\qquad
 K_a\propto\gamma_a(1-\gamma_5).
\]
Here $M_{ab}$ generate Lorentz transformations, $D$ generates dilatations,
and $P_a$ and $K_a$ generate translations and special conformal
transformations.  Thus
\[
 [\gamma_5,M_{ab}]=[\gamma_5,D]=0,
 \qquad
 \{\gamma_5,P_a\}=\{\gamma_5,K_a\}=0,
\]
and the conformal algebra $\so(4,2)$ has the grading
\[
 \g_+=\so(3,1)\oplus\mathbb R D,
 \qquad
 \g_-=\operatorname{span}\{P_a,K_a\}.
\]
On spinors, the even generators preserve chirality and the odd generators
exchange it.  This should be distinguished from the defining-space
interpretation of our $J$ splitting in terms of complex-linear and
complex-antilinear real maps.  Although $J^2=-\mathbf1$ whereas
$\gamma_5^2=+\mathbf1$, both squares are scalar, which is all that is
needed for involutive conjugation.

Standard MacDowell-Mansouri gravity uses the AdS or dS algebra, rather
than the full conformal algebra: its even sector is $\so(3,1)$ and its
odd sector consists of four translation-like coset generators, with no
independent special conformal generators $K_a$.  The simultaneous presence
of $P_a$ and $K_a$ therefore pertains to a conformal extension.  In the
compensator formulation developed in Section~\ref{sec:MM}, the covariant
version of this mechanism is conjugation by the Clifford image
$\slashed V$ of the compensator; in a Clifford representation adapted to
the Lorentz gauge, $\slashed V$ is proportional to $\gamma_5$. The associated projectors are displayed in Eq.~\eqref{eq:MM-involution-projectors}.

Finally, this common grading determines the algebraic splitting and its
projectors, not the gravitational dynamics by itself.  The
MacDowell-Mansouri action additionally uses a Lorentz internal dual,
represented in the spinor trace by a $\gamma_5$ insertion, together with
the first-order connection-coframe decomposition.  These extra structures,
and their distinction from the projected Yang-Mills kinetic term, are
explained in Section~\ref{sec:MM}.  The precise parallel is therefore a
shared involution mechanism, not an identification of the two theories.

\subsection{The defining representation and the central charge convention}
\label{sec:charge-conventions}

Charge eigenvectors are most naturally described after complexification.
Since $J^2=-\mathbf1$, its eigenvalues on $V_{\mathbb C}$ are $\pm\ii$,
and the defining representation splits into holomorphic and antiholomorphic
eigenspaces:
\begin{equation}
 V_{\mathbb C}=V^{1,0}\oplus V^{0,1},
 \qquad
 J|_{V^{1,0}}=+\ii,
 \qquad
 J|_{V^{0,1}}=-\ii.
 \label{eq:J-eigenspaces}
\end{equation}
Equivalently,
\begin{equation}
 V^{1,0}=\ker(J-\ii\mathbf1),
 \qquad
 V^{0,1}=\ker(J+\ii\mathbf1),
 \label{eq:charge-holomorphic-spaces}
\end{equation}
with vector-space projectors
\begin{equation}
 \Pi^{1,0}=\frac12(\mathbf1-\ii J),
 \qquad
 \Pi^{0,1}=\frac12(\mathbf1+\ii J).
 \label{eq:charge-holomorphic-projectors}
\end{equation}

As $U(n)$ representations, $V^{1,0}\simeq\mathbf n$ and
$V^{0,1}\simeq\overline{\mathbf n}$.  Because $J$ is anti-Hermitian, we
define the corresponding Hermitian central charge operator by
\begin{equation}
 Q=-\ii J,
 \qquad
 Q|_{\mathbf n}=+1,
 \qquad
 Q|_{\overline{\mathbf n}}=-1,
 \label{eq:charge-definition}
\end{equation}
For a vector $v_q$, defining its charge $q$ means any of the following
equivalent relations, with the last holding for every $\theta$:
\begin{equation}
 Qv_q=qv_q
 \quad\Longleftrightarrow\quad
 Jv_q=\ii q\,v_q
 \quad\Longleftrightarrow\quad
 e^{\theta J}v_q=e^{\ii q\theta}v_q.
 \label{eq:charge-defining-equivalences}
\end{equation}
Thus the central $U(1)$ subgroup of $U(n)$ acts by opposite phases on the two summands, and
our notation is
\begin{equation}
 V_{\mathbb C}=\mathbf n_{+1}\oplus
 \overline{\mathbf n}_{-1}.
 \label{eq:fundamental-branching}
\end{equation}
Since $e^{2\pi J}=\mathbf1$, this is the primitive normalization of the
embedded, $2\pi$-periodic central $U(1)$, with defining charges $\pm1$.
At the Lie-algebra level, rescaling $Q$ simply rescales all charge labels and for a fixed compact group, its periodicity must also be respected.
Replacing $J$ by $-J$ instead interchanges the two eigenspaces and reverses
all charge signs simultaneously.

\subsection{The adjoint representation and the charged coset components}
\label{sec:adjoint-branching}

The action of a generator depends on the representation.  On a defining
vector it acts by matrix multiplication, as in
Eq.~\eqref{eq:charge-defining-equivalences}; on an endomorphism or
adjoint-valued field it acts by commutator.  Thus an adjoint component $X_q$
has charge $q$ if the following equivalent relations hold:
\begin{align}
 [Q,X_q]=qX_q
 &\quad\Longleftrightarrow\quad
 -\ii[J,X_q]=qX_q
 \quad\Longleftrightarrow\quad
 [J,X_q]=\ii q\,X_q,
 \nonumber\\
 &\quad\Longleftrightarrow\quad
 e^{\theta J}X_qe^{-\theta J}=e^{\ii q\theta}X_q
 \quad\text{for all }\theta.
 \label{eq:charge-adjoint-equivalences}
\end{align}
The factor of $\ii$ is essential: $J$ is the anti-Hermitian compact
generator, whereas $Q$ is the Hermitian charge operator.

The appearance of antisymmetric tensor representations follows from the
orthogonal algebra itself.  By definition,
\begin{equation}
 \so(V)=\{X\in\operatorname{End}(V)\mid
 \langle Xu,v\rangle_V+\langle u,Xv\rangle_V=0\}.
 \label{eq:so-definition}
\end{equation}
Lowering the first index with the invariant metric turns $X^a{}_b$ into an
antisymmetric tensor $X_{ab}=-X_{ba}$.  Using the metric also to identify
$V_{\mathbb C}^*$ with $V_{\mathbb C}$ gives the canonical
$SO(2n)$-equivariant identification
\begin{equation}
 \so(2n,\mathbb C)\simeq\Lambda^2 V_{\mathbb C}.
 \label{eq:adjoint-twoform}
\end{equation}
Charges add in tensor products because the generator acts on each factor.
A tensor product containing $p$ factors from $V^{1,0}$ and $r$ from
$V^{0,1}$ therefore has charge $p-r$.  Equivalently, an endomorphism
mapping a charge-$r$ vector to a charge-$s$ vector has commutator charge
$s-r$.  Applying charge addition to
Eq.~\eqref{eq:fundamental-branching} gives
\begin{align}
 \Lambda^2 V_{\mathbb C}
 &=(\Lambda^2\mathbf n)_{+2}
 \oplus(\mathbf n\otimes\overline{\mathbf n})_0
 \oplus(\Lambda^2\overline{\mathbf n})_{-2},
 \nonumber\\
 \mathbf n\otimes\overline{\mathbf n}
 &=\operatorname{Adj}_{SU(n)}\oplus\mathbf1.
 \label{eq:wedge-derivation}
\end{align}
Thus the complexified adjoint branches as
\begin{equation}
 \operatorname{Adj}_{SO(2n)}\otimes\mathbb C
 =(\operatorname{Adj}_{SU(n)})_0\oplus\mathbf1_0
 \oplus(\Lambda^2\mathbf n)_{+2}
 \oplus(\Lambda^2\overline{\mathbf n})_{-2}.
 \label{eq:branching}
\end{equation}
The neutral mixed sector is $\uu(n)_{\mathbb C}$, and its singlet is the
complexified direction generated by $J$.  The two charged summands are
complex conjugates and together constitute the complexification of the
real $U(n)$-orthogonal complement:
\begin{equation}
 \m_{\mathbb C}\simeq
 (\Lambda^2\mathbf n)_{+2}\oplus
 (\Lambda^2\overline{\mathbf n})_{-2},
 \qquad
 \dim_{\mathbb R}\m=n(n-1).
 \label{eq:correctmodes}
\end{equation}
In particular,
\begin{equation}
 [J,X_0]=0,
 \qquad
 [J,X_\pm]=\pm2\ii X_\pm,
 \qquad
 [Q,X_\pm]=\pm2X_\pm,
 \label{eq:charge-coset-eigenvalues}
\end{equation}
where $X_0\in\uu(n)_{\mathbb C}$.  The orthogonal complement of $\su(n)$,
rather than of $\uu(n)$, also contains the real singlet $\mathbb RJ$ and
has dimension $n(n-1)+1$.

The charged representations are not generally
$\mathbf n\oplus\overline{\mathbf n}$.  For $n=3$, however, the
$SU(3)$ isomorphisms
$\Lambda^2\mathbf3\simeq\overline{\mathbf3}$ and
$\Lambda^2\overline{\mathbf3}\simeq\mathbf3$ give
\begin{equation}
 \mathbf6_{\mathbb C}
 =\mathbf3_{+1}\oplus\overline{\mathbf3}_{-1},
 \qquad
 \m_{\mathbb C}
 \simeq\overline{\mathbf3}_{+2}\oplus\mathbf3_{-2}.
 \label{eq:charge-SO6-U3}
\end{equation}
These are $SU(3)$ identifications, not identifications of the full
$U(3)$ representations: they do not change the independently assigned
central charge.  This is why the $\overline{\mathbf3}$ in the first coset
summand has charge $+2$, although the conjugate defining representation
has charge $-1$.

The decomposition into charges $\pm2$ is a decomposition of the
complexified space $\m_{\mathbb C}$, not of the real space $\m$ into two
independent complex sectors.  Indeed, for a real $X\in\m$, the defining
relation $\{J,X\}=0$ gives
\begin{equation}
 [J,X]=2JX,
 \qquad
 [J,[J,X]]=-4X.
 \label{eq:charge-real-coset}
\end{equation}
Thus $\ad_J^2=-4\mathbf1$ on $\m$.  Since the eigenvalues of $\ad_J$ are
therefore $\pm2\ii$, no nonzero real $X\in\m$ can itself be a charge
eigenvector.

After complexification, $X$ decomposes into the corresponding
eigenvectors
\begin{equation}
 X_+
 =\frac12X-\frac{\ii}{4}[J,X],
 \qquad
 X_-
 =\frac12X+\frac{\ii}{4}[J,X],
 \qquad
 X=X_++X_-.
 \label{eq:charge-coset-components}
\end{equation}
A direct calculation gives
\begin{equation}
 [J,X_\pm]=\pm2\ii X_\pm,
 \qquad
 [Q,X_\pm]=\pm2X_\pm.
\end{equation}
Hence $X_\pm$ belong to the charge-$\pm2$ eigenspaces of
$\m_{\mathbb C}$.  Because $J$ and $X$ are real,
\begin{equation}
 \overline{X_+}=X_-.
 \label{eq:coset-charge-reality}
\end{equation}
The two charged components are therefore complex conjugates describing
a single real element $X\in\m$, rather than independent complex fields.
The identity $\ad_J^2=-4\mathbf1$ on $\m$ will also underlie the explicit
projector onto $\m$ constructed in
Subsection~\ref{sec:fixed-projector-properties}.

\subsection{Adjoint fields, gauge transformations, and the vacuum}
\label{sec:adjoint-fields}

This branching is a decomposition of the Lie algebra, not of one particular
field.  It therefore applies to every adjoint-valued quantity.  In particular,
an adjoint scalar (a spacetime zero-form) and a connection (a spacetime
one-form) decompose respectively as
\begin{align}
 \Phi&=\Phi_{\su}+\phi J+\Phi_++\Phi_-,
 &\Phi_+&\in(\Lambda^2\mathbf n)_{+2},
 &\Phi_-&\in(\Lambda^2\overline{\mathbf n})_{-2},
 \label{eq:Phi-decomposition}\\
 A&=A_{\su}+aJ+W_++W_-,
 &W_+&\in\Omega^1\!\left(M,(\Lambda^2\mathbf n)_{+2}\right),
 &W_-&\in\Omega^1\!\left(M,(\Lambda^2\overline{\mathbf n})_{-2}\right).
 \label{eq:A-decomposition}
\end{align}
The scalar and connection nevertheless play different roles.  The scalar
transforms homogeneously, whereas the connection transforms affinely:
\begin{equation}
 \Phi\longmapsto g^{-1}\Phi g,
 \qquad
 A\longmapsto g^{-1}Ag+g^{-1}\dd g.
 \label{eq:Phi-A-transformations}
\end{equation}
This convention uses the inverse adjoint action.  For a central
transformation $g=e^{\theta J}$, a homogeneous adjoint component therefore
transforms as
\begin{equation}
 X_q\longmapsto g^{-1}X_qg=e^{-\ii q\theta}X_q.
 \label{eq:charge-inverse-adjoint}
\end{equation}
The inverse phase does not change the charge label defined by
Eq.~\eqref{eq:charge-adjoint-equivalences}, it reflects the choice of inverse
group action, or equivalently the sign of the gauge parameter.  For the
connection, $g^{-1}\dd g=J\,\dd\theta$ is central, so
\begin{equation}
 a\longmapsto a+\dd\theta,
 \qquad
 W_\pm\longmapsto e^{\mp2\ii\theta}W_\pm.
 \label{eq:central-connection-transformations}
\end{equation}

A vacuum $\langle\Phi\rangle=vJ$ selects the centralizer of $J$, namely
$U(n)$.  Accordingly $A_{\su}+aJ$ is the unbroken $U(n)$ connection, while
$W_\pm$ are the gauge fields along the broken directions.  Since
$[Q,W_\pm]=\pm2W_\pm$, the Higgs covariant derivative becomes
\begin{equation}
 D_A\langle\Phi\rangle=v[A,J]=v[W_++W_-,J].
 \label{eq:vacuum-covariant-derivative}
\end{equation}
In an ordinary Yang-Mills-Higgs theory, the Higgs kinetic term
therefore produces an algebraic mass quadratic form for $W_\pm$.
Together with the ordinary nondegenerate Yang-Mills kinetic term, this
makes $W_\pm$ propagating massive vector fields.  By contrast, in the
projected theories considered below, the same Lie-algebra splitting is
used to make the curvature kinetic operator degenerate.  When the kinetic
projector is $\cP_{\su}$, only the $\su(n)$ component of the curvature is
retained, while both its central $\mathbb RJ$ component and its $\m$
component are projected out.  Consequently, even when a Higgs kinetic
term supplies an algebraic quadratic term for $W_\pm$, these fields do
not possess a conventional vector kinetic term and are instead auxiliary
on the regular broken branch.  Notice also that projecting out the
central $J$ component is a choice of kinetic operator, not an additional
spontaneous breaking: the $J$-vacuum itself still has stabilizer $U(n)$.

\subsection{Explicit projectors and their orthogonality properties}
\label{sec:fixed-projector-properties}

We now construct the maps implementing Eq.~\eqref{eq:fullsplit} and verify
their properties.  A useful starting point is conjugation by $J$:
\begin{equation}
 \vartheta_J(X):=JXJ^{-1}=-JXJ,
 \qquad X\in\so(2n).
 \label{eq:theta-J-definition}
\end{equation}
Since $J^2=-\mathbf1$ and $J^{-1}=-J$, this map is an involution:
\begin{equation}
 \vartheta_J^2(X)=J\bigl(JXJ^{-1}\bigr)J^{-1}=X.
 \label{eq:theta-J-involution}
\end{equation}
Its $+1$ eigenspace is $\uu(n)$, while its $-1$ eigenspace is $\m$,
\begin{align}
 \mathfrak g_{+}
 :=&\ker(\vartheta_J-\mathbf 1)
 =\{X\in\mathfrak{so}(2n)\mid[J,X]=0\}
 =\mathfrak u(n),\\
 \mathfrak g_{-}
 :=&\ker(\vartheta_J+\mathbf 1)
 =\{X\in\mathfrak{so}(2n)\mid\{J,X\}=0\}
 =\mathfrak m .
\end{align}

Moreover, $\vartheta_J$ is a Lie-algebra automorphism,
\begin{equation}
 \vartheta_J([X,Y])
 =[\vartheta_J(X),\vartheta_J(Y)],
\end{equation}
so the eigenvalue of a bracket is the product of those of its two arguments.  This proves the symmetric-pair relations in Eq.~\eqref{eq:symmetricpair}.

These eigenspaces are orthogonal with respect to the trace inner product
defined in Eq.~\eqref{eq:compact-inner-product}.  First, conjugation by
$J$ preserves this inner product:
\begin{align}
 \langle\vartheta_J(X),\vartheta_J(Y)\rangle
 &=-\Tr\!\left(
   JXJ^{-1}JYJ^{-1}
  \right)
 \nonumber\\
 &=-\Tr\!\left(JXYJ^{-1}\right)
 =-\Tr(XY)
 =\langle X,Y\rangle .
 \label{eq:theta-J-trace-isometry}
\end{align}
Now let $X_{+}\in\g_{+}$ and $X_{-}\in\g_{-}$.  Using
$\vartheta_J(X_{+})=X_{+}$ and
$\vartheta_J(X_{-})=-X_{-}$, one obtains
\begin{align}
 \langle X_{+},X_{-}\rangle
 &=\langle
   \vartheta_J(X_{+}),\vartheta_J(X_{-})
  \rangle
 \nonumber\\
 &=\langle X_{+},-X_{-}\rangle
 =-\langle X_{+},X_{-}\rangle .
\end{align}
Consequently,
\begin{equation}
 \langle X_{+},X_{-}\rangle=0,
 \qquad
 \m=\uu(n)^\perp .
 \label{eq:theta-J-eigenspace-orthogonality}
\end{equation}
Thus the commuting and anticommuting sectors are not merely
complementary: they are the mutually orthogonal eigenspaces of the
orthogonal involution $\vartheta_J$.

The eigenspace projectors of an involution are obtained by adding or
subtracting $\vartheta_J$ from the identity.  Hence
\begin{equation}
 \cP_{\uu}=\frac12(\mathbf1+\vartheta_J),
 \qquad
 \cP_{\m}=\frac12(\mathbf1-\vartheta_J),
 \label{eq:projectors-from-involution}
\end{equation}
or, explicitly,
\begin{equation}
 \cP_{\uu}X=\frac12(X-JXJ),
 \qquad
 \cP_{\m}X=\frac12(X+JXJ).
 \label{eq:projectors-explicit}
\end{equation}
The same maps can be expressed through a double commutator.  Indeed,
\begin{equation}
 [J,[J,X]]=J^2X-2JXJ+XJ^2=-2(X+JXJ),
 \label{eq:projector-double-commutator}
\end{equation}
which gives the formulas used subsequently:
\begin{equation}
 \cP_{\m}(J)X=-\frac14[J,[J,X]],
 \qquad
 \cP_{\uu}(J)X=X+\frac14[J,[J,X]].
 \label{eq:PuPm}
\end{equation}
In particular, $\cP_{\m}$ acts as the identity on the real coset sector,
because Eq.~\eqref{eq:charge-real-coset} gives
$[J,[J,X]]=-4X$ there; it vanishes on $\uu(n)$.

Using $\vartheta_J^2=\mathbf1$, we obtain idempotence, vanishing products,
and completeness:
\begin{align}
 \cP_{\uu}^2
 &=\frac14(\mathbf1+\vartheta_J)^2=\cP_{\uu},
 &
 \cP_{\m}^2
 &=\frac14(\mathbf1-\vartheta_J)^2=\cP_{\m},
 \label{eq:projector-idempotence-u-m}\\
 \cP_{\uu}\cP_{\m}
 &=\cP_{\m}\cP_{\uu}=0,
 &
 \cP_{\uu}+\cP_{\m}&=\mathbf1.
 \label{eq:projector-completeness-u-m}
\end{align}
Their images are characterized by
\begin{equation}
 [J,\cP_{\uu}X]=0,
 \qquad
 \{J,\cP_{\m}X\}=0.
 \label{eq:projector-images-u-m}
\end{equation}
Conjugation preserves the trace inner product, and an orthogonal involution
is self-adjoint.  Thus
\begin{equation}
 \langle\vartheta_JX,Y\rangle=\langle X,\vartheta_JY\rangle,
 \qquad
 \langle\cP_sX,Y\rangle=\langle X,\cP_sY\rangle,
 \quad s=\uu,\m.
 \label{eq:projector-self-adjointness-u-m}
\end{equation}
The two images are therefore orthogonal, not merely complementary.  This
self-adjointness will also allow the projectors to be moved between the
two slots of the gauge kinetic bilinear when varying the action.

To isolate $\su(n)$, we next remove the central direction.  Its norm and
orthogonal projector are
\begin{equation}
 \langle J,J\rangle=-\Tr(J^2)=2n,
 \qquad
 \cP_JX=\frac{\langle J,X\rangle}{2n}J
       =-\frac{\Tr(JX)}{2n}J.
 \label{eq:central-projector-normalization}
\end{equation}
Because $\cP_JJ=J$ and $J\in\uu(n)$, one has
\begin{equation}
 \cP_J^2=\cP_J,
 \qquad
 \cP_{\uu}\cP_J=\cP_J\cP_{\uu}=\cP_J,
 \qquad
 \cP_J\cP_{\m}=\cP_{\m}\cP_J=0.
 \label{eq:central-projector-properties}
\end{equation}
The projector onto the traceless unitary sector is consequently
\begin{equation}
 \cP_{\su}(J)=\cP_{\uu}(J)-\cP_J.
 \label{eq:Psu}
\end{equation}
Subtracting an arbitrary projector from another would not generally
produce a projector.  It works here because $\mathbb RJ$ is a subspace of
$\uu(n)$, as encoded in Eq.~\eqref{eq:central-projector-properties}:
\begin{equation}
 \cP_{\su}^2
 =\cP_{\uu}^2-\cP_{\uu}\cP_J-\cP_J\cP_{\uu}+\cP_J^2
 =\cP_{\uu}-\cP_J
 =\cP_{\su}.
 \label{eq:su-projector-idempotence}
\end{equation}
The maps $\cP_J$ and $\cP_{\su}$ are also self-adjoint.  Together with
$\cP_{\m}$ they form the mutually orthogonal, complete system
\begin{align}
 \cP_{\su}\cP_J&=\cP_J\cP_{\su}=0,
 &
 \cP_{\su}\cP_{\m}&=\cP_{\m}\cP_{\su}=0,
 \nonumber\\
 \cP_J\cP_{\m}&=\cP_{\m}\cP_J=0,
 &
 \cP_{\su}+\cP_J+\cP_{\m}&=\mathbf1_{\so(2n)}.
 \label{eq:three-projectors}
\end{align}
Their images are precisely $\su(n)$, $\mathbb RJ$, and $\m$,
respectively.  Thus the algebraic splitting, the representation branching,
and the kinetic projection are three descriptions of the same sector
decomposition.  Inserting $\cP_{\su}$ into the action retains its image
and removes the directions $\mathbb RJ\oplus\m$ from the bare curvature
kinetic operator.

%%%%%%%%%%%%%%%%%%%%%%%%%%%%%%%%%%%%%%%%%%%%%%%%%%
\section{Theory I: fixed kinetic projector}
\label{sec:fixed}
%%%%%%%%%%%%%%%%%%%%%%%%%%%%%%%%%%%%%%%%%%%%%%%%%%

In this section we use the projector $\cP_{\su}$ defined in
Eq.~\eqref{eq:Psu} to construct a fixed projected Yang-Mills theory.
The tensors defining the $SU(n)$ structure are treated as nondynamical
background data, and hence the projector is fixed throughout the
configuration space.  The resulting action retains only the
$\su(n)$-projected component of the curvature, while the complementary
$\mathbb RJ\oplus\m$ directions lie in the kernel of the bare kinetic
operator.  We analyze the residual gauge symmetry, field equations, and
canonical structure of this theory before promoting the projector to a
field-dependent one in Section~\ref{sec:dynamical}.

%%%%%%%%%%%%%%%%%%%%%%%%%%%%%%%%%%%%%%%%%%%%%%%%%%
\subsection{Action and residual symmetry}
%%%%%%%%%%%%%%%%%%%%%%%%%%%%%%%%%%%%%%%%%%%%%%%%%%

Let $A\in\Omega^1(M,\so(2n))$ and
\begin{equation}
 F=\dd A+A\wedge A.
\end{equation}

Fix the orthogonal complex structure $J$ and consider
\begin{equation}
 S_{\rm fix}[A]
 =-\frac1{2g^2}\int_M
 \left\langle
 \cP_{\su}(J)F,\ast\cP_{\su}(J)F
 \right\rangle .
 \label{eq:fixedaction}
\end{equation}
Fixing the background pair $(J,\Omega)$ reduces its stabilizer from
$SO(2n)$ to $SU(n)$.  The action above, however, depends on these data
only through $\cP_{\su}(J)$, which is constructed from $J$ and the
invariant trace pairing.  Although the image of this projector is
$\su(n)$, it is preserved by the full stabilizer of $J$, namely $U(n)$.
The action therefore retains an accidental central $U(1)$ invariance,
while the corresponding central vector has zero bare kinetic coefficient.
An $\Omega$-dependent tensor or operator must enter the action if its
residual symmetry is to be reduced to exactly $SU(n)$.

The distinction between the $U(n)$ symmetry of the action and the
$\su(n)$ image of its kinetic projector becomes transparent at the level
of the connection components.  Using the orthogonal splitting
\begin{equation}
\so(2n)=\su(n)\oplus\mathbb RJ\oplus\m,
\end{equation}
we decompose the connection as
\begin{equation}
A=B+cJ+X,
\qquad
B\in\Omega^1(M,\su(n)),
\quad
c\in\Omega^1(M),
\quad
X\in\Omega^1(M,\m).
\label{eq:Adecomp}
\end{equation}
Here $B$ is the connection along the propagating $\su(n)$ sector, $cJ$
is the connection along the central $U(1)$ direction, and $X$ contains
the components along the complementary coset directions. 

The full curvature is
\begin{equation}
 F = F_B+\dd c\,J+D_BX+c\wedge[J,X]+X\wedge X   = \cP_{\su}F + \cP_JF + \cP_{\m}F,
 \label{eq:full-curvature-fixed-decomposition}
\end{equation}
where
\begin{equation}
 F_B:=\dd B+B\wedge B,
 \qquad
 D_BX:=\dd X+B\wedge X+X\wedge B.
\end{equation}
The terms involving both $B$ and $cJ$ vanish because $[B,J]=0$, while
$cJ\wedge cJ=0$.  Moreover, the symmetric-pair relation
$[\m,\m]\subset\uu(n)$ implies
\begin{equation}
 X\wedge X
 =
 \cP_{\su}(X\wedge X)+\cP_J(X\wedge X).
\end{equation}

The effect of
the kinetic projection on these three sectors can now be seen explicitly.
Using Eq.~\eqref{eq:symmetricpair}, the projected curvature components are
\begin{align}
\cP_{\su}F
&=F_B+\cP_{\su}(X\wedge X)
\equiv K,
\label{eq:Kfixed}\\
\cP_J F
&=\dd c \, J+\cP_J(X\wedge X),\\
\cP_{\m}F
&=D_BX+c[J,X].
\end{align}
Thus $\dd c$ and $D_BX$ are absent from
Eq.~\eqref{eq:fixedaction}.  In fact, the central field $c_\mu$ drops out
of the action completely, consistently with the accidental central
$U(1)$ invariance, whereas $X_\mu$ survives only algebraically through
$X\wedge X$.

The case $n=2$ is exceptional. The selected $\su(2)$ is one
of the two commuting ideals of $\so(4)$, while
$\mathbb RJ\oplus\m$ constitutes the other. Hence
\[
 [\su(2),\m]=0,
 \qquad
 \cP_{\su}[X,X]=0,
 \qquad
 K=F_B.
\]
The fixed action is therefore independent of both $c$ and $X$.
Since the adjoint action of $SO(4)$ preserves each ideal,
the projector remains equivariant under the full connected
$SO(4)$ gauge group, despite being fixed.

Arbitrary shifts of the complementary connection $cJ+X$
are exact nonlinear symmetries. Its vanishing canonical
momenta are first-class constraints, with no associated
nontrivial secondary constraints. The coset constraint
matrix consequently vanishes identically throughout phase
space; its vanishing at the vacuum does not signal a change
of rank. The propagating sector is locally ordinary
$SU(2)$ Yang-Mills theory, with six physical degrees of freedom.
The subsequent discussion of explicit symmetry breaking
and vacuum irregularity in the fixed theory assumes $n\ge3$.

%%%%%%%%%%%%%%%%%%%%%%%%%%%%%%%%%%%%%%%
\subsection{Field equations}
%%%%%%%%%%%%%%%%%%%%%%%%%%%%%%%%%%%%%%%

Since $\cP_{\su}$ is fixed,
\begin{equation}
\delta K=\cP_{\su}\delta F
=\cP_{\su}D_A\delta A.
\end{equation}
Using the self-adjointness of $\cP_{\su}$, the fact that it acts only
on the internal indices and hence commutes with the spacetime Hodge
operator, and $\cP_{\su}K=K$, one obtains
\begin{align}
\delta S_{\rm fix}
&=-\frac1{g^2}\int_M
\left\langle
\cP_{\su}D_A\delta A,\ast K
\right\rangle
\nonumber\\
&=-\frac1{g^2}\int_M
\left\langle
D_A\delta A,\ast K
\right\rangle
=-\frac1{g^2}\int_M
\left\langle
\delta A,D_A\ast K
\right\rangle,
\end{align}
up to a boundary term.  Since $\delta A$ is an arbitrary
$\so(2n)$-valued one-form, the full equation of motion is
\begin{equation}
D_A\ast K=0.
\label{eq:fullfixedeom}
\end{equation}

To decompose this equation, first note that
$K\in\Omega^2(M,\su(n))$ and that the Hodge operator does not change its
internal Lie-algebra component.  Using \eqref{eq:Adecomp}, one therefore finds
\begin{equation}
D_A\ast K = D_B\ast K +c\wedge[J,\ast K] +[X,\ast K] = D_B\ast K+[X,\ast K],
\label{eq:fixed-eom-decomposition}
\end{equation}
because $[J,K]=0$.  The symmetric-pair relations
in Eq.~\eqref{eq:symmetricpair} imply
\begin{equation}
D_B\ast K\in\Omega^{d-1}(M,\su(n)),
\qquad
[X,\ast K]\in\Omega^{d-1}(M,\m).
\end{equation}
These terms belong to distinct, orthogonal
summands of the decomposition
$\so(2n)=\su(n)\oplus\mathbb RJ\oplus\m$.
Consequently, Eq.~\eqref{eq:fullfixedeom} is equivalent to
\begin{align}
D_B\ast K&=0,
\label{eq:Beomfixed}\\
[X,\ast K]&=0.
\label{eq:Xeomfixed}
\end{align}
Moreover, $D_A\ast K$ has no component along $\mathbb RJ$, and $c$
appears in neither Eq.~\eqref{eq:Beomfixed} nor
Eq.~\eqref{eq:Xeomfixed}. This reflects the fact that
$S_{\rm fix}$ is independent of $c$; accordingly, variation with respect
to $c$ yields the identity $0=0$.

Equation~\eqref{eq:Xeomfixed} is algebraic in $X$.
Appendix~\ref{app:X-variation} gives a direct derivation of the $X$ equation
and explains why, before using the symmetric-pair commutation relations, it
is naturally written as the projected equation
$\cP_{\m}[X,\ast K]=0$. The branch
\begin{equation}
 X=0
 \label{eq:Xzero}
\end{equation}
is an exact consistent truncation, not merely a perturbative or weak-field
approximation.  More precisely, the $X$ equation is then identically
satisfied, while Eq.~\eqref{eq:Beomfixed} becomes the ordinary $SU(n)$
Yang-Mills equation
\begin{equation}
 D_B\ast F_B=0.
\end{equation}
Consequently, every on-shell $SU(n)$ Yang-Mills connection $B$ lifts to an
exact solution $(B,c,X=0)$ of the full projected theory, with $c$ arbitrary.
The statement does not mean that an arbitrary off-shell $B$ becomes a
solution simply by setting $X=0$.

There are also exact branches with $X\neq0$.  In general they obey the coupled
system
\begin{equation}
 D_B\ast\left(F_B+\cP_{\su}(X\wedge X)\right)=0,
 \qquad
 \left[X,\ast\left(F_B+\cP_{\su}(X\wedge X)\right)\right]=0.
 \label{eq:nonzero-X-system}
\end{equation}
A particularly transparent class is obtained by imposing the stronger
condition
\begin{equation}
 K=F_B+\cP_{\su}(X\wedge X)=0.
 \label{eq:projected-flat-branch}
\end{equation}
Both field equations are then satisfied identically.  For example, on a
trivial local patch one may take
\begin{equation}
 B=0,
 \qquad
 X=T\,\xi,
 \qquad T\in\m,
 \label{eq:rank-one-X-solution}
\end{equation}
where $T$ is fixed and $\xi$ is an arbitrary one-form.  Since
$X\wedge X=0$, this configuration has $K=0$ and therefore solves the
projected equations for any $\xi$.  Nevertheless its full curvature is
generically nonzero,
\begin{equation}
 F(A)=\dd c\,J+T\,\dd\xi+c[J,T]\wedge\xi,
 \label{eq:invisible-full-curvature}
\end{equation}
where an arbitrary $c$ may be retained; setting $c=0$ gives
$F(A)=T\,\dd\xi$.  Thus the fixed-projector theory is insensitive to the
$\m$-valued curvature $D_BX+c[J,X]$.  The nonzero-$X$ solutions include large
degenerate families with vanishing projected curvature and zero action, even
though they are not, in general, solutions of ordinary $SO(2n)$ Yang-Mills
theory.  This degeneracy is the covariant counterpart of the irregular
constraint structure encountered below.

Degenerate gauge-kinetic operators have a direct precedent in the work of
Grundberg, Lindstr\"om, and Ro\v{c}ek, who considered semidefinite kinetic
terms for which gauge fields associated with a subgroup or a coset become
auxiliary \cite{Grundberg:1996Cosets}.  This precedes the general mechanism
used here, but their scalar-dependent supersymmetric construction is not the
fixed $\cP_{\su}(J)$ theory considered above and does not isolate the
specific branch $K=0$, $F(A)\neq 0$.

MacDowell-Mansouri gravity provides a particularly close gravitational
precedent for the underlying algebraic structure
\cite{Wise:2009fu}; see Section~\ref{sec:MM}.  Indeed, under the reductive
decomposition of the $(A)dS$ connection,
\begin{equation*}
\mathcal A=\omega+\frac1\ell e,
\qquad
\mathcal F_{\rm L}
=R+\frac{\sigma}{\ell^2}e\wedge e,
\qquad
\mathcal F_{\m}=\frac1\ell T,
\qquad
T=D_\omega e,
\end{equation*}
the MacDowell-Mansouri action is constructed from the Lorentz-projected
curvature $\mathcal F_{\rm L}$, together with the internal Lorentz dual,
whereas the complementary torsion curvature $\mathcal F_{\m}$ is not
assigned an independent Yang-Mills kinetic term.  The structural
correspondence is therefore
\begin{equation*}
B\longleftrightarrow\omega,
\qquad
X\longleftrightarrow\frac1\ell e,
\qquad
D_BX\longleftrightarrow\frac1\ell T,
\qquad
\cP_{\su}(X\wedge X)
\longleftrightarrow
\frac{\sigma}{\ell^2}e\wedge e.
\end{equation*}
In particular, the rank-one ansatz $X=T \,\xi$ has the algebraic analogue
$e^a=v^a\xi$, for which $e^a\wedge e^b=0$.  Such a coframe is necessarily
degenerate, however, and this analogy does not imply that an arbitrary
$\xi$ solves the gravitational equations.  On the nondegenerate-coframe branch, however, the
MacDowell-Mansouri connection equation implies vanishing torsion,
$T^a=0$.  This conclusion is not universal across first-order gravity.
Degenerate coframes can support branches with nonvanishing torsion
\cite{Kaul:2016Degenerate}, while related conformal-supergravity models
admit torsionful black-hole and gravitational-instanton solutions
\cite{Alvarez:2022wcj,Alvarez:2023auf}.  Together, these examples provide
a broader gravitational precedent for nontrivial torsional curvature
sectors beyond the ordinary nondegenerate MacDowell-Mansouri branch.

To the best of our knowledge, the particular fixed-projector family
\eqref{eq:projected-flat-branch}-\eqref{eq:invisible-full-curvature},
especially the solutions $X=T\, \xi$ with arbitrary $\xi$, has not previously
been isolated in this form.  Its degeneracy is the covariant counterpart of
the irregular constraint structure encountered below.

%%%%%%%%%%%%%%%%%%%%%%%%%%%%%%%%%%%%%%%%%%%%%
\subsection{Gauge invariance}
%%%%%%%%%%%%%%%%%%%%%%%%%%%%%%%%%%%%%%%%%%%%%
It is useful to distinguish four logically different statements: closure of
the transformation algebra, invariance of the action, invariance restricted
to a field-space locus, and preservation of that locus.  These notions do not
coincide in the fixed-projector theory.

Let $H\subset SO(2n)$ denote the stabilizer of the fixed kinetic
projector. For $\cP_{\su}$ constructed from $J$ and the invariant
inner product, its identity component is
\[
 H^\circ =
 \begin{cases}
 U(n), & n\ge3,\\
 SO(4), & n=2.
 \end{cases}
\]
For $n\ge3$, although the projector image is $\su(n)$,
conjugation by the central $U(1)$ preserves that image
and commutes with the projector. The stabilizer is reduced to $SU(n)$ only if an $\Omega$-dependent tensor or
operator actually enters the action.  Thus, for $n\ge3$, the fixed projector explicitly reduces
the connected group of ordinary gauge transformations
from $SO(2n)$ to $U(n)$.

Indeed, under the usual infinitesimal connection transformation
\begin{equation}
 \delta_\epsilon A=D_A\epsilon,
 \qquad
 \delta_\epsilon F=[F,\epsilon],
 \label{eq:fixed-gauge-transformation}
\end{equation}
the action varies as
\begin{equation}
 \delta_\epsilon S_{\rm fix}
 =-\frac1{g^2}\int_M
 \left\langle
 \cP_{\su}[F,\epsilon],\ast K
 \right\rangle .
 \label{eq:fixed-gauge-variation}
\end{equation}
If $\epsilon\in\operatorname{Lie}(H)$, equivariance of the fixed projector gives
\begin{equation}
 \cP_{\su}[F,\epsilon]=[\cP_{\su}F,\epsilon]=[K,\epsilon],
\end{equation}
and Eq.~\eqref{eq:fixed-gauge-variation} vanishes by invariance of the inner
product.  For a parameter $\xi\in\m$, by contrast, the symmetric-pair
relations imply
\begin{equation}
 \cP_{\su}[F,\xi]
 =\cP_{\su}[F_{\m},\xi],
 \qquad
 F_{\m}=D_BX+c[J,X],
 \label{eq:broken-projected-variation}
\end{equation}
and therefore
\begin{equation}
 \delta_\xi S_{\rm fix}
 =-\frac1{g^2}\int_M
 \left\langle
 \cP_{\su}[F_{\m},\xi],\ast K
 \right\rangle .
 \label{eq:broken-action-variation}
\end{equation}
This is generically nonzero. Indeed, since $[F_{\m},\xi] \in \mathfrak{u}(n)$ and $K \in \mathfrak{su}(n)$, we can remove the projector in \eqref{eq:broken-action-variation} and use the invariance of the bilinear form to write
\begin{equation}
 \delta_\xi S_{\rm fix}
 =\frac1{g^2}\int_M
 \left\langle \xi,[F_{\m},\ast K]
 \right\rangle .
 \label{eq:broken-action-variation2}
\end{equation}
The fixed projector therefore breaks the
$\m$ transformations explicitly.

There is no contradiction with the fact that the transformations
\eqref{eq:fixed-gauge-transformation} obey the full algebra
\begin{equation}
 [\delta_{\epsilon_1},\delta_{\epsilon_2}]A
 =\delta_{[\epsilon_1,\epsilon_2]}A
 \label{eq:kinematic-gauge-algebra}
\end{equation}
without terms proportional to the equations of motion.  Equation
\eqref{eq:kinematic-gauge-algebra} is a kinematical property of transformations
on a connection; it does not imply that every such transformation is a
symmetry of the chosen action.  The gauge-symmetry algebra consists only of
the parameters that preserve the kinetic projector plus a central one-form shift symmetry.

The same distinction can be seen directly at the level of the field
equation.  Let
\begin{equation}
 \mathcal E(A):=D_A\ast\cP_{\su}F=D_A\ast K.
 \label{eq:fixed-connection-equation}
\end{equation}
We now derive its gauge variation explicitly.  For an arbitrary
$\so(2n)$-valued zero-form parameter $\eta$, Eqs.~\eqref{eq:fixed-gauge-transformation}
give
\begin{equation}
 \delta_\eta A=D_A\eta,
 \qquad
 \delta_\eta F=[F,\eta].
 \label{eq:general-fixed-gauge-transformation}
\end{equation}
The fixed projector itself does not vary.  Since the Hodge operator acts only
on spacetime indices, it commutes with the internal projector and gauge
variation.  Varying both the covariant derivative and its argument therefore
gives,
\begin{align}
 \delta_\eta\mathcal E
 &=\delta_\eta\!\left(D_A\ast K\right)
 \nonumber\\
 &=[\delta_\eta A,\ast K]+D_A\ast\delta_\eta K
 \nonumber\\
 &=[D_A\eta,\ast K]+D_A\ast\cP_{\su}[F,\eta].
 \label{eq:direct-eom-variation-fixed}
\end{align}
To isolate the covariant part, add and subtract
$D_A\ast[K,\eta]$:
\begin{align}
 \delta_\eta\mathcal E
 ={}&[D_A\eta,\ast K]+D_A\ast[K,\eta]
 \nonumber\\
 &+D_A\ast\left(\cP_{\su}[F,\eta]-[K,\eta]\right).
 \label{eq:add-subtract-covariant-part}
\end{align}
Because $\ast K$ is a two-form and $\eta$ is a zero-form, the graded Leibniz
rule gives
\begin{equation}
 D_A[\ast K,\eta]
 =[D_A\ast K,\eta]-[D_A\eta,\ast K].
 \label{eq:graded-Leibniz-fixed-eom}
\end{equation}
The first line of Eq.~\eqref{eq:add-subtract-covariant-part} is consequently
$[D_A\ast K,\eta]=[\mathcal E,\eta]$.  Hence
\begin{align}
 \delta_\eta\mathcal E
 &=[\mathcal E,\eta]+D_A\ast\Delta_\eta(F),
 \label{eq:eom-transformation-fixed}\\
 \Delta_\eta(F)
 &: =\cP_{\su}[F,\eta]-[K,\eta].
 \label{eq:projector-equivariance-defect}
\end{align}
Thus
\begin{equation}
 \Delta_\eta(F)=\cP_{\su}[F,\eta]-[\cP_{\su}F,\eta]  
 =  -[\cP_{\su},\operatorname{ad}_\eta]F
\end{equation}
measures the failure of the fixed projector to commute with the infinitesimal
adjoint action generated by $\eta$.  Equivalently, it measures the failure of
projecting and gauge-transforming to be interchangeable operations.
Consider first $\epsilon\in\uu(n)$.  Decompose the curvature according to
Eq.~\eqref{eq:fullsplit} as
\begin{equation}
 F=K+F_J+F_{\m},
 \qquad
 F_J:=\cP_JF.
 \label{eq:curvature-fullsplit-gauge-variation}
\end{equation}
Since $[K,\epsilon]\in\su(n)$, $[F_J,\epsilon]=0$, and
$[F_{\m},\epsilon]\in\m$, one finds
\begin{align}
 \cP_{\su}[F,\epsilon]
 &=\cP_{\su}\bigl([K,\epsilon]+[F_J,\epsilon]
                    +[F_{\m},\epsilon]\bigr)
 \nonumber\\
 &=[K,\epsilon].
 \label{eq:unbroken-projector-equivariance}
\end{align}
Therefore $\Delta_\epsilon(F)=0$, and
\begin{equation}
 \delta_\epsilon\mathcal E=[\mathcal E,\epsilon],
 \qquad \epsilon\in\uu(n).
 \label{eq:unbroken-eom-covariance}
\end{equation}
In particular, local $U(n)$ transformations map solutions of
$\mathcal E=0$ into solutions; the terms containing $D_A\epsilon$ have
cancelled by Eq.~\eqref{eq:graded-Leibniz-fixed-eom}.

For a broken parameter $\xi\in\m$, the same decomposition instead gives
\begin{equation}
 [K,\xi]\in\m,
 \qquad
 [F_J,\xi]\in\m,
 \qquad
 [F_{\m},\xi]\in\uu(n),
 \label{eq:broken-commutator-sectors}
\end{equation}
and hence
\begin{align}
 \cP_{\su}[F,\xi]
 &=\cP_{\su}\bigl([K,\xi]+[F_J,\xi]
                    +[F_{\m},\xi]\bigr)
 \nonumber\\
 &=\cP_{\su}[F_{\m},\xi].
 \label{eq:broken-projector-step-by-step}
\end{align}
Notice that $\cP_{\su}[F_{\m},\xi]\in\su(n)$, whereas
$[K,\xi]\in\m$; these terms cannot be identified.  Substitution into
Eq.~\eqref{eq:direct-eom-variation-fixed} gives the direct form
\begin{equation}
 \delta_\xi\mathcal E
 =[D_A\xi,\ast K]
 +D_A\ast\cP_{\su}[F_{\m},\xi],
 \label{eq:broken-eom-variation-direct}
\end{equation}
or, equivalently, the covariant-plus-obstruction form \eqref{eq:eom-transformation-fixed}
\begin{equation}
 \delta_\xi\mathcal E
 =[\mathcal E,\xi]
 +D_A\ast\left(\cP_{\su}[F_{\m},\xi]-[K,\xi]\right),
 \qquad \xi\in\m.
 \label{eq:broken-eom-variation-obstruction}
\end{equation}
Thus the equivariance defect is
\begin{equation}
 \Delta_\xi(F)=\cP_{\su}[F_{\m},\xi]-[K,\xi],
 \label{eq:broken-equivariance-defect-explicit}
\end{equation}
which is generically nonzero.  Even on $\mathcal E=0$ one therefore has
\begin{equation}
 \left.\delta_\xi\mathcal E\right|_{\mathcal E=0}
 =D_A\ast\left(\cP_{\su}[F_{\m},\xi]-[K,\xi]\right)
 \neq0.
 \label{eq:broken-eom-on-shell-noncovariance}
\end{equation}
Consequently, the broken transformations are not generic on-shell gauge
symmetries. Indeed, an on-shell symmetry must be tangent to the space of
solutions: if $A$ satisfies $\mathcal E[A]=0$, then
$A+\delta_\xi A$ must also satisfy the field equations to first order, which
requires $\left.\delta_\xi\mathcal E\right|_{\mathcal E=0}=0$. This condition fails in
Eq.~\eqref{eq:broken-eom-on-shell-noncovariance}.

Equation~\eqref{eq:broken-action-variation} does show a weaker, conditional
invariance on the locus
\begin{equation}
 F_{\m}=0:
 \qquad
 \left.\delta_\xi S_{\rm fix}\right|_{F_{\m}=0}=0.
 \label{eq:conditional-invariance}
\end{equation}
This locus is neither the general solution of the projected equations nor a
constraint implied by them.  Equations~\eqref{eq:Beomfixed} and
\eqref{eq:Xeomfixed} constrain $K$ but do not require $F_{\m}=0$; in
particular, the projected-flat solutions~\eqref{eq:projected-flat-branch} may
have $F_{\m}\neq0$.

Moreover, a broken transformation is not generally tangent to the locus
$F_{\m}=0$.  Since $\delta_\xi F=[F,\xi]$, on that locus
\begin{equation}
 \left.\delta_\xi F_{\m}\right|_{F_{\m}=0}
 =\cP_{\m}[F_{\uu},\xi]
 =[F_{\uu},\xi],
 \label{eq:constraint-not-preserved}
\end{equation}
where $F_{\uu}=K+F_J$.  A parameter preserves this locus at a given
background only if it belongs pointwise to the common centralizer
\begin{equation}
 \mathfrak z_{\m}(F_{\uu})
 :=
 \left\{
 \xi\in\m\ \middle|\
 [F_{\uu\,\mu\nu},\xi]=0
 \text{ for every }\mu,\nu
 \right\}.
\end{equation}
For generic $F_{\uu}$ this space is trivial.  It can be nonzero on special
backgrounds, for example when $F_{\uu}=0$ or when the curvature takes values
in a proper subalgebra of $\uu(n)$ possessing invariant vectors in $\m$.
Such parameters are background-dependent stabilizers and do not define a
broken gauge symmetry on the full locus $F_{\m}=0$.  Hence
Eq.~\eqref{eq:conditional-invariance} expresses an infinitesimal conditional
vanishing, not a symmetry of a theory constrained by $F_{\m}=0$.

The failure also appears at finite order.  Starting from $F_{\m}=0$,
\begin{equation}
 e^{-t\xi}F_{\uu}e^{t\xi}
 =F_{\uu}+t[F_{\uu},\xi]
 +\frac{t^2}{2}[[F_{\uu},\xi],\xi]+O(t^3).
 \label{eq:finite-broken-transformation}
\end{equation}
The first-order term lies in $\m$ and is removed by the fixed projector, but
the second-order term returns to $\uu(n)$ and generically changes the
projected action.  The conditional first-order invariance therefore does not
exponentiate to a finite broken gauge symmetry.

The situation should be distinguished from the familiar ``on-shell symmetry''
mechanism in MacDowell-Mansouri formulations of supergravity.  In those
theories, the obstruction to a broken transformation, most notably
supersymmetry, is controlled by a specific supertorsion equation.  After the
dependent spin connection or auxiliary fields are eliminated, the induced
transformations act on the solution space and close modulo the field
equations.  No corresponding mechanism is available for the $\m$
transformations of Eq.~\eqref{eq:fixedaction}; the comparison is developed
in Sec.~\ref{sec:MM}.

Finally, this explicit breaking should be contrasted with Theory II of section \ref{sec:dynamical}.  When
the projector is built equivariantly from a transforming field $\Phi$, its
variation supplies precisely the term missing in
Eq.~\eqref{eq:projector-equivariance-defect}:
\begin{equation}
 (\delta_\epsilon\cP_{\su})F
 +\cP_{\su}(\Phi)[F,\epsilon]
 =[\cP_{\su}(\Phi)F,\epsilon].
 \label{eq:dynamic-projector-cancellation}
\end{equation}
The complete $SO(2n)$ gauge symmetry is then exact off shell.  The
$\m$ directions are not explicitly broken; rather, they move the Higgs field
along its vacuum orbit and become the spontaneously broken gauge directions
about $\Phi=vJ$.

%%%%%%%%%%%%%%%%%%%%%%%%%%%%%%%%%%%%%%%%%%%%%
\subsection{Canonical formulation}
%%%%%%%%%%%%%%%%%%%%%%%%%%%%%%%%%%%%%%%%%%%%%

Let us consider Minkowski spacetime with metric $\eta_{\mu\nu}=\operatorname{diag}(-,+,+,+)$, and use $i,j,\ldots$ to denote spatial indices. Write
\begin{equation}
 K_{0i}=\dot B_i-D_i^{(B)}B_0+\cP_{\su}[X_0,X_i],
\end{equation}
and
\begin{equation}
 K_{ij}=F_{ij}(B)+\cP_{\su}[X_i,X_j].
\end{equation}
The Lagrangian density is
\begin{equation}
 \cL_{\rm fix}
 =\frac1{2g^2}\langle K_{0i},K_{0i}\rangle
 -\frac1{4g^2}\langle K_{ij},K_{ij}\rangle. \label{eq:L_fix}
\end{equation}
Let $a=1,\ldots,h$, with $h=n^2-1$, label $\su(n)$, and let $\alpha=1,\ldots,r$, with $r=n(n-1)$, label $\m$.  The nonzero spatial momenta are
\begin{equation}
 \pi_a^{i}:= \frac{\partial \cal{L}_\mathrm{fix}}{\partial \dot{A}^a_i}=\frac1{g^2}K_{0i\, a}.
\end{equation}
The primary constraints are
\begin{equation}
 \pi_a^0\approx0,
 \qquad p_c^\mu\approx0,
 \qquad p_\alpha^\mu\approx0,
 \label{eq:fixedprimary}
\end{equation}
where $p_c^\mu$ is the momentum conjugate to $c_\mu$, and
$p_\alpha^\mu$ is the momentum conjugate to $X^\alpha_\mu$.

The canonical Hamiltonian density can be written as
\begin{align}
 \cH_{\rm fix}
 =&\frac{g^2}{2}\langle\pi^i,\pi^i\rangle
 +\frac1{4g^2}\langle K_{ij},K_{ij}\rangle
 -\langle B_0,D_i^{(B)}\pi^i\rangle
 -\langle\pi^i,\cP_{\su}[X_0,X_i]\rangle.
 \label{eq:Hfixed}
\end{align}
The third term in Eq.~\eqref{eq:Hfixed} is the standard Gauss-law contribution associated with the unbroken $\su(n)$ gauge symmetry. It arises after integrating by parts the term
$\langle\pi^i,D_i^{(B)}B_0\rangle$ in the Legendre transform. Since $B_0$ has no time derivative, its conjugate momentum satisfies the primary constraint $\pi_a^0\approx0$. Preservation of $\pi_a^0$ first yields the secondary constraint
\begin{equation}
\cG_a^{(0)}=(D_i^{(B)}\pi^i)_a\approx0.
\end{equation}
Since $p_\alpha^\mu\approx0$ are primary constraints, this may be replaced by the weakly equivalent representative
\begin{equation}
\cG_a=\cG_a^{(0)}
+f_{a\beta}{}^{\gamma}X_\mu^\beta p_\gamma^\mu
\approx0.
\end{equation}
The two expressions define the same constraint surface, since they differ only by a term proportional to primary constraints. They are nevertheless distinct as phase-space generators: the additional $Xp$ term is required for $\cG_a$ to generate the correct $SU(n)$ transformation of $X_\mu$.

Because $c_\mu$ is absent from the action, all four $p_c^\mu$ are first class and generate an accidental local shift
\begin{equation}
 c_\mu\longmapsto c_\mu+\zeta_\mu.
 \label{eq:cshift}
\end{equation}
This is stronger than an ordinary $U(1)$ gauge transformation and shows that the central connection is a null variable in the pure projected theory.

Since $X_\mu^\alpha$ enters the fixed-projector action without derivatives,
its canonical momenta vanish identically,
$p_\alpha^\mu\approx0$. Their preservation,
\begin{equation}
\dot p_\alpha^\mu
=-\frac{\partial\mathcal H_{\rm fix}}{\partial X_\mu^\alpha}
\approx0,
\label{eq:fixedsecondaryX}
\end{equation}
yields the secondary constraints
\begin{align}
\cC_\alpha^0
&=
\left\langle T_\alpha,[\pi^i,X_i]\right\rangle
\approx0,
\label{eq:fixedsecondaryX0}\\
\cC_\alpha^i
&=
\left\langle T_\alpha,
[X_0,\pi^i]
+\frac{1}{g^2}[X_j,K_{ij}]
\right\rangle
\approx0.
\label{eq:fixedsecondaryXi}
\end{align}
Here the projector may be omitted inside the inner products because
$\pi^i$ and $K_{ij}$ are $\su(n)$-valued. Using
$\pi^i=g^{-2}K_{0i}$, the temporal and spatial equations combine into
the $3+1$ decomposition of the covariant complementary-sector
equation~\eqref{eq:Xeomfixed}. More explicitly, $\cC^\mu$ is the $3+1$ form of the $\mathfrak m$-valued covariant equation $
\cP_{\m}[X_\nu,K^{\mu\nu}]=0$, which up to the sign conventions for the Hodge dual and the
Levi-Civita tensor, this vector equation is the Hodge dual of the
three-form equation $\cP_{\m}[X,\ast K]=0.$ With the conventions $\epsilon_{0123}=+1$, and $(\ast K)_{\mu\nu} =\tfrac12\epsilon_{\mu\nu\rho\sigma}K^{\rho\sigma}$, one finds
\begin{equation}
 \frac{1}{3!}\epsilon^{\mu\nu\rho\sigma}
 \bigl([X,\ast K]\bigr)_{\nu\rho\sigma}
 =
 -[X_\nu,K^{\mu\nu}].
\end{equation}

At a generic phase-space configuration, the $X$-sector constraint
matrix is
\begin{equation}
 \Delta_{\alpha\beta}^{\mu\nu}(x,y)
 =\{p_\alpha^\mu(x),\cC_\beta^\nu(y)\}
 =-\frac{\delta\cC_\beta^\nu(y)}
         {\delta X_\mu^\alpha(x)}
 =\mathscr D_{\alpha\beta}^{\mu\nu}(y)\,
   \delta^3(x-y).
 \label{eq:fixedhessian}
\end{equation}
For orthonormal generators, define
$f_{a\alpha\beta}:=\langle T_a,[T_\alpha,T_\beta]\rangle$.
Holding the canonical momentum $\pi_a^i$ fixed when taking the
derivative, the blocks of $\mathscr D$ are
\begin{align}
 \mathscr D_{\alpha\beta}^{00}
 &=0, \nonumber\\
 \mathscr D_{\alpha\beta}^{0j}
 &=f_{a\alpha\beta}\pi_a^j = - \mathscr D_{\alpha\beta}^{j0}\nonumber\\
 \mathscr D_{\alpha\beta}^{ij}
 &=-\frac1{g^2}\left[
    f_{a\alpha\beta}K_{ij}^{a}
    +\delta^{ij}f_{a\beta\gamma}f_{a\alpha\delta}
       X_k^\gamma X_k^\delta
    -f_{a\beta\gamma}f_{a\alpha\delta}
       X_i^\gamma X_j^\delta
   \right].
 \label{eq:fixedhessianblocks}
\end{align}
Here
$K_{ij}^{a}=F_{ij}^{a}(B)
 +f_{a\gamma\delta}X_i^\gamma X_j^\delta$.
The matrix may be invertible at a generic configuration, depending
on the values of $\pi^i$, $K_{ij}$, and $X_i$. Invertibility must be assessed on the constraint surface.
The residual central $U(1)$ rotation acts as
\[
 \delta_\theta X=\theta[X,J],
 \qquad \delta_\theta B=0,
 \qquad \delta_\theta\pi^i=0.
\]
Holding $B$ and $\pi$ fixed, covariance under the central
rotation implies
\[
 \sum_{\nu,\beta}
 \frac{\partial\cC_\alpha^\mu}{\partial X_\nu^\beta}
 [X_\nu,J]^\beta
 =[\cC^\mu,J]_\alpha
 \approx0.
\]
For $X\neq0$, the variation $[X,J]$ is nonzero because
$\operatorname{ad}_J$ is invertible on $\m$.
Consequently, the Jacobian of the secondary constraints
with respect to $X$, at fixed $B$ and $\pi$, is necessarily
singular at any solution with $X\neq0$ before fixing the
residual central $U(1)$ symmetry.
This null direction reflects the residual gauge freedom
and does not, by itself, imply irregularity of the full
constraint system. The classification of constraints on
these branches requires a separate analysis.

The regular elimination considered here therefore concerns
the branch
\[
 X_\mu=0,
 \qquad
 \left.\det\mathscr D\right|_{X=0}\neq0.
\]
Whenever this condition holds, the implicit-function theorem
makes $X=0$ the unique nearby solution of the auxiliary
constraints at fixed $B,\pi$. The pairs
$(p_\alpha^\mu,\cC_\alpha^\mu)$ then form $8r$ second-class
constraints and eliminate the coset canonical variables.
The central rotation acts trivially on the resulting reduced
variables and supplies no additional independent constraint.

On this regular $X=0$ branch, the degree-of-freedom count is

\begin{equation}
 \begin{split}
 \dim\Gamma&=8(h+1+r),\\
 N_{\rm FC}&=2h+4,\\
 N_{\rm SC}&=8r,
 \end{split}
\end{equation}
and hence
\begin{equation}
 N_{\rm phys}
 =\frac12\left[8(h+1+r)-2(2h+4)-8r\right]
 =2h=2(n^2-1).
 \label{eq:fixedcount}
\end{equation}
Only the two transverse polarizations of the $SU(n)$ connection propagate.

At the zero-field phase-space background
$B_\mu=X_\mu=0$, $\pi_a^i=0$, every block of
$\Delta_{\alpha\beta}^{\mu\nu}$ vanishes. Hence the pairs
$(p_\alpha^\mu,\mathcal C_\alpha^\mu)$ do not form a regular
second-class system at this background. Using the notation $b$ for the fluctutation of $B$ around the vacuum configuration, the action gives
\begin{equation}
 S_{\rm fix}^{(2)}[b]
 =-\frac1{2g^2}\int\langle\dd b,\ast\dd b\rangle.
 \label{eq:fixedquadratic}
\end{equation}
Neither $c$ nor $X$ appears in this quadratic action,
which is invariant under arbitrary local shifts of both
fields. The shift of $c$ is an exact symmetry of the full
action, whereas the shift of $X$ is present only at
quadratic order. The caveat is that the zero-field background does not define a
separate regular nonlinear sector with a well-defined Dirac count.
At this point the $X$-sector Dirac matrix loses rank. The apparent
shift symmetry of $X_\mu^\alpha$ is only a symmetry of the quadratic
action: interactions involving $X$, schematically
$(\dd b)X^2$, $b^2X^2$, and $X^4$, break it. In the quadratic theory, the $4r$ vanishing momenta conjugate
to $X_\mu^\alpha$ are first-class constraints generating
arbitrary local shifts of $X$. Together with the four
central shift constraints and the $2h$ constraints of the
linearized Yang-Mills sector, this gives
\[
 N_{\rm FC}^{(2)}=2h+4+4r,
 \qquad N_{\rm SC}^{(2)}=0.
\]
The quadratic theory therefore propagates
\begin{equation}
 N_{\rm phys}^{(2)}
 =\frac12\left[
 8(h+1+r)-2(2h+4+4r)
 \right]
 =2h=2(n^2-1).
\end{equation}
This agrees numerically with the regular nonlinear count
in Eq.~\eqref{eq:fixedcount}, but the constraint classifications
differ: the quadratic theory has $4r$ first-class coset
shift constraints, whereas the regular nonlinear branch
has $8r$ second-class coset constraints.
For $n\ge3$, interactions break the arbitrary $X$-shift
symmetry, so the quadratic count does not establish
regularity of the nonlinear constraint system at the
zero-field background.

%%%%%%%%%%%%%%%%%%%%%%%%%%%%%%%%%%%%%%%%%%%%%%%%%%%%%%%%%%%%%
\section{Theory II: adjoint Higgs and equivariant kinetic map}
\label{sec:dynamical}
%%%%%%%%%%%%%%%%%%%%%%%%%%%%%%%%%%%%%%%%%%%%%%%%%%%%%%%%%%%%%

We now promote the fixed internal structure to an adjoint field $\Phi$
with the finite potential~\eqref{eq:potential}, allowing fluctuations
both tangent and normal to the complex-structure vacuum orbit.
The field-dependent maps are exact projectors on this orbit but
generally act as equivariant kinetic maps away from it. We introduce the equivariant action, discuss the local unitary-gauge
description, and determine the quadratic vacuum spectrum. We then
derive the covariant field equations and examine the limitations
associated with changes in the off-orbit kinetic rank.

%%%%%%%%%%%%%%%%%%%%%%%%%%%%%%%%%%%
\subsection{Adjoint vacuum}
%%%%%%%%%%%%%%%%%%%%%%%%%%%%%%%%%%%

To restore the full $SO(2n)$ covariance of the construction, we replace the fixed complex structure by an adjoint-valued scalar field. We therefore introduce a zero-form
\begin{equation}
\Phi\in\Omega^0(M,\so(2n)),
\end{equation}
transforming in the adjoint representation,
\begin{equation}
\Phi\mapsto g^{-1}\Phi g,
\end{equation}
while the gauge connection transforms in the usual way,
\begin{equation}
A\mapsto g^{-1}Ag+g^{-1}\dd g.
\end{equation}
The field $\Phi$ plays the role of an adjoint Higgs field: its value determines locally the internal splitting that was fixed in Theory~I. In particular, the relevant vacuum configurations are taken to lie on the adjoint orbit of a reference element proportional to the complex structure $J$. Choosing the representative
\begin{equation}
\Phi_v=vJ
\end{equation}
selects the stabilizer of $J$ as the unbroken subgroup, $H \simeq U(n)$. More generally, every vacuum on the same orbit can be written as
\begin{equation}
\Phi=g^{-1} \Phi_v g,  \label{eq:vacuum-orbit}
\end{equation}
so different representatives correspond to different gauge choices rather than physically distinct symmetry-breaking patterns. The associated projector will therefore be constructed directly from $\Phi$ in such a way that it transforms equivariantly under the full gauge group and reduces, in the gauge $\Phi=vJ$, to the fixed projector introduced in the previous section.

Since $\Phi$ is antisymmetric, $v\mathbf1$ is not an allowed adjoint vacuum.  Consider instead
\begin{equation}
 V(\Phi)=\frac{\lambda_H}{4} \Tr(\Phi^2+v^2\mathbf1)^2.
 \label{eq:potential}
\end{equation}
with $\lambda_H > 0$. A real antisymmetric matrix can be brought to skew-diagonal form
\begin{equation}
 \Phi=g^{-1}\operatorname{diag}(\lambda_1\epsilon,\ldots,\lambda_n\epsilon)g,
 \qquad
 \epsilon=\begin{pmatrix}0&1\\-1&0\end{pmatrix}.
\end{equation}
The absolute minima satisfy $\lambda_r^2=v^2$, and hence
\begin{equation}
 \Phi^2=-v^2\mathbf1,
 \qquad
 \vev{\Phi}=vJ
 \label{eq:phivacuum}
\end{equation}
in a local vacuum-adapted gauge.  The component containing $vJ$ is the orbit $SO(2n)/U(n)$. The constraint surface has in fact two disconnected components, distinguished
by the sign of the Pfaffian. We work in the component containing $vJ$,
whose elements are $\Phi=g^{-1}(vJ)g$ with $g\in SO(2n)$. The second orientation component is discussed in Sec.~\ref{subsec:constrained-geometry}.

On the vacuum orbit define
\begin{align}
 \cP_{\uu}(\Phi)X
 &=X+\frac1{4v^2}[\Phi,[\Phi,X]],\\
 \cP_{\Phi}X
 &=\frac{\langle\Phi,X\rangle}{\langle\Phi,\Phi\rangle}\Phi,\\
 \cP_{\su}(\Phi)&=\cP_{\uu}(\Phi)-\cP_{\Phi}.
 \label{eq:dynamicprojector}
\end{align}
These maps transform equivariantly:
\begin{equation}
 \cP_{\su}(g^{-1}\Phi g)(g^{-1}Xg)
 =g^{-1}\cP_{\su}(\Phi)Xg.
 \label{eq:equivariance}
\end{equation}

These maps are exact orthogonal projectors throughout the vacuum
orbit $\Phi=g^{-1}(vJ)g$, since every point on that orbit satisfies
$\Phi^2=-v^2\mathbf1$. Their images rotate with $\Phi$, while their
ranks remain fixed. These algebraic properties hold pointwise even
for a spacetime-dependent Higgs orientation and do not require
$D_A\Phi=0$.

The same formulas extend equivariantly to arbitrary nonzero
$\Phi\in\so(2n)$. Away from the constraint surface
$\Phi^2=-v^2\mathbf1$, however, they generally satisfy
\begin{equation}
 \cP_{\uu}(\Phi)^2\neq\cP_{\uu}(\Phi),
 \qquad
 \cP_{\su}(\Phi)^2\neq\cP_{\su}(\Phi).
\end{equation}
The map $\cP_\Phi$ remains an exact orthogonal projector onto
$\mathbb R\Phi$ for every $\Phi\neq0$, but is undefined at $\Phi=0$.

The corresponding action therefore remains $SO(2n)$ gauge
invariant wherever these maps are defined, although
$\cP_{\su}(\Phi)$ must generally be interpreted off the constraint
surface as a field-dependent kinetic map rather than an orthogonal
projector of fixed rank. There are three natural ways of treating
this distinction:
\begin{enumerate}
 \item[(i)] \emph{Exactly constrained theory.}
 Restricting the field to the orbit of $vJ$ removes the normal
 fluctuations and makes the projectors exact off shell.  This separate
 model, with either an intrinsic parametrization or a matrix Lagrange
 multiplier, is developed in Sec.~\ref{sec:constrained}.

 \item[(ii)] \emph{Near-orbit effective theory.}
 One may retain the polynomial potential
 \begin{equation}
  V(\Phi)=\frac{\lambda_{\!H}}4
  \Tr\!\left(\Phi^2+v^2\mathbf1\right)^2
 \end{equation}
 with large but finite $\lambda_{\!H}$.  The fluctuations normal to the
 vacuum orbit then acquire masses of order
 $m_\rho^2\sim\lambda_{\!H}v^2/Z_\Phi$ for fixed positive $Z_\Phi$.
 An effective description at energies well below $m_\rho$ would require
 integrating them out, including curvature-dependent sources and induced
 operators.  At finite
 $\lambda_{\!H}$ the map $\cP_{\su}(\Phi)$ is not exactly idempotent,
 and the directions projected out on the vacuum orbit can acquire
 small kinetic terms.  Because their vanishing changes the kinetic rank,
 a controlled expansion in $E^2/m_\rho^2$ is not established merely by
 choosing a large $\lambda_{\!H}$.  The exactly constrained model may be viewed formally as the
$\lambda_H\to\infty$ limit, in which normal Higgs fluctuations are
suppressed. However, the gauge kinetic rank also changes in this
limit. Establishing the constrained theory as a controlled
low-energy description therefore requires integrating out the
heavy modes and checking the resulting interactions and corrections.

 \item[(iii)] \emph{Spectral projectors for different eigenvalue patterns.}
One may instead construct exact projectors from the spectral
decomposition of $\operatorname{ad}_{\Phi}^{\,2}$.  Such a construction
is smooth within a region of field space where the relevant spectral
gaps remain open.  The skew eigenvalues may vary within this region,
provided their multiplicity pattern and the dimension of the
centralizer remain unchanged.  When eigenvalues coincide, split,
or pass through zero, the centralizer can change, and a projector
onto it may change rank and fail to extend smoothly across the
transition.  Moreover, a generic antisymmetric $\Phi$ has centralizer
$U(1)^n$, rather than $U(n)$, so projection onto the centralizer
does not by itself provide an $\su(n)$ projector away from the
vacuum orbit.  The projector and its target algebra must therefore
be specified for each eigenvalue pattern under consideration.
This possibility is mentioned only as an alternative construction. We do not develop its action, field equations, or constraint analysis
in this paper.

\end{enumerate}

Two formulations are considered in this paper: the unrestricted
adjoint-Higgs model (ii) and the exactly constrained model (i). Below we use the unrestricted finite-$\lambda_H$ theory.  Its spectrum
is established only at quadratic order about the vacuum.  Interpretation
(i) and its nonlinear canonical analysis belong to
Sec.~\ref{sec:constrained}.  No controlled off-orbit effective expansion is assumed and the limitations of interpretation (ii) are detailed in
Sec.~\ref{subsec:off-orbit-scope}.

%%%%%%%%%%%%%%%%%%%%%%%%%%%%%%%%%%%%%%%%%
\subsection{Equivariant action}
\label{subsec:Higgsprojectorgaugecovariance}
%%%%%%%%%%%%%%%%%%%%%%%%%%%%%%%%%%%%%%%%%

The dynamical action is
\begin{align}
 S[A,\Phi]=&-\frac1{2g^2}\int
 \left\langle \cP_{\su}(\Phi)F,
 \ast\cP_{\su}(\Phi)F\right\rangle
 \nonumber\\
 &-\frac{Z_\Phi}{2}\int
 \langle D_A\Phi,\ast D_A\Phi\rangle
 -\int V(\Phi)\vol .
 \label{eq:dynamicalaction}
\end{align}
Because $\Phi$ transforms in the adjoint representation and the
kinetic map is equivariant, this action is invariant under the full
local $SO(2n)$ gauge group. The present formulation is defined for
$\Phi\neq0$, as required by the denominator in $\cP_\Phi$. The vacuum $\Phi=vJ$ has stabilizer $U(n)$. In a local gauge aligned
with this vacuum, write $A=B+cJ+X$, with $B\in\su(n)$ and $X\in\m$.
Since $\cP_{\su}(vJ)J=0$, the central connection $cJ$ has no
curvature kinetic term on the vacuum orbit. It also receives no
Higgs mass term, because $[cJ,vJ]=0$. The central $U(1)$ therefore
remains unbroken, although its connection is nonpropagating in
the exactly projected theory.

% \subsection{Gauge covariance of the Higgs-dependent projector}
% \label{subsec:Higgsprojectorgaugecovariance}

It is useful to make explicit how the full gauge symmetry is realized before
and after choosing a representative of the vacuum orbit.  Let
$g=\exp(t\eta)$, with $\eta\in\so(2n)$.  With the finite-transformation
conventions assumed here, the infinitesimal transformations are
\begin{equation}
 \delta_\eta A=D_A\eta,
 \qquad
 \delta_\eta F=[F,\eta],
 \qquad
 \delta_\eta\Phi=[\Phi,\eta].
 \label{eq:generalinfinitesimalgauge}
\end{equation}
Equation~\eqref{eq:equivariance}, applied to
$K=\cP_{\su}(\Phi)F$, gives
\begin{equation}
 K(\Phi^g,F^g)=g^{-1}K(\Phi,F)g,
 \qquad
 \delta_\eta K=[K,\eta].
 \label{eq:Kgeneralcovariance}
\end{equation}
On varying the definition of $K$ directly, one instead obtains
\begin{equation}
 \delta_\eta K
 =(\delta_\eta\cP_{\su})F
 +\cP_{\su}(\Phi)[F,\eta].
 \label{eq:Kvariationchainrule}
\end{equation}
Comparison with Eq.~\eqref{eq:Kgeneralcovariance} yields the infinitesimal
equivariance identity
\begin{equation}
 (\delta_\eta\cP_{\su})F
 =[K,\eta]-\cP_{\su}(\Phi)[F,\eta].
 \label{eq:projectorinfinitesimalequivariance}
\end{equation}
Thus the variation of the projector compensates for the failure of the
projector at a fixed value of $\Phi$ to commute with a general adjoint
transformation.

The projected kinetic term is consequently invariant:
\begin{align}
 \delta_\eta S_{\rm proj}
 &=-\frac1{g^2}\int_M
 \langle[K,\eta],\ast K\rangle=0.
 \label{eq:projectedkineticgaugeinvariance}
\end{align}
The last equality follows from invariance of the bilinear form, together with
the fact that the spacetime Hodge operator does not act on gauge indices.
Moreover,
\begin{equation}
 \delta_\eta(D_A\Phi)=[D_A\Phi,\eta],
\end{equation}
and $V(g^{-1}\Phi g)=V(\Phi)$.  Hence the Higgs kinetic and potential terms are
also invariant, proving the full $SO(2n)$ gauge invariance of
Eq.~\eqref{eq:dynamicalaction}.

To identify the unbroken and broken transformations, now choose the local
vacuum representative $\Phi_v=vJ$ and decompose
\begin{equation}
 \eta=\epsilon+\chi,
 \qquad
 \epsilon\in\uu(n),
 \qquad
 \chi\in\m.
\end{equation}
For $\epsilon\in\uu(n)$ one has $[J,\epsilon]=0$, and therefore
\begin{equation}
 \delta_\epsilon\Phi_v=0,
 \qquad
 \delta_\epsilon\cP_{\su}(\Phi_v)=0.
 \label{eq:unbrokenprojectorvariation}
\end{equation}
The fixed vacuum projector is equivariant under the stabilizer $U(n)$:
\begin{equation}
 \cP_{\su}[F,\epsilon]
 =[\cP_{\su}F,\epsilon]=[K,\epsilon].
 \label{eq:unbrokenprojectorequivariance}
\end{equation}
Indeed, writing $F=K+F_J+F_{\m}$, the relations
$[\su(n),\uu(n)]\subset\su(n)$,
$[J,\uu(n)]=0$, and
$[\m,\uu(n)]\subset\m$ leave only $[K,\epsilon]$ after projection.
Writing further $\epsilon=\epsilon_{\su}+\alpha J$, the $SU(n)$ part acts in
the usual way, whereas
\begin{equation}
 \delta_{\alpha J}K=[K,\alpha J]=0.
\end{equation}
Thus the central $U(1)$ is an unbroken local gauge symmetry even though its
connection has no curvature kinetic term of its own in the rank-deficient
action.  Gauge symmetry does not by itself require a nondegenerate kinetic
operator for every connection component.

The situation is different for $\chi\in\m$.  In general,
\begin{equation}
 \delta_\chi\Phi_v=v[J,\chi]\ne0,
 \label{eq:brokenHiggsvacuumvariation}
\end{equation}
so an $\m$ transformation moves the chosen vacuum representative along the
orbit $SO(2n)/U(n)$.  Nevertheless, it remains an exact gauge transformation
of the Higgs-dependent theory.  Equations~\eqref{eq:Kvariationchainrule} and
\eqref{eq:Kgeneralcovariance} give
\begin{equation}
\delta_\chi K=
 (\delta_\chi\cP_{\su})F
 +\cP_{\su}[F,\chi]=[K,\chi].
 \label{eq:brokentransformationcompensation}
\end{equation}
At the vacuum, the symmetric-space brackets imply
$[\uu(n),\m]\subset\m$ and $[\m,\m]\subset\uu(n)$, whence
\begin{equation}
 \cP_{\su}[F,\chi]=\cP_{\su}[F_{\m},\chi]
\end{equation}
and
\begin{equation}
 (\delta_\chi\cP_{\su})F
 =[K,\chi]-\cP_{\su}[F_{\m},\chi].
 \label{eq:brokenprojectorvariationexplicit}
\end{equation}
The first term describes the rotation of the projected subspace itself; the
second cancels the component obtained by applying the old projector to the
transformed curvature.  In particular, there is no contradiction in
$\delta_\chi K = [K,\chi]\in\m$ relative to the original splitting: after the transformation,
$K$ belongs to the rotated $\su(n)$ subspace selected by
$\Phi_v+\delta_\chi\Phi_v$, rather than to the subspace selected by $\Phi_v$.

This also isolates the distinction from the fixed-projector theory.  If
$\Phi_v$ is frozen and its transformation is suppressed, then
$\delta_\chi\cP_{\su}=0$ and the compensation in
Eq.~\eqref{eq:brokentransformationcompensation} is lost.  The resulting
fixed-projector variation is
\begin{equation}
 \delta_\chi^{\rm fix}K
 =\cP_{\su}[F,\chi]
 =\cP_{\su}[F_{\m},\chi],
\end{equation}
which differs in general from the covariant law $\delta_\chi K=[K,\chi]$.
Thus the $\m$ transformations are spontaneously broken gauge transformations
in the Higgs-dependent theory: they move the vacuum and act on the Goldstone
directions.  They are absent as residual transformations after imposing a
local unitary gauge, and become explicitly broken only when the Higgs field
and the associated projector variation are removed from the theory.

%%%%%%%%%%%%%%%%%%%%%%%%%%%%%%%%%%%%%%%%
\subsection{Local unitary gauge and global qualifications}
%%%%%%%%%%%%%%%%%%%%%%%%%%%%%%%%%%%%%%%%

Let us clarify the meaning of unitary gauge in the present context. In a regular
neighborhood of the chosen vacuum orbit, and on a sufficiently small spacetime
patch, the adjoint Higgs may be parametrized as
\begin{equation}
\Phi=e^{-\tau}(vJ+\rho)e^\tau,
\qquad
\tau\in\m,
\qquad
\rho\in\uu(n).
\label{eq:local-Higgs-slice}
\end{equation}
Here $\tau$ parametrizes the directions tangent to the orbit
$SO(2n)/U(n)$, while $\rho$ describes fluctuations normal to it.  A broken
gauge transformation can then remove $\tau$, giving the local unitary gauge
\begin{equation}
\Phi_{\mathrm{UG}}=vJ+\rho.
\label{eq:local-unitary-general}
\end{equation}
This is directly analogous to the removal of the Goldstone fields from the
electroweak Higgs doublet.  Only after restricting to the exact vacuum orbit,
for which $\rho=0$, does Eq.~\eqref{eq:local-unitary-general} reduce to
\begin{equation}
\Phi_{\mathrm{UG}} \xrightarrow{\ \rho=0\ } vJ.
\label{eq:localunitary}
\end{equation}

The distinction is important.  Conjugation changes the internal orientation
of $\Phi$ but preserves its skew eigenvalues, or equivalently its adjoint
invariants.  Consequently, a general off-shell configuration cannot be
transformed to $vJ$; the departures of these invariants from their vacuum
values belong to the normal sector $\rho$ and cannot be removed by gauge
transformations.  This does not obstruct the local elimination of the orbit
coordinate $\tau$ on a regular field-space patch.  Rather, it shows that the
condition $\Phi=vJ$ is stronger than unitary gauge: it combines gauge fixing
with restriction to the vacuum orbit.

The comparison with the electroweak Higgs also makes clear why unequal field
components and unequal skew eigenvalues have different meanings.  A single
electroweak Higgs is a fundamental doublet, transformed by left multiplication
rather than by conjugation.  At any point where it is nonzero, a configuration
of the form
\begin{equation}
 \phi(x)=
 \begin{pmatrix}
  v_1\\ v_2+h(x)
 \end{pmatrix}
 = g^{-1}
 \begin{pmatrix}
  0\\ r(x)
 \end{pmatrix},
 \label{eq:EW-doublet-unitary-comparison}
\end{equation}
where $r(x)=\sqrt{|v_1|^2+|v_2+h(x)|^2}$ and
\begin{equation}
g(x)=
\frac{1}{\sqrt{|v_1|^2+|v_2+h(x)|^2}}
\begin{pmatrix}
v_2+h(x) & -v_1\\
v_1^* & v_2^*+h(x)^*
\end{pmatrix}
\in SU(2).
\end{equation}
Thus the two entries of the doublet are not separately gauge invariant; for
one doublet the only independent pointwise invariant is its norm
$\phi^\dagger\phi$.  The physical unitary-gauge fluctuation is the single
radial variable $r(x)-v$, where $v$ is the vacuum value of $r$, whereas a
varying relative orientation of the two entries belongs to the Goldstone
sector.  By contrast, the adjoint field can
be brought pointwise to the skew-diagonal form
\begin{equation}
 \Phi\sim\bigoplus_{s=1}^{n}\lambda_s
 \begin{pmatrix}
  0&1\\ -1&0
 \end{pmatrix},
 \label{eq:adjoint-skew-unitary-comparison}
\end{equation}
and the $\lambda_s$ are invariant under conjugation (up to the action of the Weyl group).
Unitary gauge rotates the corresponding two-planes but cannot turn unequal
skew eigenvalues into equal ones.  Their departures from the vacuum values
are therefore genuine normal fluctuations encoded in $\rho$, rather than
additional Goldstone orientations.

There is also a global qualification.  Even when the vacuum
condition holds locally, the orbit-valued Higgs field, represented in a
trivialization by a map
\begin{equation}
\Phi:M\longrightarrow SO(2n)/U(n),
\end{equation}
need not admit a globally smooth alignment with $J$.  Local unitary gauges
may always be chosen on suitable patches, but on their overlaps they are
related by $U(n)$ transition functions.  A single global choice
$\Phi=vJ$ exists only when the resulting $U(n)$ reduction admits a global
adapted frame; equivalently, for a trivial $SO(2n)$ bundle, the pullback of
\begin{equation}
U(n)\longrightarrow SO(2n)\longrightarrow SO(2n)/U(n)
\end{equation}
must be a trivial $U(n)$ bundle.  Nontrivial topology, Higgs zeros, or other
nonregular configurations at which the orbit type changes may therefore
require several gauge patches or make unitary gauge singular.

These limitations are generic to Higgs theories and are not peculiar to the
present construction.  In perturbative electroweak theory, for example, one
normally works in a topologically trivial sector with a nowhere-vanishing
Higgs field, where unitary gauge is well defined. It can nevertheless become
singular at Higgs zeros or in nonperturbative sectors.  Likewise, the quadratic analysis below uses a local vacuum-adapted
chart.  The nonlinear constrained analysis in
Sec.~\ref{subsec:constrained-Hamiltonian} uses such a chart within its
restricted field space.  The covariant theory does not assume that one
unitary-gauge chart covers the complete configuration space.

%%%%%%%%%%%%%%%%%%%%%%%%%%%%%%%%%%%%%%
\subsection{Quadratic spectrum of the unrestricted model near the vacuum}
\label{subsec:unconstrained-spectrum}
%%%%%%%%%%%%%%%%%%%%%%%%%%%%%%%%%%%%%%

This subsection concerns the quadratic expansion about $A=0$,
$\Phi=vJ$, with $Z_\Phi>0$.  The $n^2$ normal fluctuations are retained in this finite-$\lambda_H$
model.  The constrained theory of Sec.~\ref{sec:constrained} has no
such fields.  The expansion does not determine the nonlinear off-orbit
spectrum, for the reasons in Sec.~\ref{subsec:off-orbit-scope}.
In a local vacuum chart write
\begin{equation}
 A=B+cJ+X,
 \qquad
 \Phi=vJ+\varphi.
\end{equation}
Decompose the Higgs fluctuation into tangent and normal parts,
\begin{equation}
 \varphi=\varphi_{\m}+\varphi_{\uu}.
\end{equation}
The tangent fields $\varphi_{\m}$ are the Goldstone fluctuations of $SO(2n)/U(n)$.  Locally they can be removed from the quadratic action.  The normal components $\varphi_{\uu}$ are scalar fluctuations in the finite-$\lambda_H$ quadratic theory; the Hessian of Eq.~\eqref{eq:potential}, divided by $Z_\Phi$, gives their squared masses of order $\lambda_H v^2/Z_\Phi$.

At quadratic order in the local vacuum gauge,
\begin{align}
 \cL^{(2)}=&-\frac1{4g^2}F^a_{\mu\nu}(B)F_a^{\mu\nu}(B)
 -\frac{Z_\Phi}{2}\langle\partial_\mu\varphi_{\uu},
 \partial^\mu\varphi_{\uu}\rangle
 \nonumber\\
 &-\frac12M^2_{\alpha\beta}X_\mu^\alpha X^{\mu\beta}
 -\frac12\varphi_{\uu}\cdot\cM_H^2\varphi_{\uu}.
 \label{eq:quadraticspectrum}
\end{align}
where $\cM_H^2$ denotes the Hessian of $V$ on the normal subspace,
with canonical squared-mass matrix $\cM_H^2/Z_\Phi$.
Here
$F_{\mu\nu}(B)$ is understood to linear order when reading
$\cL^{(2)}$ as a strictly quadratic action.
There is no $D_{[\mu}X_{\nu]}D^{[\mu}X^{\nu]}$ term, and no $(\dd c)^2$ term
at this order. Choosing an orthonormal basis $T_\alpha\in\mathfrak m$ and using
the invariance of the inner product, we obtain
\begin{equation}
 \langle[T_\alpha,J],[T_\beta,J]\rangle
 =-\langle T_\alpha,[J,[J,T_\beta]]\rangle
 =4\delta_{\alpha\beta},
\end{equation}
where we used $[J,[J,T_\beta]]=-4T_\beta$ on $\mathfrak m$.
The algebraic coset mass matrix is therefore
\begin{equation}
 M^2_{\alpha\beta}
 =Z_\Phi v^2
 \langle[T_\alpha,J],[T_\beta,J]\rangle
 =4Z_\Phi v^2\delta_{\alpha\beta}.
 \label{eq:Xmassmatrix}
\end{equation}

Calling this a ``mass'' must not obscure the
fact that $X_\mu$ has no kinetic term at this order: it is an auxiliary vector, not a Proca
particle.  Appendix~\ref{app:goldstone-auxiliary} derives the corresponding
Stueckelberg combination and shows explicitly why neither $X$ nor the tangent
Goldstone field propagates.

The propagating perturbative spectrum is therefore
\begin{equation}
 2(n^2-1)\quad\text{$SU(n)$ vector polarizations}
\end{equation}
plus $n^2$ normal scalar modes in the finite-$\lambda_H$ quadratic
extension.  These scalars are absent if the orbit constraint is imposed
exactly.  Neither the central $U(1)$ connection nor the
$\Lambda^2\mathbf n\oplus\Lambda^2\overline{\mathbf n}$ vectors propagate
in this free vacuum spectrum.  Higher-order off-orbit derivative terms
prevent promoting this result to a generic nonlinear degree count.

%%%%%%%%%%%%%%%%%%%%%%%%%%%%%%%%%%%%%%%%
\subsection{Covariant field equations}
%%%%%%%%%%%%%%%%%%%%%%%%%%%%%%%%%%%%%%%%

For $\Phi\ne0$, set
\begin{equation}
 \mathsf P(\Phi):=\cP_{\su}(\Phi),
 \qquad K:=\mathsf P(\Phi)F,
 \qquad \mathcal H_\Phi:=\mathsf P(\Phi)K=\mathsf P(\Phi)^2F.
 \label{eq:K-and-H-definitions}
\end{equation}
The map in Eq.~\eqref{eq:dynamicprojector} is self-adjoint even away
from the vacuum orbit, although it is then generally not idempotent.
Holding $\Phi$ fixed and using $\delta_A F=D_A\delta A$ gives
\begin{align}
 \delta_A S_{\rm proj}
 &=-\frac1{g^2}\int
 \left\langle
 \mathsf P(\Phi)D_A\delta A,\ast K
 \right\rangle
 \nonumber\\
 &=-\frac1{g^2}\int
 \left\langle D_A\delta A,\ast\mathsf P(\Phi)K\right\rangle
 \nonumber\\
 &=-\frac1{g^2}\int
 \left\langle\delta A,D_A\ast\mathcal H_\Phi\right\rangle ,
 \label{eq:off-orbit-A-variation}
\end{align}
up to a boundary term.  The Higgs kinetic term contributes
\begin{equation}
 \delta_A S_\Phi
 =-Z_\Phi\int
 \left\langle\delta A,
 [\Phi,\ast D_A\Phi]\right\rangle .
\end{equation}
Combining both contributions gives the general connection equation
\begin{equation}
 D_A\ast\mathcal H_\Phi
 +g^2Z_\Phi[\Phi,\ast D_A\Phi]=0.
 \label{eq:Aeomdynamic}
\end{equation}
On the exact complex-structure orbit, $\mathcal H_\Phi=K$.
The corresponding constrained connection equation is displayed in
Eq.~\eqref{eq:Aeomdynamic-exact-orbit}.  The curvature-specific
simplification under covariant constancy is discussed in
Sec.~\ref{subsec:off-orbit-scope}.

For unrestricted adjoint variations, variation with respect to $\Phi$ gives
\begin{equation}
 Z_\Phi D_A\ast D_A\Phi
 -\frac1{g^2}\cJ_\Phi(F,K)
 -\frac{\partial V}{\partial\Phi}\vol=0.
 \label{eq:phieomdynamic}
\end{equation}
Here $\cJ_\Phi$ is defined by
\begin{equation}
 \left\langle\delta\Phi,\cJ_\Phi(F,K)\right\rangle
 =\left\langle(\delta_\Phi\cP_{\su})F,\ast K\right\rangle.
 \label{eq:Jsource}
\end{equation}
This source definition does not use idempotence.  All adjoint
variations, including normal variations, are allowed here.  Restricting
the variations to an orbit gives a different scalar equation, derived
in Sec.~\ref{subsec:constrained-equations}, simply setting $\Phi=vJ$
in Eq.~\eqref{eq:phieomdynamic} does not implement that restriction.

The polynomial part of Eq.~\eqref{eq:dynamicprojector} contributes to
the source in Eq.~\eqref{eq:Jsource} through
\begin{equation}
 (\delta_\Phi\cP_{\uu})X
 =\frac1{4v^2}\left(
 [\delta\Phi,[\Phi,X]]+[\Phi,[\delta\Phi,X]]
 \right),
 \label{eq:deltaprojector}
\end{equation}
with additional terms from $\delta\cP_\Phi$.
Equations~\eqref{eq:Aeomdynamic} and \eqref{eq:phieomdynamic} display a
reciprocal coupling between the gauge and Higgs sectors. When $\Phi$ lies on the vacuum orbit, $\mathsf P(\Phi)$ is an exact
projector of fixed rank. Its covariant derivative $D_A\mathsf P$
describes changes in the orientation of the projected subspace
and its orthogonal complement. Off the vacuum orbit, it can also describe changes in kinetic weights.
Conversely, $\delta_\Phi\mathsf P$ supplies a curvature-dependent Higgs
source.  In unitary gauge on the orbit, the curvature part of the action
agrees with the fixed-projector expression, but for $Z_\Phi\ne0$ the Higgs
kinetic term still supplies the algebraic coset-vector term.  In particular,
$\Phi=vJ$ does not imply $D_A\Phi=0$.  Covariant constancy is a separate
restriction, not a requirement for idempotence, as explained in
Sec.~\ref{subsec:off-orbit-scope}.

This statement requires some care because non-Abelian charge is a
Lie-algebra-valued, gauge-covariant local quantity, and its decomposition into
separate gauge-sector and Higgs-sector contributions is not itself gauge
invariant.  There are in fact two
related but distinct effects.  First, when $Z_\Phi\ne0$, the Higgs kinetic term
defines the adjoint-valued Higgs current
\begin{equation}
 j_\Phi:=Z_\Phi[\Phi,\ast D_A\Phi],
 \label{eq:Higgs-gauge-current}
\end{equation}
and Eq.~\eqref{eq:Aeomdynamic} takes the Yang-Mills-with-matter form
\begin{equation}
 D_A\ast\mathcal H_\Phi=-g^2j_\Phi.
 \label{eq:projected-flux-sourced}
\end{equation}
The flux is $\mathcal H_\Phi$, reducing to $K$ on the exact orbit.
Its covariant divergence is sourced by the Higgs current.  A nonzero
$Z_\Phi$ permits this current but does not, by itself, imply independent
physical Higgs particles: the exactly constrained model has only orbit
coordinates, whereas the unrestricted theory also has normal scalars.  The
reciprocal action of the gauge field on the Higgs is the curvature-dependent
source $\cJ_\Phi(F,K)$ in Eq.~\eqref{eq:phieomdynamic}.

Second, the Higgs field determines the internal decomposition of the
gauge algebra. On the vacuum orbit,
\begin{equation}
 \so(2n)=\mathfrak h_\Phi\oplus\mathfrak h_\Phi^\perp,
 \qquad K=\mathsf P(\Phi)F,
 \label{eq:local-Higgs-splitting}
\end{equation}
where $\mathfrak h_\Phi=\operatorname{im}\mathsf P(\Phi)\simeq\su(n)$.
Both on and off the orbit, the Bianchi identity $D_A F = 0$ gives
\begin{equation}
 D_AK=\mathsf P(\Phi)D_AF+(D_A\mathsf P)F  =(D_A\mathsf P)F.
 \label{eq:projector-rotation-mixing}
\end{equation}
On the orbit, $D_A\mathsf P$ describes the covariant rotation of
the projected subspace and its orthogonal complement. Indeed, differentiating $\mathsf P^2=\mathsf P$ gives
\begin{equation}
 (D_A\mathsf P)\mathsf P+\mathsf P(D_A\mathsf P)
 =D_A\mathsf P.
\end{equation}
Multiplying on both sides by $\mathsf P$, or by $1-\mathsf P$,
respectively, yields
\begin{equation}
 \mathsf P(D_A\mathsf P)\mathsf P=0,
 \qquad
 (1-\mathsf P)(D_A\mathsf P)(1-\mathsf P)=0.
 \label{eq:projector-derivative-off-diagonal}
\end{equation}
Thus $D_A\mathsf P$ maps the projected subspace into its orthogonal
complement and vice versa. Off the orbit,
it also includes changes in kinetic weights, and the image of
$\mathsf P$ need not be an $\su(n)$ subalgebra.
Writing $N_\Phi:=\langle\Phi,\Phi\rangle$, the covariant derivative
of the kinetic map is explicitly
\begin{align}
 (D_A\mathsf P)F
 ={}&\frac{1}{4v^2}
 \left(
 [D_A\Phi,[\Phi,F]]
 +[\Phi,[D_A\Phi,F]]
 \right)
 \nonumber\\
 &-\frac{1}{N_\Phi}
 \left(
 D_A\Phi\wedge\langle\Phi,F\rangle
 +\Phi\,\langle D_A\Phi,F\rangle
 \right)
 \nonumber\\
 &+\frac{2}{N_\Phi^2}
 \langle\Phi,D_A\Phi\rangle
 \wedge\langle\Phi,F\rangle\,\Phi .
 \label{eq:covariant-derivative-kinetic-map}
\end{align}
The pairings include exterior multiplication.
On the vacuum orbit, $N_\Phi=2nv^2$ is constant and the last
term vanishes
\begin{equation}
 \langle\Phi,D_A\Phi\rangle
 =\frac12\,\dd N_\Phi=0.
\end{equation}
Covariant constancy
$D_A\Phi=0$ implies $D_A\mathsf P=0$, but does not ensure
$\mathsf P^2=\mathsf P$ on the full algebra. These effects enter the connection equation through the derivative
of the curvature flux $\mathcal H_\Phi=\mathsf P(\Phi)K$:
\begin{equation}
 D_A\ast\bigl(\mathsf P(\Phi)K\bigr)
 =\mathsf P(\Phi)D_A\ast K
 +(D_A\mathsf P)\wedge\ast K.
 \label{eq:kinetic-flux-product-rule}
\end{equation}
Since $K=\mathsf P(\Phi)F$, this can also be written as
\begin{equation}
 D_A\ast\mathcal H_\Phi
 =\mathsf P^2D_A\ast F
 +\bigl[(D_A\mathsf P)\mathsf P
       +\mathsf P(D_A\mathsf P)\bigr]\wedge\ast F.
 \label{eq:off-orbit-flux-expansion}
\end{equation}
Unlike $D_AF$, the quantity $D_A\ast F$ does not vanish by the
Bianchi identity.

On the vacuum orbit, $\mathsf PK=K$, so
Eq.~\eqref{eq:kinetic-flux-product-rule} implies
\begin{equation}
 (1-\mathsf P)D_A\ast K
 =(D_A\mathsf P)\wedge\ast K.
 \label{eq:complementary-flux-rotation}
\end{equation}
Thus the covariant derivative of a projected curvature can have
a component outside the instantaneous projected subspace.
This component is determined by the covariant variation of the
projector.

For example, consider a common radial displacement
\begin{equation}
 \Phi=a(x)J,\qquad a(x)\neq0,\qquad
 q(x):=1-\frac{a(x)^2}{v^2},
 \label{eq:radial-kinetic-weight}
\end{equation}
where $q$ is the dimensionless eigenvalue of $\mathsf P$ on $\m$.
Decomposing the full curvature relative to $J$ as
$F=F_{\su}+F_J+F_{\m}$ gives
\begin{equation}
 K=F_{\su}+qF_{\m},
 \qquad
 \mathcal H_\Phi=F_{\su}+q^2F_{\m}.
 \label{eq:radial-projected-flux}
\end{equation}
Here $F_{\su}$ includes the contribution quadratic in the
complementary connection. The connection equation becomes
\begin{equation}
 D_A\ast F_{\su}
 +q^2D_A\ast F_{\m}
 +\dd(q^2)\wedge\ast F_{\m}
 =-g^2j_\Phi.
 \label{eq:radial-flux-equation}
\end{equation}
At $a=v$, the complementary curvature drops out of the flux.
For $a^2\neq v^2$, the coefficient $q^2$ activates its kinetic term,
while $\dd(q^2)$ couples it to gradients of the Higgs amplitude.

Gauge invariance remains intact, and self-adjointness ensures
nonnegative internal kinetic weights:
\begin{equation}
 \langle Y,\mathsf P^2Y\rangle
 =\langle\mathsf PY,\mathsf PY\rangle\geq0.
\end{equation}
The resulting changes in kinetic rank and constraint structure
are discussed in Sec.~\ref{subsec:off-orbit-scope}.

The compatibility of the two field equations is encoded in the gauge Noether
identity.  Introduce the unrescaled field equations
\begin{align}
 \widehat{\mathcal E}_A
 &:=\frac1{g^2}D_A\ast\mathcal H_\Phi+Z_\Phi[\Phi,\ast D_A\Phi],\\
 \mathcal E_\Phi
 &:=Z_\Phi D_A\ast D_A\Phi
 -\frac1{g^2}\cJ_\Phi(F,K)
 -\frac{\partial V}{\partial\Phi}\vol.
\end{align}
Invariance under $\delta_\eta A=D_A\eta$ and
$\delta_\eta\Phi=[\Phi,\eta]$ implies, with the variation convention
$\delta S=-\int\langle\delta A,\widehat{\mathcal E}_A\rangle
+\int\langle\delta\Phi,\mathcal E_\Phi\rangle$,
\begin{equation}
 D_A\widehat{\mathcal E}_A-[\Phi,\mathcal E_\Phi]=0.
 \label{eq:dynamic-Noether-identity}
\end{equation}
Accordingly, the covariant divergence of the connection equation and the
component of the Higgs equation tangent to the gauge orbit are not independent.
Equation~\eqref{eq:dynamic-Noether-identity} expresses the covariant
charge balance.  In the constrained model it also shows that the tangent
Higgs equation follows from the connection equation.

There is an analogous qualification for energy.  Since the gauge kinetic
operator depends on $\Phi(x)$, the stress tensor obtained from the projected
curvature term is not generally conserved separately.  On use of the gauge
equation, its divergence contains schematically
\begin{equation}
 \nabla_\mu T_{\rm proj}^{\mu\nu}
 \sim-\left\langle\cJ_\Phi(F,K),D^\nu\Phi\right\rangle,
 \qquad
 \nabla_\mu T_{\Phi}^{\mu\nu}
 \sim+\left\langle\cJ_\Phi(F,K),D^\nu\Phi\right\rangle,
 \label{eq:sector-energy-exchange}
\end{equation}
with additional potential terms arranged by the Higgs equation.  Only the
total stress tensor is conserved on shell; in general the exchange also
includes the gauge-force contribution involving $F$ and $j_\Phi$.
For the unrestricted theory with $Z_\Phi>0$, this describes coupling
to dynamical normal Higgs fields as well as to orbit coordinates.
In the exactly constrained model it does not imply an additional
propagating Higgs sector after gauge fixing and auxiliary-field elimination.

If the Higgs kinetic term is omitted, $Z_\Phi=0$, the theory nevertheless
remains fully $SO(2n)$ gauge invariant: invariance follows from the
equivariance of $\cP_{\su}(\Phi)$ and from the invariance of $V(\Phi)$, neither
of which requires $D_A\Phi$.  The field equations reduce to
\begin{equation}
 D_A\ast\mathcal H_\Phi=0,
 \qquad
 \frac1{g^2}\cJ_\Phi(F,K)
 +\frac{\partial V}{\partial\Phi}\vol=0.
 \label{eq:auxiliary-Higgs-equations}
\end{equation}
These are the unrestricted equations.  On the exact orbit the first uses
$\mathcal H_\Phi=K$, while only the tangent projection of the second is
imposed, as in Eq.~\eqref{eq:constrained-Higgs-equation}.
The second equation is pointwise algebraic in $\Phi$ and the curvature.  The
Higgs is an auxiliary compensator that selects or constrains the local
orientation of the kinetic projector; it has neither an independent wave
equation nor the current~\eqref{eq:Higgs-gauge-current}.  Hence in this limit
it is not accurate to describe the Higgs as a dynamical reservoir that absorbs
and emits gauge charge or energy.  What remains is the algebraic backreaction
of the curvature on the projector and the corresponding Higgs-mediated
selection or rotation of gauge-algebra components.  Moreover, without the
term $\langle D_A\Phi,\ast D_A\Phi\rangle$, expansion about $\Phi=vJ$ produces
no algebraic Higgs mass matrix $Z_\Phi v^2[X,J]^2$ for the coset connection.
The $Z_\Phi=0$ model is therefore a gauge-invariant projected theory with an
auxiliary compensator, not an ordinary Yang-Mills-Higgs phase.

The Higgs-mediated rotation between the instantaneous projected and
complementary gauge-algebra directions, exhibited in
Eq.~\eqref{eq:projector-rotation-mixing}, provides an additional motivation
for the full-rank kinetic operator discussed in
Sec.~\ref{sec:fullrank}.  In the rank-deficient theory, the kernel of the
curvature kinetic operator depends on the spacetime-dependent field $\Phi$.
Consequently, as the Higgs orientation varies,
$D_A\cP_{\mathfrak u}(\Phi)$ mixes directions with nonzero kinetic weight
with directions whose kinetic coefficient vanishes.  Assigning finite
kinetic coefficients to both
$\cP_{\mathfrak u}(\Phi)F$ and
$\cP_{\perp}(\Phi)F$ makes the curvature kinetic operator nondegenerate on
the regular vacuum orbit, so that this rotation takes place between two
dynamical curvature sectors.  The singular kinetic limit
$g_{\perp}^{-2}\rightarrow0$, equivalently $g_{\perp}\rightarrow\infty$,
gives a rank-deficient $\uu(n)$-projected theory; recovering the
$\su(n)$-projected model also requires removing the central kinetic
term, as detailed in Sec.~\ref{sec:fullrank}.  Neither limit is a
generic consistent truncation of the complementary curvature sector.

%%%%%%%%%%%%%%%%%%%%%%%%%%%%%%%%%%%%%%%%
\subsection{Off-orbit kinetic rank and scope of the analysis}
\label{subsec:off-orbit-scope}
%%%%%%%%%%%%%%%%%%%%%%%%%%%%%%%%%%%%%%%%

The continuation of Eq.~\eqref{eq:dynamicprojector} to arbitrary nonzero $\Phi$ defines a dynamical theory distinct from the exactly constrained model. The double-commutator term is quadratic in $\Phi$,
whereas the central subtraction $\cP_\Phi$ contains the denominator
$\langle\Phi,\Phi\rangle$ and is undefined at $\Phi=0$.  Its curvature kinetic operator is
\begin{equation}
 \mathcal Q_\Phi:=\mathsf P(\Phi)^\dagger\mathsf P(\Phi)
 =\mathsf P(\Phi)^2,
 \qquad
 S_{\rm proj}=-\frac1{2g^2}\int
 \langle F,\ast\mathcal Q_\Phi F\rangle .
 \label{eq:off-orbit-kinetic-operator}
\end{equation}
Equivariance ensures full $SO(2n)$ gauge invariance, while
$\langle Y,\mathcal Q_\Phi Y\rangle
=\langle\mathsf P Y,\mathsf P Y\rangle\geq0$ for the compact inner product.
Thus the loss of idempotence does not introduce wrong-sign vector kinetic
terms.  It does, however, change the rank of the gauge kinetic term.

Away from the vacuum orbit, it is sufficient to consider $\Phi$ written in a local skew-diagonal frame
\begin{equation}
 \Phi(x)=g(x)^{-1}
 \operatorname{diag}\!\bigl(
 \lambda_1(x)\epsilon,\ldots,\lambda_n(x)\epsilon
 \bigr)g(x),
 \qquad
 \epsilon=\begin{pmatrix}0&1\\-1&0\end{pmatrix},
 \label{eq:off-orbit-skew-form}
\end{equation}
where $g(x)\in SO(2n)$ and the $\lambda_r(x)$ are real
skew eigenvalues. The matrix $\Phi$ is real antisymmetric, with
complex eigenvalues $\pm i\lambda_r$. For $r<s$, the root spaces of the complexified algebra have adjoint
frequencies $\pm(\lambda_r+\lambda_s)$ and
$\pm(\lambda_r-\lambda_s)$.  They are orthogonal to $\Phi$, so
$\cP_\Phi$ vanishes on them.  Consequently $\mathsf P$ has eigenvalues
\begin{equation}
 p^{(+)}_{rs}=1-\frac{(\lambda_r+\lambda_s)^2}{4v^2},
 \qquad
 p^{(-)}_{rs}=1-\frac{(\lambda_r-\lambda_s)^2}{4v^2},
 \label{eq:off-orbit-root-weights}
\end{equation}
and $\mathcal Q_\Phi$ has the corresponding squared eigenvalues.
Near the representative $\lambda_r=v$, the plus-sign root spaces are the
complexified coset directions and the minus-sign root spaces belong to
$\uu(n)$.  At this representative, $p^{(+)}_{rs}=0$ and
$p^{(-)}_{rs}=1$.  On the Cartan subalgebra, $\mathsf P=1-\cP_\Phi$:
it kills $\Phi$ and is the identity on the remaining $n-1$ directions.
Thus its on-orbit rank is $n^2-1$, as required.

Generically, neither off-orbit root weight vanishes.  The only null
direction is then $\Phi$, giving kinetic rank $\dim\so(2n)-1$.
Additional null directions occur when
$|\lambda_r\pm\lambda_s|=2v$.  Even the common radial displacement
$\lambda_r=v+s$ gives $p^{(+)}_{rs}=1-(v+s)^2/v^2$, so unequal
skew eigenvalues are not necessary for activating coset kinetic terms.
The resulting image need not be an $\su(n)$ subalgebra.

The canonical momentum conjugate to the spatial connection $A_i$
in the full unrestricted theory is
\begin{equation}
 \Pi^i=\frac1{g^2}\mathcal Q_\Phi F_{0i}.
 \label{eq:off-orbit-canonical-momentum}
\end{equation}
The vanishing coset momenta of the orbit theory therefore cease to be
primary constraints on generic off-orbit configurations.  The vector
constraint analysis in Sec.~\ref{subsec:constrained-Hamiltonian} cannot be extended there by simply adding
normal scalar canonical pairs.  Establishing a full physical
degree-of-freedom count would require a separate Dirac analysis on each
region of constant kinetic rank.

Let us now return to why the quadratic vacuum spectrum does not
detect the change in kinetic rank. In a local unitary-gauge chart,
write $\Phi=vJ+\rho$, with $\rho\in\uu(n)$, and expand
$\mathsf P=\mathsf P_0+\delta\mathsf P(\rho)+\cdots$.
Since $\mathsf P_0D_{A_H}X=0$, the leading term quadratic in
the covariant derivatives of the connection in the complement of $\uu(n)$ has the form
\begin{equation}
 S_{\rm proj}\supset-\frac1{2g^2}\int
 \left\langle
 \delta\mathsf P(\rho)D_{A_H}X,
 \ast\delta\mathsf P(\rho)D_{A_H}X
 \right\rangle ,
 \label{eq:off-orbit-induced-X-kinetic}
\end{equation}
with $\delta\mathsf P=O(\rho/v)$. When both $\rho$ and $X$ are treated as small fluctuations about $\Phi=vJ$, $A=0$, this term is of order $\rho^2(\dd X)^2$ and therefore does not appear in the quadratic action. However, a nonzero background value of $\rho$ generally gives a nonzero
coefficient for $(\dd X)^2$, providing a kinetic term for fluctuations of $X$. Absence of this term in the free vacuum action therefore does not establish absence of the corresponding modes in the nonlinear
finite-$\lambda_H$ theory.

For a locally constant scalar background
$\bar\Phi=vJ+\bar\rho$, with $\bar\rho\in\uu(n)$ constant,
the residual $U(n)$ freedom allows us to choose a skew-diagonal frame,
\begin{equation}
 \bar\rho=\operatorname{diag}(\rho_1\epsilon,\ldots,\rho_n\epsilon),
 \qquad \lambda_r=v+\rho_r.
\end{equation}
Consider a connection component along a root generator $E$
satisfying $[\bar\Phi,E]=i\mu E$ and
$\mathsf P(\bar\Phi)E=pE$. The root directions split into those
belonging to the original coset complement, with adjoint frequencies
$\mu=\pm(\lambda_r+\lambda_s)$, and those within $\uu(n)$,
with $\mu=\pm(\lambda_r-\lambda_s)$, where $r<s$.
Their respective kinetic weights are
\begin{equation}
 p^{(+)}_{rs}
 =1-\frac{(2v+\rho_r+\rho_s)^2}{4v^2},
 \qquad
 p^{(-)}_{rs}
 =1-\frac{(\rho_r-\rho_s)^2}{4v^2},
 \qquad r<s.
\end{equation}
Thus $p$ depends on both the background displacement $\bar\rho$
and the root direction. For a component of the coset connection $X$
associated with the pair $(r,s)$, we take $p=p^{(+)}_{rs}$.
For $p\ne0$, the quadratic kinetic coefficient is $p^2/g^2$. 

For the coset directions, the weights $p^{(+)}_{rs}$ vanish on
the vacuum orbit and are generally linear in a small background
displacement, making the kinetic coefficients
$(p^{(+)}_{rs})^2/g^2$ quadratic in that displacement.
Even a common radial displacement $\bar\rho=sJ$ gives
\begin{equation}
 p^{(+)}_{rs}=1-\frac{(v+s)^2}{v^2},
 \qquad p^{(-)}_{rs}=1,
\end{equation}
and hence generally activates the coset kinetic terms while
leaving the minus-family root modes massless.

Here the scalar background is prescribed and frozen. A
physical particle interpretation additionally requires a
supported background or a solution of the complete field
equations. For a vanishing connection and constant
$\bar\Phi$, the background condition is
$\partial V/\partial\Phi|_{\bar\Phi}=0$.

Explicitly, for one real coset component on a vanishing background
connection, the quadratic Lagrangian is
\begin{equation}
 \mathcal L_X^{(2)}
 =-\frac{p^2}{4g^2}
  (\partial_\mu X_\nu-\partial_\nu X_\mu)
  (\partial^\mu X^\nu-\partial^\nu X^\mu)
  -\frac{Z_\Phi\mu^2}{2}X_\mu X^\mu.
\end{equation}
Rescaling the field as $X_{\rm can}=|p|X/g$ brings its kinetic
term to the standard normalization,
\begin{equation}
 \mathcal L_X^{(2)}
 =-\frac14
  (\partial_\mu X_{{\rm can},\nu}
   -\partial_\nu X_{{\rm can},\mu})
  (\partial^\mu X_{\rm can}^{\nu}
   -\partial^\nu X_{\rm can}^{\mu})
  -\frac12m_X^2X_{{\rm can},\mu}X_{\rm can}^{\mu}.
\end{equation}
from which we read off the squared mass parameter of the frozen-background
\begin{equation}
 m_X^2=\frac{g^2Z_\Phi\mu^2}{p^2}.
 \label{eq:off-orbit-canonical-mass}
\end{equation}
The same mass formula applies separately to each root direction
with nonzero kinetic weight. Directions with $\mu=0$ remain massless,
whereas those with $p=0$ have no quadratic kinetic term and
cannot be assigned a mass by this normalization.

For the coset components, $|\mu|\to2v$ and $p\to0$ as the
vacuum orbit is approached, so the quadratic mass diverges.
Within this quadratic description, a mode with $m_X$ much larger
than a fixed probe energy cannot be produced as a real particle,
and its exchange has a range of order $m_X^{-1}$ \cite{deBlas:2017xtg}.
Here, however, the divergence arises from the vanishing kinetic
coefficient, while the algebraic coefficient $Z_\Phi\mu^2$
remains finite. At exactly $p=0$, the linearized equation for $X$
is algebraic: in the absence of a source it sets $X=0$, whereas
a source can induce an algebraic response and local interactions.

The disappearance of the propagating mode does not by itself
imply that its low-energy effects vanish. For illustration,
an interaction $hX_\mu j^\mu$, with $h$ finite in the original
variables, would give
\begin{equation}
 h_{\rm can}=\frac{hg}{|p|},
 \qquad
 \frac{h_{\rm can}^2}{m_X^2}
 =\frac{h^2}{Z_\Phi\mu^2}.
\end{equation}
Thus an induced interaction can remain finite even as
$m_X\to\infty$. Whether this occurs in the present theory must
be determined from its actual interaction vertices.
The coset modes therefore leave the finite-energy quadratic
particle spectrum as the orbit is approached, but establishing
a controlled decoupling limit requires an interaction analysis.
The singular canonical normalization alone also does not
determine a strong-coupling scale.

On a separate but related point, covariant constancy, \(D_A\Phi=0\), and exact projector algebra are distinct conditions. For $\Phi\ne0$, $D_A\Phi=0$ implies $D_A\mathsf P=0$, but it does not
fix the skew eigenvalues to their vacuum values.  For example,
$A=0$ and any constant off-orbit $\Phi$ satisfy $D_A\Phi=0$ although the
weights in Eq.~\eqref{eq:off-orbit-root-weights} need not equal zero or
one.  The operator is therefore not generally idempotent on the full
algebra.  Nevertheless, integrability gives
\begin{equation}
 D_A^2\Phi=[F,\Phi]=0,
 \qquad F\in\mathfrak z_\Phi:=\ker\ad_\Phi .
 \label{eq:parallel-Higgs-integrability}
\end{equation}
On this centralizer, $\cP_{\uu}(\Phi)$ acts as the identity, and
$\mathsf P=1-\cP_\Phi$ is an exact orthogonal projection.  In particular,
\begin{equation}
 \mathsf P^2F=\mathsf PF=K,
 \qquad D_A\ast K=0.
 \label{eq:parallel-Higgs-curvature-projection}
\end{equation}
The second relation uses the connection equation and $D_A\Phi=0$.
This is idempotence on the compatible curvature, not on arbitrary
algebra-valued variations.  For generic nonzero, distinct skew-eigenvalue
magnitudes, $\mathfrak z_\Phi\simeq\uu(1)^n$ and the projected curvature
lies in an $(n-1)$-dimensional Abelian subspace, not an $\su(n)$ sector.
For $\Phi=vJ$, instead, $\mathfrak z_\Phi=\uu(n)$ and its orthogonal
complement to $\Phi$ is $\su(n)$.

The condition $D_A\Phi=0$ is not sufficient to solve the unrestricted
Higgs equation: it still requires
\begin{equation}
 \frac1{g^2}\cJ_\Phi(F,K)
 +\frac{\partial V}{\partial\Phi}\vol=0.
 \label{eq:parallel-Higgs-normal-condition}
\end{equation}
Consequently it is not, by itself, a proof of a consistent truncation.
Nor can the rank of the unrestricted fluctuation problem be inferred
from $\mathcal H_\Phi=K$ on the background.  Perturbations need not obey
$D_A\delta\Phi+[\delta A,\Phi]=0$ and can probe the remaining directions
of $\mathcal Q_\Phi$.  Conversely, an orbit-valued field has an exact
projector even when $D_A\Phi\ne0$: in unitary gauge,
$D_A\Phi=v[X,J]$.  We do not impose covariant constancy as an extra
off-shell constraint in the Hamiltonian analysis.

Let us clarify the scope of the present analysis and its possible
extensions. The exact nonlinear constraint count applies to the
separate orbit-valued model of Sec.~\ref{sec:constrained}. At finite $\lambda_H$, the particle spectrum established
here is the quadratic spectrum about $\Phi=vJ$, $A=0$,
with $Z_\Phi>0$. The off-orbit analysis additionally
characterizes the kinetic operator and frozen-background
quadratic coefficients. The present analysis does not establish a nonlinear
degree of freedom count in an open neighborhood of the
vacuum in the full adjoint field space. A controlled matching
to a low-energy effective theory and a nonlinear Hamiltonian analysis
away from the vacuum orbit remain subjects for future investigation.

%%%%%%%%%%%%%%%%%%%%%%%%%%%%%%%%%%%%%%%%%%%%%
\section{Theory III: exactly constrained Higgs and exact projectors}
\label{sec:constrained}
%%%%%%%%%%%%%%%%%%%%%%%%%%%%%%%%%%%%%%%%%%%%%

We now define a separate theory in which the Higgs field is restricted
exactly to the complex-structure orbit.  This restriction is part of the
off-shell field space, not a gauge choice and not an approximation justified
solely by a large finite $\lambda_H$.  The unrestricted action and its
off-orbit caveats remain those of Sec.~\ref{sec:dynamical}.  Here we give
two equivalent classical implementations of the constraint, derive their
equations, and analyze the regular local dynamics.

%%%%%%%%%%%%%%%%%%%%%%%%%%%%%%%%%%%%%%%%%%%%%
\subsection{Field space, symmetry, and exact projector identities}
\label{subsec:constrained-geometry}
%%%%%%%%%%%%%%%%%%%%%%%%%%%%%%%%%%%%%%%%%%%%%

The constraint is a matrix equation in the defining real representation:
\begin{equation}
 C(\Phi):=\Phi^2+v^2\mathbf1=0,
 \qquad
 \Phi^I{}_K\Phi^K{}_L=-v^2\delta^I{}_L,
 \qquad \Phi^T=-\Phi.
 \label{eq:orbit-matrix-constraint}
\end{equation}
The square denotes ordinary matrix multiplication, not a Lie bracket.
Taking its trace yields $-\Tr\Phi^2=2nv^2$, but this single norm equation
does not imply the matrix constraint.  The field $J_\Phi=\Phi/v$ is an
orthogonal complex structure.  We choose the connected orbit containing
$vJ$,
\begin{equation}
 \mathcal O_J=\{v g^{-1}Jg:g\in SO(2n)\}\simeq SO(2n)/U(n).
 \label{eq:constrained-orbit}
\end{equation}
The full matrix constraint has two $SO(2n)$ orbits, distinguished by the
sign of the Pfaffian relative to that of $vJ$; choosing
$\operatorname{Pf}\Phi=\operatorname{Pf}(vJ)$ selects the one used here.
On a connected spacetime a smooth constrained field cannot pass between
these components.  A multiplier imposing only $C=0$ includes both unless
the component is specified separately.

For a local gauge transformation $g(x)\in SO(2n)$,
\begin{equation}
 \Phi\mapsto g^{-1}\Phi g,
 \qquad A\mapsto g^{-1}Ag+g^{-1}\dd g,
 \qquad C(\Phi)\mapsto g^{-1}C(\Phi)g.
 \label{eq:constrained-gauge-symmetry}
\end{equation}
Both the constraint and the chosen orbit are preserved. Thus the gauge
group remains the full $SO(2n)$, while the coset $SO(2n)/U(n)$ is the
Higgs target space. The stabilizer of each point on this orbit is
conjugate to $U(n)$.

At each allowed field value the algebra splits orthogonally into
\begin{equation}
 \uu(n)_\Phi=\{Y:[\Phi,Y]=0\},
 \qquad \m_\Phi=\{Y:\{\Phi,Y\}=0\}.
\end{equation}
For $Y\in\m_\Phi$, $\Phi Y\Phi=v^2Y$, and hence
\begin{equation}
 [\Phi,[\Phi,Y]]
 =\Phi^2Y-2\Phi Y\Phi+Y\Phi^2=-4v^2Y.
 \label{eq:constrained-double-commutator}
\end{equation}
It vanishes instead on $\uu(n)_\Phi$.  Therefore
\begin{align}
 \cP_\m(\Phi)&=-\frac{\ad_\Phi^2}{4v^2},&
 \cP_\uu(\Phi)&=1+\frac{\ad_\Phi^2}{4v^2},\\
 \cP_\Phi Y&=\frac{\langle\Phi,Y\rangle}{2nv^2}\Phi,&
 \mathsf P(\Phi)&=\cP_\su(\Phi)=\cP_\uu(\Phi)-\cP_\Phi
 \label{eq:constrained-projectors}
\end{align}
are exact orthogonal projectors, with
\begin{equation}
 \mathsf P^2=\mathsf P=\mathsf P^\dagger,
 \qquad \operatorname{rank}\mathsf P=n^2-1,
 \qquad 1=\mathsf P+\cP_\Phi+\cP_\m.
 \label{eq:constrained-projector-identities}
\end{equation}
These are off-shell identities everywhere on the allowed field space.
They do not require $D_A\Phi=0$.  The complementary space must be defined
relative to the actual $\Phi$, or transported with it under conjugation.

%%%%%%%%%%%%%%%%%%%%%%%%%%%%%%%%%%%%%%%%%%%%%
\subsection{Intrinsic formulation: the constraint built into the field}
\label{subsec:constrained-intrinsic}
%%%%%%%%%%%%%%%%%%%%%%%%%%%%%%%%%%%%%%%%%%%%%

To distinguish the Higgs parametrization from gauge transformations,
introduce a group-valued field $k(x)$ and locally set
\begin{equation}
 \Phi= k(x)^{-1} \, (vJ) \, k(x),\qquad k(x)\in SO(2n).
 \label{eq:scope-exact-orbit}
\end{equation}
The group-valued field is redundant: $k\mapsto h k$, $h(x)\in U(n)$,
leaves $\Phi$ unchanged.  With this convention its equivalence classes
form $U(n)\backslash SO(2n)$, identified by inversion with the orbit
$SO(2n)/U(n)$.  Thus the number of independent scalar functions is
\begin{equation}
 n(2n-1)-n^2=n(n-1)=r.
 \label{eq:orbit-coordinate-count}
\end{equation}
This counts independent field components before gauge fixing,
not physical propagating degrees of freedom. A local representative $k=e^\tau$, with $\tau\in\m$, gives $\Phi=e^{-\tau}vJe^\tau$. Since $\Phi$ is restricted to the orbit, there is no independent normal field $\rho$.
Because the orbit is curved, the expansion of
$e^{-\tau}vJe^\tau-vJ$ contains, beyond linear order, components
normal to the orbit at $vJ$. These are determined by $\tau$ and maintain the constraint
$\Phi^2=-v^2\mathbf1$. They do not represent independent fluctuations
away from the orbit. Indeed, expanding around the reference point gives
\begin{equation}
 \Phi=vJ+\delta\Phi^{(1)}+\delta\Phi^{(2)}+O(\tau^3),
 \qquad
 \delta\Phi^{(1)}=v[J,\tau],
 \qquad
 \delta\Phi^{(2)}=\frac v2[[J,\tau],\tau].
\end{equation}
The linear term is generally nonzero and belongs to $\m$, so it
describes a tangent displacement along the orbit. Its contribution
to the linearized constraint nevertheless vanishes:
\begin{equation}
 \{vJ,\delta\Phi^{(1)}\}
 =v^2\{J,[J,\tau]\}
 =v^2[J^2,\tau]=0.
\end{equation}
At second order, the constraint requires
\begin{equation}
 \{vJ,\delta\Phi^{(2)}\}  +\bigl(\delta\Phi^{(1)}\bigr)^2=0.
\end{equation}
Thus the quadratic term, which belongs to $\uu(n)$, compensates
for the square of the linear tangent displacement. This normal
component relative to the reference point is fixed by $\tau$;
it is not an independent fluctuation away from the orbit. Under the full gauge group, $k\mapsto kg$ implements
Eq.~\eqref{eq:constrained-gauge-symmetry}.

Define the action directly on this field space by
\begin{equation}
 S_{\mathcal O}[A,k]=
 -\frac1{2g^2}\int_M\langle K,\ast K\rangle
 -\frac{Z_\Phi}{2}\int_M\langle D_A\Phi,\ast D_A\Phi\rangle,
 \qquad K=\mathsf P(\Phi)F.
 \label{eq:constrained-action}
\end{equation}
Here the constant $g$ in the prefactor is the gauge coupling, whereas
$k(x)$ parametrizes the orbit.  Equivariance makes this action fully
$SO(2n)$ gauge invariant.

The constrained scalar sector has a direct interpretation as a gauged
nonlinear sigma model. The orbit $\mathcal O_J\simeq SO(2n)/U(n)$ is
a compact Hermitian symmetric space of type DIII, familiar from
supersymmetric and gauged sigma-model
constructions~\cite{Higashijima:1999ki,GrootNibbelink:2000hu}.
The same target also appears in effective world-sheet theories of
non-Abelian vortices~\cite{Gudnason:2010rm}.
To make this identification explicit, introduce local coordinates
$q^i$, $i=1,\ldots,n(n-1)$, on the orbit and write $\Phi=\Phi(q)$.
The invariant trace pairing induces the target-space metric
\begin{equation}
 G_{ij}(q)
 :=
 \left\langle \partial_i\Phi,\partial_j\Phi\right\rangle
 =
 -\operatorname{Tr}\!\left(\partial_i\Phi\,\partial_j\Phi\right),
 \qquad
 \partial_i\Phi:=\frac{\partial\Phi}{\partial q^i}.
 \label{eq:orbit-sigma-metric}
\end{equation}
This metric is positive definite on the tangent space to the orbit
and agrees, up to overall normalization, with its standard
$SO(2n)$-invariant metric.

To express the gauging in these coordinates, let $T_\alpha$ be a basis
of $\so(2n)$ and write $A=A^\alpha T_\alpha$. Define the Killing vector
fields $\xi_\alpha=\xi_\alpha^i\partial_i$ by
\begin{equation}
 \xi_\alpha^i(q)\,\partial_i\Phi=[\Phi,T_\alpha].
 \label{eq:orbit-sigma-killing}
\end{equation}
With our convention $\Phi\mapsto g^{-1}\Phi g$, an infinitesimal
transformation gives $\delta q^i=\epsilon^\alpha\xi_\alpha^i$.
The covariant derivative therefore takes the form
\begin{equation}
 \mathcal Dq^i:=dq^i-A^\alpha\xi_\alpha^i(q),
 \qquad
 D_A\Phi=\partial_i\Phi\,\mathcal Dq^i,
 \label{eq:orbit-sigma-covariant-derivative}
\end{equation}
and the scalar contribution to the action becomes
\begin{equation}
 S_\Phi
 =
 -\frac{Z_\Phi}{2}\int_M
 G_{ij}(q)\,\mathcal Dq^i\wedge *\mathcal Dq^j.
 \label{eq:orbit-sigma-action}
\end{equation}
For $Z_\Phi>0$, this is the standard gauged nonlinear sigma-model
kinetic term on $\mathcal O_J$. Here the full $SO(2n)$ action on the
target is gauged. The target-space dimension counts scalar coordinates
before gauge fixing; the physical degrees of freedom must be determined
from the complete action, including its degenerate curvature kinetic
term.

In this setup no constraining potential is necessary.  Every gauge-invariant potential
depending only on $\Phi$ is constant on a single orbit, and the quartic
potential~\eqref{eq:potential} vanishes identically there.  A constant
vacuum-energy term may be retained independently, but it has no Higgs
equation.  The parameter $v$ specifies the orbit rather than being
determined dynamically by minimization in this theory.

%%%%%%%%%%%%%%%%%%%%%%%%%%%%%%%%%%%%%%%%%%%%%
\subsection{Matrix-multiplier formulation and equivalence}
\label{subsec:constrained-multiplier}
%%%%%%%%%%%%%%%%%%%%%%%%%%%%%%%%%%%%%%%%%%%%%

Alternatively, let $\Phi$ be an unrestricted antisymmetric matrix
and introduce a symmetric matrix-valued zero-form
$\Lambda=\Lambda^T$, transforming as
$\Lambda\mapsto g^{-1}\Lambda g$. Write
\begin{equation}
 S_{\rm LM}[A,\Phi,\Lambda]
 =S_0[A,\Phi]
 +\frac12\int_M\Tr\!\left[\Lambda(\Phi^2+v^2\mathbf1)\right]\vol.
 \label{eq:constrained-multiplier-action}
\end{equation}
Here $S_0$ is any gauge-invariant action defined smoothly near the
chosen orbit that agrees with Eq.~\eqref{eq:constrained-action}
on the orbit. For definiteness, we use
Eq.~\eqref{eq:dynamicalaction} without its potential, with the
maps of Eq.~\eqref{eq:dynamicprojector} defined for $\Phi\ne0$.
These maps need not be exact projectors away from the constraint
surface. The multiplier action remains fully $SO(2n)$ invariant.

Variation of $\Lambda$ imposes $C(\Phi)=0$. A single scalar
multiplier coupled to $\Tr C$ would enforce only the norm and
is insufficient. Write the variation of $S_0$ with respect to
the unrestricted Higgs field as
\begin{equation}
 \delta_\Phi S_0
 =\int_M\langle\delta\Phi,\mathcal E_\Phi^{(0)}\rangle,
 \qquad
 \mathcal E_\Phi^{(0)}
 =Z_\Phi D_A\ast D_A\Phi-\frac1{g^2}\cJ_\Phi(F,K)
 =:e_\Phi\vol,
 \label{eq:constrained-ambient-E}
\end{equation}
where the source convention is Eq.~\eqref{eq:Jsource}.
Using $\langle Y,Z\rangle=-\Tr(YZ)$, the variation of the
multiplier term is
\begin{equation}
 \delta_\Phi S_{\rm mult}
 =-\frac12\int_M
 \langle\delta\Phi,\{\Lambda,\Phi\}\rangle\vol.
\end{equation}
The Higgs equation is therefore
\begin{equation}
 e_\Phi=\tfrac12\{\Lambda,\Phi\}=:N,
 \label{eq:constrained-multiplier-E}
\end{equation}
or, explicitly,
\begin{equation}
 Z_\Phi D_A\ast D_A\Phi-\frac1{g^2}\cJ_\Phi(F,K)
 =\tfrac12\{\Lambda,\Phi\}\vol.
 \label{eq:constrained-ambient-Ev2}
\end{equation}

To compare this equation with the intrinsic formulation, decompose
an unrestricted Higgs variation at a point on the constraint
surface into tangent and normal parts:
\begin{align}
 \delta\Phi
 &=\delta\Phi_{\rm tan}+\delta\Phi_{\rm nor},
 \\
 \delta\Phi_{\rm tan}
 &=-\frac1{4v^2}[\Phi,[\Phi,\delta\Phi]],
 &
 \{\Phi,\delta\Phi_{\rm tan}\}&=0,
 \\
 \delta\Phi_{\rm nor}
 &=\delta\Phi+\frac1{4v^2}[\Phi,[\Phi,\delta\Phi]],
 &
 [\Phi,\delta\Phi_{\rm nor}]&=0.
\end{align}
The tangent variations can equivalently be written as
$\delta\Phi_{\rm tan}=[\Phi,\eta]$. Their effect on the constraint
is
\begin{equation}
 \delta_{\rm tan}C=0,
 \qquad
 \delta_{\rm nor}C=2\Phi\,\delta\Phi_{\rm nor}.
\end{equation}
At the reference point $\Phi=vJ$, these variations take the form
\begin{equation}
 \delta\Phi_{\rm tan}=v[J,\delta\tau],
 \qquad \delta\tau\in\m,
 \qquad
 \delta\Phi_{\rm nor}=\delta\rho,
 \qquad \delta\rho\in\uu(n).
\end{equation}
Only tangent variations are allowed in the intrinsic formulation.
The multiplier formulation allows both, with the constraint
imposed by variation of $\Lambda$.

On $C=0$, the multiplier force satisfies
\begin{equation}
 [\Phi,N]=\tfrac12[\Phi^2,\Lambda]=0.
\end{equation}
It is therefore normal to the orbit:
\begin{equation}
 \langle N,[\Phi,\eta]\rangle
 =\langle[N,\Phi],\eta\rangle=0.
\end{equation}
Consequently, the tangent part of
Eq.~\eqref{eq:constrained-multiplier-E} is
$[\Phi,e_\Phi]=0$, precisely the Higgs equation obtained by
tangent variations of Eq.~\eqref{eq:constrained-action}.
The normal part determines the required multiplier force.

Conversely, suppose $A$ and $\Phi$ satisfy
$\Phi^2=-v^2\mathbf1$ and $[\Phi,e_\Phi]=0$, where $e_\Phi$
is calculated from $S_0$ independently of $\Lambda$. Define
\begin{equation}
 \Lambda=-\frac{\Phi e_\Phi}{v^2}.
 \label{eq:constrained-multiplier-solution}
\end{equation}
Because $\Phi$ and $e_\Phi$ are commuting antisymmetric matrices,
this $\Lambda$ is symmetric. Substitution gives
\begin{equation}
 \tfrac12\{\Lambda,\Phi\}
 =-\frac{\Phi e_\Phi\Phi+\Phi^2e_\Phi}{2v^2}
 =e_\Phi,
\end{equation}
which verifies Eq.~\eqref{eq:constrained-multiplier-E} and
establishes the equivalence of the two Higgs equations on
the chosen orbit.

This proves local classical equivalence with the intrinsic variation.
Symmetric components of $\Lambda$ anticommuting with $\Phi$ do not enter
the force on the constraint surface; they reflect redundant multiplier
components and do not add propagating fields.  A canonical analysis in
ambient coordinates must remove these redundancies rather than treating
all matrix entries of $C$ as independent constraints.  On the constraint surface, the linearized map
$\delta\Phi\mapsto\delta C=\{\Phi,\delta\Phi\}$ has kernel
$\m_\Phi$. By the rank-nullity theorem,
\begin{equation}
 \operatorname{rank}(\delta C)
 =\dim\so(2n)-\dim\m_\Phi
 =n(2n-1)-n(n-1)=n^2.
\end{equation}
Thus the matrix constraint imposes $n^2$ independent conditions
locally, removing the normal Higgs fluctuations.

Different choices of $S_0$ that agree on the orbit give the same
equations for the constrained fields. Their Higgs equations can
differ in the normal directions, but this difference can be
compensated by changing $\Lambda$.

Including the quartic potential~\eqref{eq:potential} in $S_0$
would not change these equations, since both the potential and
its first variation vanish on $C=0$.

In both constrained formulations, only the tangent part of the
Higgs equation from $S_0$ must vanish. Its normal part is balanced
by the multiplier force. This differs from imposing the full
unrestricted Higgs equation at $\Phi=vJ$, which would also require
the normal part to vanish.

%%%%%%%%%%%%%%%%%%%%%%%%%%%%%%%%%%%%%%%%%%%5
\subsection{Covariant equations and the orbit Noether identity}
\label{subsec:constrained-equations}
%%%%%%%%%%%%%%%%%%%%%%%%%%%%%%%%%%%%%%%%%%%5

A variation of the Higgs field within the constrained orbit,
with the connection held fixed, satisfies
\begin{equation}
 \delta\Phi=[\Phi,\eta],\qquad
 \{\Phi,\delta\Phi\}=0.
\end{equation}
All tangent variations are generated this way, because
$\ad_\Phi^2=-4v^2$ on $\m_\Phi$.  The field equations are
\begin{align}
 D_A\ast K+g^2Z_\Phi[\Phi,\ast D_A\Phi]&=0,
 \label{eq:Aeomdynamic-exact-orbit}\\
 \cP_\m(\Phi)\mathcal E_\Phi^{(0)}=0
 \quad\Longleftrightarrow\quad
 [\Phi,\mathcal E_\Phi^{(0)}]&=0.
 \label{eq:constrained-Higgs-equation}
\end{align}
Normal components of $\mathcal E_\Phi^{(0)}$ are not extra equations;
they determine the constraint force in the multiplier description.
With
\begin{equation}
 \widehat{\mathcal E}_A^{\mathcal O}
 =g^{-2}D_A\ast K+Z_\Phi[\Phi,\ast D_A\Phi],
\end{equation}
gauge invariance gives, on the constrained field space but off shell,
\begin{equation}
 D_A\widehat{\mathcal E}_A^{\mathcal O}
 -[\Phi,\mathcal E_\Phi^{(0)}]=0.
 \label{eq:constrained-Noether}
\end{equation}
Thus the orbit equation follows from the connection equation.  It does
not introduce an independent scalar wave equation after gauge fixing.
This is the restricted-field-space counterpart of
Eq.~\eqref{eq:dynamic-Noether-identity}.

The constraint differentiated covariantly also gives
\begin{equation}
 \{\Phi,D_A\Phi\}=0.
\end{equation}
Thus $D_A\Phi$ takes values in the tangent space $\mathfrak m_\Phi$
to the orbit at $\Phi$. This does not require $D_A\Phi=0$.
In particular, idempotence of the projectors is an algebraic
off-shell identity throughout the constrained field space,
whereas covariant constancy is a further condition that may
follow on a particular solution branch.

%%%%%%%%%%%%%%%%%%%%%%%%%%%%%%%%%%%%%%%%%%%%%%%%%%%%%%%%%%%%%%%%%%
\subsection{Dressed connection, local unitary gauge, and null directions}
\label{subsec:constrained-unitary}
%%%%%%%%%%%%%%%%%%%%%%%%%%%%%%%%%%%%%%%%%%%%%%%%%%%%%%%%%%%%%%%%%%

The transformation $g=k^{-1}$ sets $\Phi=vJ$ locally and sends the
connection to
\begin{equation}
 \mathcal A=kAk^{-1}+k\dd k^{-1}
 =B+cJ+X,
 \qquad B\in\su(n),\quad X\in\m.
 \label{eq:constrained-dressed-A}
\end{equation}
Under the original gauge transformation, accompanied by $k\mapsto kg$,
$\mathcal A$ is invariant.  Under the representative redundancy
$k\mapsto hk$ it transforms as
$\mathcal A\mapsto h\mathcal A h^{-1}+h\dd h^{-1}$, with $h\in U(n)$.
The residual local symmetry in this chart is therefore $U(n)$.
Globally, the construction uses representatives $k$ defined on
spacetime patches and related by $U(n)$ transformations on their
overlaps. The unitary-gauge action written below is understood
in these local frames. Its Lagrangian is invariant under the
transition functions and therefore defines a global action,
even when no single smooth unitary gauge exists on all of spacetime. There are no Higgs zeros or changes of orbit type within this
fixed-$v$ constrained field space, although its reduced bundle can be
topologically nontrivial.

The identities
\begin{equation}
 kD_A\Phi k^{-1}=v[X,J],
 \qquad
 K_J:=kKk^{-1}=F_B+\cP_\su(X\wedge X)
 \label{eq:constrained-dressed-K}
\end{equation}
give the exact local unitary-gauge action
\begin{equation}
 S_{\mathcal O}^{\rm UG}
 =-\frac1{2g^2}\int_M\langle K_J,\ast K_J\rangle
 -2Z_\Phi v^2\int_M\langle X,\ast X\rangle.
 \label{eq:constrained-UG-action}
\end{equation}
No derivative of $X$ and no central connection $c$ occurs in this
action. The Higgs kinetic term therefore supplies an algebraic
quadratic term for $X$, without a vector kinetic term.
The orbit constraint removes the normal scalar fluctuations,
and unitary gauge removes the tangent fields.

In the gauged nonlinear sigma-model description, the full $SO(2n)$
gauge symmetry allows the tangent Higgs coordinates to be removed
locally by unitary gauge, while the exact orbit restriction excludes
independent normal fluctuations. The sigma-model kinetic term then
becomes an algebraic quadratic term for the coset connection $X$.
Since the projected curvature kinetic term contains no derivatives
of $X$, this connection is auxiliary, and the central connection $c$
drops out of the action entirely. With an ordinary nondegenerate
Yang-Mills kinetic term, the coset connection would also acquire
derivative terms, and the same scalar sector would supply masses
to propagating coset gauge fields through the conventional Higgs
mechanism. Thus the sigma-model target determines the geometry of
the scalar sector, while the projected gauge kinetic operator
changes its dynamical consequences.

Equation~\eqref{eq:constrained-UG-action} is the local unitary-gauge
form of the fully $SO(2n)$-invariant action, with manifest residual
gauge symmetry $U(n)$. Before gauge fixing, the dressed connection
$\mathcal A$ is invariant under simultaneous $SO(2n)$ transformations
of $A$ and $k$. Arbitrary $SO(2n)$ connection transformations applied
directly to $\mathcal A$ with $J$ held fixed are not symmetries
of the displayed action.

The absence of $c$ gives an additional exact local symmetry,
$c\mapsto c+\zeta$, for an arbitrary one-form $\zeta$.
This shift is independent of the ordinary residual $U(n)$ gauge
transformations. In the original variables it reads
\begin{equation}
 A\mapsto A+\zeta J_\Phi,
 \qquad
 \Phi\mapsto\Phi.
\end{equation}
It preserves $D_A\Phi$, and the induced change in $F$ is annihilated
by $\mathsf P$ because it lies in the central and tangent directions.
This additional symmetry explains why the central connection
contributes no local propagating degrees of freedom.

Projecting the connection equation in this chart gives
\begin{equation}
 D_B\ast K_J=0,\qquad
 [X,\ast K_J]+4g^2Z_\Phi v^2\ast X=0.
 \label{eq:constrained-split-E}
\end{equation}
The second equation is algebraic in $X$.  This equation always admits the solution $X=0$. Its linearization
about that solution is invertible when the curvature-dependent term
is sufficiently small compared with the algebraic scale
$g^2Z_\Phi v^2$.
Substituting this solution into
Eq.~\eqref{eq:constrained-UG-action} gives
\begin{equation}
 S_{\rm red}[B]
 =-\frac1{2g^2}\int_M\langle F_B,\ast F_B\rangle.
 \label{eq:constrained-reduced-YM}
\end{equation}
Moreover, Eq.~\eqref{eq:constrained-dressed-K} implies
$D_A\Phi=0$ on this branch. Covariant constancy thus follows
from the field equations on the regular branch, rather than
being imposed as an additional constraint.

This establishes an exact local classical reduction to $SU(n)$
Yang-Mills theory on the regular branch. The field $X$ is eliminated
through its own equation, and the undetermined central connection is
removed by the additional local shift symmetry. This result applies
near $X=0$ where the linearized algebraic operator is invertible.
Possible nonzero branches and the conditions under which they can
meet the $X=0$ branch are examined in
Sec.~\ref{subsec:nontrivial-X-branches}.

%%%%%%%%%%%%%%%%%%%%%%%%%%%%%%%%%%%%%%%%%%%%%%%%%%%%%%%%%%%%%%%%%%%%%
\subsection{Nontrivial branches of the auxiliary coset connection}
\label{subsec:nontrivial-X-branches}
%%%%%%%%%%%%%%%%%%%%%%%%%%%%%%%%%%%%%%%%%%%%%%%%%%%%%%%%%%%%%%%%%%%%%

The regular solution $X=0$ need not exhaust the algebraic solution
space. To examine possible nonzero branches, write the $X$ equation as
\begin{equation}
 [X,\ast K_J]+\mu^2\ast X=0,
 \qquad
 \mu^2:=4g^2Z_\Phi v^2,
 \qquad
 K_J=F_B+\cP_\su(X\wedge X).
 \label{eq:X-algebraic-branch-equation}
\end{equation}
This equation is pointwise algebraic in $X$ and contains a cubic
term because $K_J$ depends on $X\wedge X$. A solution of the full
field equations must also satisfy the remaining connection equation
\begin{equation}
 D_B\ast K_J=0.
 \label{eq:nontrivial-branch-B-equation}
\end{equation}

\subsubsection{Cartan-valued quadratic ansatz}

A useful class of configurations is obtained when $X\wedge X$ lies
in a Cartan subalgebra of $\uu(n)$. In the complexified algebra,
consider a pair of opposite coset-root generators satisfying
\begin{equation}
 [E_\alpha,E_{-\alpha}]=H_\alpha,
\end{equation}
where $H_\alpha$ belongs to the complexified Cartan subalgebra.
With the convention $E_{-\alpha}=\overline{E_\alpha}$, the reality
condition on $X$ gives the ansatz
\begin{equation}
 X=uE_\alpha+\bar{u} E_{-\alpha}.
 \label{eq:cartan-X-ansatz}
\end{equation}
Here $u$ is a complex one-form. To make the reality condition explicit,
choose real generators $T_1,T_2\in\m$ spanning the corresponding
root plane and write
\begin{equation}
 E_\alpha=\frac12(T_1-\ii T_2),
 \qquad
 E_{-\alpha}=\frac12(T_1+\ii T_2),
 \qquad
 u=a+\ii b,
\end{equation}
where $a$ and $b$ are real one-forms. Then
\begin{equation}
 X=aT_1+bT_2,
 \qquad
 u\wedge\bar u=-2\ii\,a\wedge b.
\end{equation}
With this convention,
\begin{equation}
 H_\alpha=[E_\alpha,E_{-\alpha}]
 =\frac{\ii}{2}[T_1,T_2],
 \qquad
 \overline{H_\alpha}=-H_\alpha.
\end{equation}
Although $H_\alpha$ is purely imaginary, the quadratic expression
\begin{equation}
 X\wedge X
 =u\wedge\bar u\,H_\alpha
 =a\wedge b\,[T_1,T_2]
\end{equation}
is real and need not vanish. The imaginary factors in
$u\wedge\bar u$ and $H_\alpha$ combine to give a real
$\uu(n)$-valued two-form. The corresponding expressions entering
Eq.~\eqref{eq:X-algebraic-branch-equation} are
\begin{equation}
 X\wedge X=u\wedge \bar{u}\,H_\alpha,
 \qquad
 \cP_\su(X\wedge X)
 =u\wedge \bar{u}\,\widehat H_\alpha,
 \qquad
 \widehat H_\alpha:=\cP_\su H_\alpha.
 \label{eq:cartan-X-square}
\end{equation}
The central component of $H_\alpha$, when present, is removed by
$\cP_\su$.

Suppose in addition that $F_B$ takes values in the same Cartan
subalgebra. Then $K_J$ is also Cartan-valued, and its adjoint action
preserves each root space. Evaluating the root $\alpha$ on this
curvature defines a complex two-form through
\begin{equation}
 \kappa_\alpha:=\alpha(K_J)\,, \quad [K_J,E_{\pm\alpha}]
 =\pm\kappa_\alpha E_{\pm\alpha},
\end{equation}
where the evaluation acts only on the Lie-algebra component.
The first term in Eq.~\eqref{eq:X-algebraic-branch-equation}
therefore becomes
\begin{equation}
 [X,\ast K_J]
 =-u\wedge\ast\kappa_\alpha\,E_\alpha
 +\bar u\wedge\ast\kappa_\alpha\,E_{-\alpha}.
\end{equation}
Applying the Hodge operator gives
\begin{equation}
 \ast[X,\ast K_J]
 =-\ast(u\wedge\ast\kappa_\alpha)\,E_\alpha
 +\ast(\bar u\wedge\ast\kappa_\alpha )\,E_{-\alpha}.
\end{equation}
This suggests defining the spacetime endomorphism
\begin{equation}
 \mathscr R_{\kappa_\alpha}(w)
 :=\ast\!\left(w\wedge\ast\kappa_\alpha\right)
\end{equation}
on complex one-forms. Applying the Hodge operator to
Eq.~\eqref{eq:X-algebraic-branch-equation} and using $\ast^2=1$
on one-forms in four-dimensional Lorentzian signature, we obtain
\begin{align}
 \mathscr R_{\kappa_\alpha}(u)&=\mu^2u, \label{eq:cartan-root-branch-equations1}
 \\
 \mathscr R_{\kappa_\alpha}(\bar{u})&=-\mu^2 \bar{u}.
 \label{eq:cartan-root-branch-equations2}
\end{align}
Since $K_J$ takes values in the compact real Cartan algebra,
$\overline{\kappa_\alpha}=-\kappa_\alpha$. Consequently,
\begin{equation}
 \overline{\mathscr R_{\kappa_\alpha}(u)}
 =-\mathscr R_{\kappa_\alpha}(\bar u).
\end{equation}
Thus Eqs.~\eqref{eq:cartan-root-branch-equations1} and
\eqref{eq:cartan-root-branch-equations2} are complex conjugates
and are compatible with the reality condition. They remain
nonlinear because $\kappa_\alpha$ depends on $u\wedge\bar{u}$
through
\begin{equation}
 K_J=F_B+u\wedge \bar{u}\,\widehat H_\alpha.
\end{equation}
For a nonzero solution, $u$ and $\bar{u}$ must therefore belong
to eigenspaces with opposite eigenvalues of the spacetime
endomorphism determined by this same configuration.

Within the Cartan ansatz, the algebraic $X$ equation is thus
equivalent to a nonlinear eigenform equation and its complex
conjugate. This characterization does not guarantee nonzero
solutions for an arbitrary background curvature. A solution
of the full theory must also satisfy the remaining connection
equation~\eqref{eq:nontrivial-branch-B-equation}.

For $n=2$, the coset directions are $SU(2)$ singlets, so
$\kappa_\alpha=0$. With $\mu^2>0$,
Eqs.~\eqref{eq:cartan-root-branch-equations1} and
\eqref{eq:cartan-root-branch-equations2} therefore imply $u=0$.
Nonzero solutions within this ansatz can occur only for $n\geq3$.

%%%%%%%%%%%%%%%%%%%%%%%%%%%%%%%%%%%%%%%%%%
\subsubsection{Fixed-point formulation}
%%%%%%%%%%%%%%%%%%%%%%%%%%%%%%%%%%%%%%%%%%

Another approach to finding nontrivial branches is to rewrite the
algebraic equation as a fixed-point equation. Applying the Hodge
operator to Eq.~\eqref{eq:X-algebraic-branch-equation}, gives
\begin{equation}
 X=-\frac1{\mu^2}\mathcal T_{K_J}(X),
 \qquad
 \mathcal T_{K_J}(X):=\ast[X,\ast K_J].
 \label{eq:X-fixed-point}
\end{equation}
For a given solution, repeated substitution gives
\begin{equation}
 X=
 \left(-\frac1{\mu^2}\right)^r
 \mathcal T_{K_J}^{\,r}(X).
 \label{eq:X-fixed-point-iteration}
\end{equation}
Here $K_J$ is evaluated on that solution and held fixed when taking
powers of the operator. This identity does not itself construct a
solution, since both $X$ and the curvature $K_J(X)$ remain unknown.

Equation~\eqref{eq:X-fixed-point} can also be written as the
eigenvalue equation
\begin{equation}
 \mathcal T_{K_J}(X)=-\mu^2X.
 \label{eq:X-nonlinear-eigenvalue}
\end{equation}
A nonzero solution therefore requires $-\mu^2$ to be an eigenvalue
of $\mathcal T_{K_J}$. Since $K_J$ itself depends on $X$, this is
a nonlinear self-consistency condition. Equivalently, define
\begin{equation}
 \mathcal L_{K_J}(Y)
 :=\mu^2Y+\mathcal T_{K_J}(Y).
 \label{eq:X-full-operator}
\end{equation}
The exact equation is then
\begin{equation}
 \mathcal L_{K_J(X)}(X)=0.
 \label{eq:X-full-operator-equation}
\end{equation}
For fixed $K_J$, the operator $\mathcal L_{K_J}$ is linear in its
argument $Y$. The equation for $X$ is nevertheless nonlinear because
\begin{equation}
 K_J(X)=F_B+\cP_\su(X\wedge X).
\end{equation}

At $X=0$, the linearized operator is
\begin{equation}
 \mathcal L_B(\delta X)
 :=
 \mu^2\delta X+\mathcal T_{F_B}(\delta X)
 =
 \mu^2\delta X+\ast[\delta X,\ast F_B].
 \label{eq:X-linearized-operator}
\end{equation}
Separating the linear and cubic terms, the exact equation becomes
\begin{equation}
 \mathcal L_B(X)+\mathcal N(X)=0,
 \qquad
 \mathcal N(X):=
 \ast\!\left[X,\ast\cP_\su(X\wedge X)\right].
 \label{eq:X-linear-cubic-equation}
\end{equation}
Thus $\mathcal L_B(X)=0$ is the linearized equation, while
$\mathcal L_{K_J(X)}(X)=0$ includes the cubic interaction.

At each spacetime point, $\mathcal L_B$ is a finite-dimensional
linear operator whose invertibility is characterized by the
basis-independent condition
\begin{equation}
 \det\mathcal L_B\neq0
 \quad\Longleftrightarrow\quad
 -\mu^2\notin\operatorname{spec}(\mathcal T_{F_B}).
\end{equation}
Whenever this condition holds, the implicit-function theorem
ensures that $X=0$ is the unique solution in a neighborhood
of that configuration.

%%%%%%%%%%%%%%%%%%%%%%%%%%%%%%%%%%%%%%%%%%%%%%%%%%
\subsubsection{Bifurcation of a nonzero branch}
%%%%%%%%%%%%%%%%%%%%%%%%%%%%%%%%%%%%%%%%%%%%%%%%%%

A nonzero branch can meet the $X=0$ branch only at a background
for which the linearized operator becomes singular:
\begin{equation}
 \ker\mathcal L_{B_{\rm c}}\neq0,
 \qquad
 -\mu^2\in\operatorname{spec}
 \bigl(\mathcal T_{F_{B_{\rm c}}}\bigr).
 \label{eq:X-critical-background}
\end{equation}
Let $F_{\rm c}:=F_{B_{\rm c}}$ denote the curvature of the
critical background connection and write
$\mathcal L_{\rm c}:=\mathcal L_{B_{\rm c}}$.
For a bifurcation of solutions of the full theory, the critical
background must also satisfy $D_{B_{\rm c}}\ast F_{\rm c}=0$.

Let $\psi$ be a zero mode,
\begin{equation}
 \mathcal L_{\rm c}\psi=0,
\end{equation}
and consider a one-parameter deformation of the background,
\begin{equation}
 F_B=F_{\rm c}+\varepsilon\,\delta F+\cdots,
 \qquad
 X=a\psi+X_{\rm rem}.
 \label{eq:X-critical-expansion}
\end{equation}
Here $\varepsilon$ is a constant parameter labeling a family of
backgrounds. The amplitude $a=a(x;\varepsilon)$ is generally a
scalar function, since the algebraic equation for $X$ is solved
pointwise. The zero mode $\psi=\psi(x)$ is an $\m$-valued one-form,
and $X_{\rm rem}=X_{\rm rem}(x;\varepsilon)$ belongs to a chosen
complement of the critical kernel within the space of
$\m$-valued one-forms.

Since $\mathcal L_{\rm c}$ commutes with $\operatorname{ad}_J$,
its real zero modes occur in pairs. The scalar reduction below
assumes a consistent ansatz with a single independent real
critical amplitude after fixing the residual phase. If additional
critical amplitudes remain, the reduction instead gives coupled
amplitude equations. All projections below are understood within
the chosen ansatz.

Keeping the spacetime metric and $\mu^2$ fixed, the deformation
of the linear operator is
\begin{equation}
 \mathcal L_B
 =\mathcal L_{\rm c}
 +\varepsilon\mathcal L_1+O(\varepsilon^2),
 \qquad
 \mathcal L_1(Y)=\ast[Y,\ast\delta F].
 \label{eq:X-linear-operator-deformation}
\end{equation}
For a connection family
$B=B_{\rm c}+\varepsilon b+\cdots$, the curvature variation is
$\delta F=D_{B_{\rm c}}b$.

Assume that the projected linearized equation is invertible in
the complementary directions. The corresponding equations then
determine $X_{\rm rem}$ smoothly as a function of $a$ and
$\varepsilon$. Their leading forcing terms are proportional to
$\varepsilon a$ and $a^3$, so
\begin{equation}
 X_{\rm rem}=O(\varepsilon a,a^3).
 \label{eq:X-remainder-order}
\end{equation}
Substitution into Eq.~\eqref{eq:X-linear-cubic-equation} gives
\begin{equation}
 0=\mathcal L_{\rm c}X_{\rm rem}
 +\varepsilon a\,\mathcal L_1\psi
 +a^3\mathcal N(\psi)
 +O(\varepsilon^2a,\varepsilon a^3,a^5).
 \label{eq:X-critical-expanded-equation}
\end{equation}

Choose a left zero mode $\ell$, namely a linear functional
satisfying $\ell\circ\mathcal L_{\rm c}=0$.
Applying it to Eq.~\eqref{eq:X-critical-expanded-equation}
eliminates the first term and gives
\begin{equation}
 c_1\varepsilon a+c_3a^3
 +O(\varepsilon^2a,\varepsilon a^3,a^5)=0,
 \label{eq:X-amplitude-equation}
\end{equation}
with
\begin{align}
 c_1
 &=\ell\!\left(\mathcal L_1\psi\right)
 =\ell\!\left(\ast[\psi,\ast\delta F]\right),
 \\
 c_3
 &=\ell\!\left(\mathcal N(\psi)\right)
 =\ell\!\left(
 \ast[\psi,\ast\cP_\su(\psi\wedge\psi)]
 \right).
 \label{eq:X-amplitude-coefficients}
\end{align}
These coefficients are evaluated pointwise and may depend on
spacetime. Their values also depend on the chosen normalization
of $\psi$ and $\ell$. The coefficient $c_1$ measures the effect
of the background deformation, while $c_3$ comes from the cubic
interaction of the critical mode.

For $c_1c_3\neq0$, a nonzero branch may satisfy
\begin{equation}
 a^2=-\frac{c_1}{c_3}\varepsilon+O(\varepsilon^2),
 \label{eq:X-bifurcating-branch}
\end{equation}
provided the right-hand side is positive. If either coefficient
vanishes, higher-order terms must be examined.

The coefficients in Eq.~\eqref{eq:X-amplitude-coefficients}
refer to a prescribed background family $B(\varepsilon)$.
Constructing a solution of the full theory also requires
Eq.~\eqref{eq:nontrivial-branch-B-equation}. If solving that
equation induces a curvature correction $a^2F_2$, its contribution
must be included in the cubic coefficient:
\begin{equation}
 c_3^{\rm coupled}
 =\ell\!\left(
 \mathcal N(\psi)+\ast[\psi,\ast F_2]
 \right).
 \label{eq:X-coupled-cubic-coefficient}
\end{equation}
The algebraic amplitude equation alone therefore does not
establish the existence of a branch of the full theory.

Equation~\eqref{eq:X-critical-background} is necessary only for
a nonzero branch that joins the $X=0$ branch. Other solutions
at finite $X$ are not excluded by invertibility of $\mathcal L_B$
at $X=0$. Any such branch must solve both
Eqs.~\eqref{eq:X-algebraic-branch-equation} and
\eqref{eq:nontrivial-branch-B-equation}. The reduced Yang-Mills
action~\eqref{eq:constrained-reduced-YM} therefore describes
the regular local branch without excluding other branches
of the constrained theory.

%%%%%%%%%%%%%%%%%%%%%%%%%%%%%%%%%%%%%%%%%%%%%%%
\subsection{Hamiltonian analysis on the regular constrained branch}
\label{subsec:constrained-Hamiltonian}
%%%%%%%%%%%%%%%%%%%%%%%%%%%%%%%%%%%%%%%%%%%%%%%

The purpose of this subsection is to show that the nonlinear
degree-of-freedom count is valid at a zero-field background of the
orbit-constrained theory and in a neighborhood of that background.
Throughout, the Higgs field is restricted to the orbit of $vJ$, with $v\ne 0$ and $Z_\Phi>0$, so that $\mathsf P^2=\mathsf P$ on the entire allowed field
space. This restriction does not require the gauge fields to vanish or
$D_A\Phi=0$. In unitary gauge we examine the background
$\Phi=vJ$, $B_\mu=X_\mu=0$, $\pi^i=0$; the null central connection
$c_\mu$ can be set to zero by its shift symmetry. The neighborhood
considered below lies within the constrained phase space of the orbit-constrained theory.

The momenta, primary constraints, $SU(n)$ Gauss chain, and accidental
shift symmetry of $c_\mu$ follow the canonical analysis starting with
Eq.~\eqref{eq:fixedprimary}. The remaining question is the rank of the
$X$-sector constraint matrix. In the fixed-projector theory, this
matrix vanishes at $B_\mu=X_\mu=0$, $\pi^i=0$, making that background
an irregular point of its constraint system. In the orbit-constrained
theory, the Higgs kinetic term supplies the algebraic quadratic term
in Eq.~\eqref{eq:constrained-UG-action}. We show that its nondegenerate
mass matrix makes the chosen background a regular point of the
$X$-sector constraint system.

In the local unitary gauge $\Phi=vJ$, the projected-curvature
density has the same form as $\cL_{\rm fix}$. The Higgs kinetic term
in Eq.~\eqref{eq:constrained-UG-action} adds an algebraic quadratic
term for $X_\mu$, so the Lagrangian density is
\begin{equation}
 \cL=\cL_{\rm fix}
 +\frac12M^2_{\alpha\beta}X_0^\alpha X_0^\beta
 -\frac12M^2_{\alpha\beta}X_i^\alpha X_i^\beta,
 \label{eq:dynLham}
\end{equation}
where $\cL_{\rm fix}$ is given in \eqref{eq:L_fix} and $M^2_{\alpha\beta}$ is given in Eq.~\eqref{eq:Xmassmatrix}.
Because this term contains no velocities, it leaves the primary
constraints unchanged. It does, however, modify the secondary
constraints obtained by preserving $p_\alpha^\mu\approx0$.

Preservation of $p_\alpha^0$ gives
\begin{equation}
 \cC_\alpha^0
 =\cC_{\alpha; \mathrm{fix}}^0 - M^2_{\alpha\beta}X_0^\beta
 \approx0 , \qquad \cC_{\alpha; \mathrm{fix}}^0 = -f_{a\alpha\beta}\pi_a^{i}X_i^\beta
 \label{eq:C0dynamic}
\end{equation}
Preservation of $p_\alpha^i$ gives
\begin{equation}
 \cC_\alpha^i
 =\cC_{\alpha; \mathrm{fix}}^i +M^2_{\alpha\beta}X_i^\beta \approx0 ,\qquad \cC_{\alpha; \mathrm{fix}}^i = -f_{a\beta\alpha}\pi_a^{i}X_0^\beta
 +\frac1{g^2}f_{a\alpha\beta}K_{ij}^{a}X_j^\beta
 .
 \label{eq:Cidynamic}
\end{equation}
% \begin{equation}
%  \cC_\alpha^i
%  =M^2_{\alpha\beta}X_i^\beta
%  -f_{a\beta\alpha}\pi_a^{i}X_0^\beta
%  +\frac1{g^2}f_{a\alpha\beta}K_{ij}^{a}X_j^\beta
%  \approx0.
%  \label{eq:Cidynamic}
% \end{equation}
The field-dependent contributions to the $X$-sector constraint
matrix follow from Eq.~\eqref{eq:fixedhessianblocks}. The Higgs
kinetic term adds the mass matrix to the diagonal blocks. With
the secondary-constraint conventions of
Eqs.~\eqref{eq:C0dynamic}-\eqref{eq:Cidynamic} one obtains
\begin{equation}
 \{p_\alpha^\mu(x),\cC_\beta^\nu(y)\}
 =\Delta_{\alpha\beta}^{\mu\nu}(x,y)
 -M^2_{\alpha\beta}\eta^{\mu\nu}\delta^3(x-y),
 \label{eq:SCmatrixdynamic}
\end{equation}
where $\Delta_{\alpha\beta}^{\mu\nu}$ is the fixed-theory
matrix given in Eq.~\eqref{eq:fixedhessianblocks}.
At the chosen background,
$\Delta_{\alpha\beta}^{\mu\nu}=0$.
The nondegenerate mass matrix therefore makes the block
$\{p,\cC\}$ invertible. This property persists for sufficiently
small $\pi^i$, $F_{ij}(B)$, and $X_\mu$.
Thus $(p_\alpha^\mu,\cC_\alpha^\mu)$ form $8r$
second-class constraints, with the local solution
\begin{equation}
 X_\mu=0,
 \qquad p_\alpha^\mu=0.
\end{equation}
Preservation of $\cC_\alpha^\mu$ determines the Lagrange
multipliers enforcing the primary constraints
$p_\alpha^\mu\approx0$. No further constraints arise in
the $X$ sector, so the Dirac algorithm closes. The coset fields are regular auxiliary
variables: the algebraic mass term does not give them propagating
vector modes.

The remaining constraints have the classification used in
Eq.~\eqref{eq:fixedcount}. Hence
\begin{equation}
 N_{\rm phys}
 =\frac12\left[8(h+1+r)-2(2h+4)-8r\right]
 =2h=2(n^2-1).
 \label{eq:dyncount}
\end{equation}
The value of the count agrees with that on the regular branch of the
fixed-projector theory. The result established here is that the count
remains valid at the zero-field background of the orbit-constrained
theory, because the Higgs kinetic term gives the $X$-sector Dirac
matrix a nondegenerate, field-independent contribution. It is stable
in a neighborhood within the constrained phase space and agrees
with the perturbative spectrum.

This degree-of-freedom count applies to the exactly
orbit-constrained model. The unrestricted theory with a finite
Higgs potential has a different field space and requires a
separate nonlinear constraint analysis, because its vector
kinetic rank can change away from the orbit; see
Eq.~\eqref{eq:off-orbit-canonical-momentum}.

%%%%%%%%%%%%%%%%%%%%%%%%%%%%%%%%%%%%%%%%%%%%%%%%%%%%%%%

The count above was obtained from the local unitary-gauge action
\eqref{eq:constrained-UG-action}. We now explain why it also applies
to the covariant action \eqref{eq:constrained-action}, accounting
for the Higgs variables and the broken gauge constraints.
In the covariant formulation, the orbit restriction already excludes
independent normal Higgs fluctuations. The remaining
$r=n(n-1)$ scalar coordinates describe the orientation of $\Phi$
within the orbit. Setting $\Phi=vJ$, with the connection transformed
simultaneously as in Eq.~\eqref{eq:constrained-dressed-A}, fixes these
$r$ gauge directions locally.

The momentum conjugate to the spatial connection $A_i$ is
\begin{equation}
 \Pi^i:=\frac{\partial\cL}{\partial\dot A_i}
 =\frac1{g^2}\cP_{\su}(\Phi)F_{0i}
 =\frac1{g^2}K_{0i},
 \qquad
 F_{0i}=\dot A_i-D_iA_0,
\end{equation}
where the derivative is understood using the invariant inner product
on $\so(2n)$. This expression uses the self-adjointness and
idempotence of $\cP_{\su}(\Phi)$ on the constrained orbit.
%The Higgs kinetic term contains no $\dot A_i$ and therefore does not contribute to $\Pi^i$.
Since $\Pi^i$ lies in the
image of the projector, it satisfies the primary constraints
\begin{equation}
 (1-\cP_{\su}(\Phi))\Pi^i\approx0.
 \label{eq:covprimary}
\end{equation}
and preservation of the temporal primary constraints
$\Pi_A^0\approx0$ gives the full Gauss law
\begin{equation}
 \cG=D_i\Pi^i+[\Phi,\Pi_\Phi]\approx0,
 \qquad
 \Pi_\Phi=Z_\Phi D_0\Phi.
 \label{eq:fullgauss}
\end{equation}
Here $\Pi_\Phi$ is tangent to the orbit. Indeed, A matrix $Y$ is tangent to the orbit at $\Phi$ if it is the velocity of a curve that remains on the orbit:
\[
 \Phi(s)\in\mathcal O_J,\qquad
 \Phi(0)=\Phi,\qquad
 Y=\left.\frac{d\Phi(s)}{ds}\right|_{s=0}.
\]
For an adjoint orbit, every tangent vector can be obtained from
a curve of the form
\[
 \Phi(s)=e^{-s\eta}\Phi e^{s\eta},
 \qquad \eta\in\so(2n),
\]
so that
\[
 Y=[\Phi,\eta].
\]
Now, in the constrained orbit $\Phi^2=-v^2\mathbf1$, this condition is equivalent to $\{\Phi,Y\} = 0$. To see this we can substitute $Y=[\Phi,\eta]$ into the anticommutator:
\[
\begin{aligned}
 \{\Phi,Y\}
 &=\{\Phi,[\Phi,\eta]\}\\
 &=\Phi(\Phi\eta-\eta\Phi)
   +(\Phi\eta-\eta\Phi)\Phi\\
 &=\Phi^2\eta-\eta\Phi^2\\
 &=[\Phi^2,\eta]=0,
\end{aligned}
\]
Therefore the Higgs momentum
\[
 \{\Phi,\Pi_\Phi\}
 =Z_\Phi\{\Phi,D_0\Phi\}
 =Z_\Phi D_0(\Phi^2)=0.
\]
Thus, in the local unitary gauge $\Phi=vJ$, this gives $\Pi_\Phi\in\m$.

The broken components of $\cG$ generate changes in the orientation
of the Higgs field. For $v\ne0$, the map
$\chi\mapsto[vJ,\chi]$ is invertible on $\m$, since
\begin{equation}
\left.\operatorname{ad}_{vJ}^{\,2}\right|_{\m}
 =-4v^2\mathbf1_{\m} \quad \Rightarrow \quad
\left(\left.\operatorname{ad}_{vJ}\right|_{\m}\right)^{-1}
 =-\frac{1}{4v^2}\left.\operatorname{ad}_{vJ}\right|_{\m}
\end{equation}
The gauge conditions fixing the orbit coordinates therefore pair
locally with these $r$ Gauss constraints. Together, they eliminate
the $r$ Higgs canonical pairs.

The Higgs momentum must nevertheless be accounted for when making
this reduction. Preserving $\Phi=vJ$ in time sets
$\partial_0\Phi=0$, but leaves
\begin{equation}
 \Pi_\Phi=Z_\Phi[X_0,vJ].
\end{equation}
Thus unitary gauge does not by itself impose $\Pi_\Phi=0$ or
$D_A\Phi=0$. In this gauge, Eq.~\eqref{eq:covprimary} places
$\Pi^i=\pi^i$ in $\su(n)$. Substituting into the $\m$ component
of Eq.~\eqref{eq:fullgauss} gives
\begin{equation}
 (\cG_{\m})_\alpha
 =f_{a\alpha\beta}\pi_a^iX_i^\beta
  +M^2_{\alpha\beta}X_0^\beta
 =-\cC_\alpha^0
 \approx0,
\end{equation}
where we used
$[vJ,[X_0,vJ]]=4v^2X_0$ and replaced the mass matrix Eq.~\eqref{eq:Xmassmatrix}. The temporal auxiliary constraint
\eqref{eq:C0dynamic} obtained from the unitary-gauge action
therefore retains precisely the broken Gauss equation.
Likewise, the spatial degeneracy conditions
\eqref{eq:covprimary} become $p_c^i\approx0$ and
$p_\alpha^i\approx0$; preservation of the latter gives
Eq.~\eqref{eq:Cidynamic}. The tangent Higgs equation is also
accounted for, since the orbit Noether identity
\eqref{eq:constrained-Noether} makes it a consequence of the
connection equation.

Consequently, imposing $\Phi=vJ$ before the Hamiltonian analysis
correctly accounts for the Higgs variables and the broken gauge
constraints, provided all components of $B_\mu$, $c_\mu$, and
$X_\mu$ are retained and varied in
Eq.~\eqref{eq:constrained-UG-action}. In particular, $X_\mu=0$
follows from the auxiliary constraints on the regular branch;
it is not part of the gauge choice. The two actions therefore
give the same local physical count,
$N_{\rm phys}=2(n^2-1)$, on that branch. This argument requires
only a local unitary-gauge chart.

%%%%%%%%%%%%%%%%%%%%%%%%%%%%%%%%%%%%%%%%%%
%\subsection{Fate of the orthogonal modes}
%%%%%%%%%%%%%%%%%%%%%%%%%%%%%%%%%%%%%%%%%%

The following table compares the regular constrained branch with the
quadratic vacuum spectrum of ordinary adjoint-Higgs Yang-Mills theory,
for $Z_\Phi>0$. The finite-potential projected model and its normal
scalars were treated separately in Sec.~\ref{subsec:unconstrained-spectrum}:
\begin{center}
\begin{tabular}{@{}p{0.20\textwidth}p{0.34\textwidth}p{0.34\textwidth}@{}}
\toprule
Sector & Ordinary adjoint-Higgs YM & Exactly constrained projected theory \\
\midrule
$\su(n)$ & massless, propagating & massless, propagating \\[0.5ex]
$\uu(1)$ & massless, propagating & null/nonpropagating \\[0.5ex]
$\Lambda^2\mathbf n\oplus\Lambda^2\overline{\mathbf n}$
& massive, propagating & auxiliary, zero local degrees \\[0.5ex]
tangent Higgs modes ($\m$)
& Goldstone modes; supply the longitudinal polarizations of the
massive coset vectors
& removed by unitary gauge; no propagating scalar or longitudinal
vector modes \\[0.5ex]
normal Higgs modes & massive scalars & absent from the field space \\[0.5ex]
\bottomrule
\end{tabular}
\end{center}
The longitudinal polarizations associated with the tangent Higgs
modes in the ordinary theory are already included in the
massive-vector row.

In the exactly constrained theory the coset vector equations are algebraic.
On the regular weak-curvature branch they set $X_\mu=0$.  Nonzero branches may appear when
\begin{equation}
 \det\!\left(
 \frac{\partial\cC_\alpha^\mu}{\partial X_\nu^\beta}
 \right)=0.
 \label{eq:singularlocus}
\end{equation}
This condition signals failure of the regular elimination used here; it
is necessary for a nearby bifurcation from the $X=0$ branch, not proof
that a new branch exists.  The full consistency algorithm, including
possible further constraints, must be reconsidered there.  No generic
strong-field particle count follows from the weak-curvature analysis.

We conclude this section by examining the case $Z_\Phi=0$, for $n\ge 3$, and
clarifying the scope of the regularity result. Setting $Z_\Phi=0$ leaves a fully gauge-invariant constrained compensator
model with the same exact projectors.  Its unitary-gauge action is precisely
the fixed-projector action of Theory~I.  The orbit equation still follows
from the connection equation, but the field-independent algebraic matrix
$M^2$ vanishes.  The weak-curvature regularity argument above then fails
at the trivial vacuum: $[X,\ast K_J]=0$ has the irregular constraint
structure discussed in Sec.~\ref{sec:fixed}.  In particular, restoring
gauge covariance with a constrained compensator does not by itself
regularize the auxiliary sector; that result uses $Z_\Phi>0$.

The intrinsic and multiplier formulations describe the same local
classical dynamics on a specified orbit, for either choice of $Z_\Phi$.
The equivalence established here is classical. Quantum equivalence
requires matching the path-integral measures and treating gauge
fixing and redundant multiplier components consistently. The present analysis establishes neither quantum stability of the
degenerate kinetic operator nor a generic strong-field degree count.
Complementary curvature terms allowed by $SO(2n)$ covariance change
the dynamics if included; their quantum generation requires a separate
analysis.  For example, the exact central shift of this action supplies
an additional restriction beyond $SO(2n)$ covariance alone.

The exactly constrained theory has no normal Higgs particles.
Relating it to the finite-potential model requires integrating out
the normal fields, including their couplings to curvature, and
accounting for the change in vector kinetic rank. The intrinsic
and multiplier formulations both describe Theory~III. A finite
Higgs potential defines the distinct Theory~II.

%%%%%%%%%%%%%%%%%%%%%%%%%%%%%%%%%%%%%%%%%%%%%%%%%%%%%%%%%%%%%%%%%%%
\section{Comparison to MacDowell-Mansouri gravity}
\label{sec:MM}
%%%%%%%%%%%%%%%%%%%%%%%%%%%%%%%%%%%%%%%%%%%%%%%%%%%%%%%%%%%%%%%%%%%

MacDowell-Mansouri gravity provides a useful comparison because it also selects the curvature of a subalgebra, but combines that projection with an internal dual and a first-order gravitational decomposition. These additional structures make the equation obtained by varying the Lorentz connection an algebraic torsion constraint.

%%%%%%%%%%%%%%%%%%%%%%%%%%%%%%%%%%%%%%%%%%%%%%%%%%%%%%%%%%%%%%%%%%%
\subsection{Lorentz and coset projectors from the compensator}
%%%%%%%%%%%%%%%%%%%%%%%%%%%%%%%%%%%%%%%%%%%%%%%%%%%%%%%%%%%%%%%%%%%

Let $A,B=0,\ldots,4$ be ambient (A)dS indices and let $V^A$ be a non-null
zero-form compensator,
\begin{equation}
 V^AV_A=V^2=-\sigma\ell^2,
 \qquad \sigma=\pm1.
 \label{eq:MM-compensator-norm}
\end{equation}
We use an ambient metric whose restriction to the Lorentz directions is
$\eta_{ab}=\operatorname{diag}(-,+,+,+)$ and whose fifth entry is
$\eta_{44}=-\sigma$.  Accordingly, in the convention used throughout this
section,
\begin{equation}
 \sigma=+1:\quad SO(3,2)\ \text{(anti-de Sitter)},
 \qquad
 \sigma=-1:\quad SO(4,1)\ \text{(de Sitter)}.
 \label{eq:MM-sigma-convention}
\end{equation}
In the gauge $V^A=\ell\delta^A{}_4$ one therefore has
$V^2=\eta_{44}\ell^2=-\sigma\ell^2$.  With the generator convention used
below, $[P_a,P_b]=\sigma J_{ab}$, and hence the same $\sigma$ appears in the
$e\wedge e$ contribution to Eq.~\eqref{eq:MM-curvature-components}.  We retain
$\sigma$ so that the formulas cover both cases \eqref{eq:MM-sigma-convention}. The Stelle-West and
anti-de Sitter MacDowell-Mansouri theory is obtained by setting
$\sigma=+1$.

Its stabilizer and the associated reductive decomposition are
\begin{equation}
 H_V=\operatorname{Stab}_{G_\sigma}(V)\simeq SO(3,1),
 \qquad
 \mathfrak g_\sigma=\h_V\oplus\m_V,
 \qquad
 \h_V=\{X^{AB}\mid X^{AB}V_B=0\},
 \label{eq:MM-stabilizer-decomposition}
\end{equation}
where $G_{+1}=SO(3,2)$ and $G_{-1}=SO(4,1)$. In order to see this, first define the vector projector
\begin{equation}
 h^A{}_B(V)=\delta^A{}_B-\frac{V^AV_B}{V^2}.
 \label{eq:MM-vector-projector}
\end{equation}
Clearly the tensor $h^A{}_B(V)$ annihilates the compensator:
\begin{align}
 h^A{}_B(V)V^B
 &=
 \left(
 \delta^A{}_B-\frac{V^AV_B}{V^2}
 \right)V^B
 \nonumber\\
 &=
 V^A-\frac{V^A(V_BV^B)}{V^2}
 =V^A-V^A=0.
\end{align}
Thus $h^A{}_B(V)$ projects onto the hyperplane orthogonal to $V^A$:
\begin{equation}
 \operatorname{Im}h=V^\perp,
 \qquad
 \ker h=\operatorname{span}\{V\}.
\end{equation}

Since the adjoint is represented by bivectors $X^{AB}=-X^{BA}$, the induced
Lorentz and coset projectors are
\begin{align}
 \bigl(\cP_{\rm L}(V)X\bigr)^{AB}
 &=h^A{}_Ch^B{}_DX^{CD}
 =X^{AB}+\frac{2}{V^2}V^{[A}X^{B]C}V_C,
 \label{eq:MM-Lorentz-projector-tensor}\\
 \bigl(\cP_\m(V)X\bigr)^{AB}
 &=-\frac{2}{V^2}V^{[A}X^{B]C}V_C,
 \qquad
 \cP_\m=\mathbf1-\cP_{\rm L}.
 \label{eq:MM-coset-projector-tensor}
\end{align}
For notational simplicity, we will henceforth suppress the explicit
dependence of the projectors on the compensator $V$. They obey
\begin{equation}
 \cP_{\rm L}^2=\cP_{\rm L},
 \qquad
 \cP_\m^2=\cP_\m,
 \qquad
 \cP_{\rm L}\cP_\m=0,
 \qquad
 \cP_{\rm L}+\cP_\m=\mathbf1.
 \label{eq:MM-projector-identities}
\end{equation}
If $w^A:=X^{AB}V_B$, then $w\cdot V=0$ and one obtains
\begin{equation}
 (\cP_{\m}X)^{AB}
 =
 -\frac1{V^2}\left(V^Aw^B-V^Bw^A\right).
\end{equation}
The minus sign ensures that the projected bivector reproduces the same vector
$w^A=X^{AB}V_B$ upon contraction with $V_B$:
\begin{align}
 (\cP_{\m}X)^{AB}V_B
 &=-\frac1{V^2}
 \left(V^Aw^B-V^Bw^A\right)V_B
 \nonumber\\
 &=-\frac1{V^2}
 \left[V^A(w\cdot V)-V^2w^A\right]
 =w^A.
\end{align}
Thus $\cP_{\m}X$ is the unique bivector with one
internal leg along $V^A$ and the other along $w^A\in V^\perp$.  In the gauge
$V^A=\ell\delta^A{}_4$, this becomes
\begin{equation}
 w^a=-\sigma \ell X^{a4},
 \qquad
 w^4=0,
 \qquad
 (\cP_{\m}X)^{a4}=X^{a4},
 \qquad
 (\cP_{\m}X)^{ab}=0.
\end{equation}
Hence $\cP_{\m}$ retains precisely the coset components $X^{a4}$ and removes
the Lorentz components $X^{ab}$.
Thus $\m_V$ is the tangent representation of the coset
$G_\sigma/SO(3,1)$.

With the convention
\begin{equation}
 (\cP X)^{AB}=(\cP)^{AB}{}_{CD}X^{CD},
\end{equation}
the identity on the bivector representation is
\begin{equation}
 \mathbf1^{AB}{}_{CD}
 =\delta^{[A}{}_C\delta^{B]}{}_D
 =\frac12\left(
 \delta^A{}_C\delta^B{}_D
 -\delta^A{}_D\delta^B{}_C
 \right).
\end{equation}
Hence
\begin{equation}
 (\cP_{\rm L})^{AB}{}_{CD}
 =
 \frac12\left(
 h^A{}_Ch^B{}_D-h^A{}_Dh^B{}_C
 \right)
 =
 \mathbf1^{AB}{}_{CD}
 -(\cP_{\m})^{AB}{}_{CD},
\end{equation}
where
\begin{equation}
 (\cP_{\m})^{AB}{}_{CD}
 =
 \frac{2}{V^2}\,
 V^{[A}V_{[C}\delta^{B]}{}_{D]}.
\end{equation}
% \begin{equation}
%   (\cP_{\m})^{AB}{}_{CD}
%  =
%  \frac{1}{2V^2}
%  \left[
%  V^A\left(V_C\delta^B{}_D-V_D\delta^B{}_C\right)
%  -
%  V^B\left(V_C\delta^A{}_D-V_D\delta^A{}_C\right)
%  \right].
% \end{equation}

% The bivector projector is antisymmetric in both index pairs and therefore
% annihilates a symmetric tensor $S^{CD}=S^{DC}$:
% \begin{equation}
%  (\cP_{\rm L})^{AB}{}_{CD}S^{CD}=0.
% \end{equation}
% For a symmetric rank-two tensor, the appropriate induced projector is instead
% \begin{equation}
%  \boxed{
%  (\cP_{\rm L}^{\mathrm{sym}})^{AB}{}_{CD}
%  =
%  \frac12\left(
%  h^A{}_Ch^B{}_D+h^A{}_Dh^B{}_C
%  \right).}
% \end{equation}
% It acts according to
% \begin{equation}
%  (\cP_{\rm L}^{\mathrm{sym}}S)^{AB}
%  =h^A{}_Ch^B{}_DS^{CD},
% \end{equation}
% and projects both tensor indices onto the hyperplane orthogonal to $V^A$.

The same maps admit a double-commutator representation closely analogous to
the Yang-Mills projector~\eqref{eq:dynamicprojector}.  Embed $V$ into the
Clifford algebra by
\begin{equation}
 \slashed V:=V^A\Gamma_A,
 \qquad
 X=\frac14X^{AB}\Gamma_{AB},
 \qquad
 \label{eq:MM-Clifford-embedding}
\end{equation}
where
\begin{equation}
 \Gamma_{AB}:=\frac12[\Gamma_A,\Gamma_B], \quad \{\Gamma_A,\Gamma_B\}=2\eta_{AB}\mathbf1.
\end{equation}
one finds
\begin{align}
 \slashed V^{\,2}
 &=V^AV^B\Gamma_A\Gamma_B
 \nonumber\\
 &=\frac12V^AV^B\{\Gamma_A,\Gamma_B\}
 +\frac12V^AV^B[\Gamma_A,\Gamma_B].
\end{align}
The second term vanishes because $V^AV^B$ is symmetric in $A,B$, whereas
$[\Gamma_A,\Gamma_B]$ is antisymmetric, therefore one obtains
\begin{equation}
 \slashed V^{\,2}
 =V^AV^B\eta_{AB}\mathbf1
 =V^2\mathbf1.
\end{equation}
Consequently, for $V^2\neq0$,
\begin{equation}
 \slashed V^{-1}=\frac{\slashed V}{V^2}.
\end{equation}

On the bivector subspace one then has
\begin{equation}
 \cP_\m(\slashed V) X
 =\frac1{4V^2}[\slashed V,[\slashed V, X]],
 \qquad
 \cP_{\rm L}(\slashed V) X
 =X-\frac1{4V^2}[\slashed V,[\slashed V, X]],
 \label{eq:MM-double-commutator-projectors}
\end{equation}
which can be checked by direct computation, using $[\Gamma_A,\Gamma_{BC}] = 4 \eta_{A[B}\Gamma_{C]}$.

Equivalently, using the involution,
\begin{equation}
 \vartheta_{ V}( X):=\slashed V  X \slashed V^{-1},
 \qquad
 \vartheta_{ V}^2( X)=\vartheta_{ V}(\vartheta_{ V}( X))= X,
\end{equation}
where the two eigenspaces of $\vartheta_V$ correspond to the stabilizer and coset
subspaces.  For a stabilizer bivector $X_{\rm L}\in\mathfrak h_V$,
\begin{equation}
 \vartheta_V(X_{\rm L})=+X_{\rm L},
 \qquad
 [\slashed V,X_{\rm L}]=0.
\end{equation}
For a coset bivector $X_{\m}\in\m_V$,
\begin{equation}
 \vartheta_V(X_{\m})=-X_{\m},
 \qquad
 \{\slashed V,X_{\m}\}=0.
\end{equation}
Consequently, the projectors onto the two eigenspaces are
% \begin{equation}
%  \cP_{\rm L}
%  =\frac12\left(\mathbf1+\vartheta_V\right),
%  \qquad
%  \cP_{\m}
%  =\frac12\left(\mathbf1-\vartheta_V\right).
% \end{equation}
 \begin{equation}
 \cP_\m=\frac12(\mathbf1-\vartheta_V),
 \qquad
 \cP_{\rm L}=\frac12(\mathbf1+\vartheta_V).
 \label{eq:MM-involution-projectors}
\end{equation}
The stabilizer is the Clifford commutant
\begin{equation}
 \h_V=\{X\in \mathfrak g_\sigma\mid[\slashed V,X]=0\}.
 \label{eq:MM-Clifford-stabilizer}
\end{equation}
The Clifford embedding is essential here: $V^A$ is a vector, not an element of
$\mathfrak g_\sigma$, so $[V,X]$ is not intrinsically defined in the Lie algebra.

The parallel with the preceding Yang-Mills construction is explicit.
There the adjoint Higgs $\Phi\in\so(2n)$ acts directly through
$\ad_\Phi^2$, and its commutant is $\uu(n)$.  Here the vector compensator
acts through $\ad_{\slashed V}^2$ on Clifford bivectors, and its commutant is
$\so(3,1)$.  In both cases the double commutator distinguishes the stabilizer
from the tangent space to the vacuum orbit.

% Under $SO(3,2)$,
% \begin{equation}
%  V\mapsto V^g=g^{-1}V,
%  \qquad
%  X\mapsto X^g=g^{-1}Xg,
%  \label{eq:MM-compensator-transformation}
% \end{equation}
% and the projector is equivariant:
% \begin{equation}
%  \cP_{\rm L}(V^g)(g^{-1}Xg)
%  =g^{-1}\cP_{\rm L}(V)(X)g.
%  \label{eq:MM-projector-equivariance}
% \end{equation}
% This is the direct analogue of Eq.~\eqref{eq:equivariance}.  Freezing $V$
% reduces the symmetry to its stabilizer, whereas transforming $V$ allows the
% projector itself to move covariantly.

Under $G_\sigma$, the vector-valued compensator transforms in the defining
representation,
\begin{equation}
 (V^g)^A=(g^{-1})^A{}_B V^B,
 \qquad
 X^g=g^{-1}Xg.
 \label{eq:MM-compensator-transformation}
\end{equation}
After the Clifford embedding $\slashed V=V^A\Gamma_A$, the same transformation
is represented by conjugation,
\begin{equation}
 \slashed V^g
 =
 S(g)^{-1}\slashed V S(g),
 \qquad
 X^g=S(g)^{-1}XS(g),
\end{equation}
where $S(g)$ denotes the spin representative of $g$.  Consequently, the
projector is equivariant:
\begin{equation}
 \cP_{\rm L}(V^g)\bigl(g^{-1}Xg\bigr)
 =
 g^{-1}\cP_{\rm L}(V)(X)g.
 \label{eq:MM-projector-equivariance}
\end{equation}

%%%%%%%%%%%%%%%%%%%%%%%%%%%%%%%%%%%%%%%%%%%%%%%%%%%%%%%%%
\subsection{Projected curvature and the MacDowell-Mansouri action}
%%%%%%%%%%%%%%%%%%%%%%%%%%%%%%%%%%%%%%%%%%%%%%%%%%%%%%%%%

The compensator $V^A$ determines the covariant splitting
$\mathfrak g_\sigma=\h_V\oplus\m_V$, with
$\h_V\simeq\mathfrak{so}(3,1)$.
To express the connection and curvature in Lorentz and coset
components, we locally choose an internal frame adapted to $V^A$,
so that $V^A=\ell\delta^A{}_4$.
Writing $P_a:=J_{a4}$, the generators obey the (A)dS algebra
\begin{align}
 [J_{ab},J_{cd}]
 &=-\eta_{ac}J_{bd}+\eta_{bc}J_{ad}+\eta_{ad}J_{bc} -\eta_{bd}J_{ac},
 \nonumber\\
 [J_{ab},P_c]
 &=-\eta_{ac}P_b +\eta_{bc}P_a,
 \nonumber\\
 [P_a,P_b]&=\sigma J_{ab},
 \label{eq:MM-AdS-algebra}
\end{align}
%where $\sigma=+1$ gives the AdS algebra $\mathfrak{so}(3,2)$ and $\sigma=-1$ gives the dS algebra $\mathfrak{so}(4,1)$.
In this gauge, the connection and curvature take the form
\begin{align}
 \mathcal A
 &=\frac12\omega^{ab}J_{ab}+\frac1\ell e^aP_a,
 \label{eq:MM-A)dS-connection}\\
 \mathcal F
 &=\dd\mathcal A+\mathcal A\wedge\mathcal A
 =\frac12\mathcal F^{ab}J_{ab}+\frac1\ell T^aP_a,
 \label{eq:MM-A)dS-curvature}\\
 \mathcal F^{ab}
 &=R^{ab}+\frac{\sigma}{\ell^2}e^a\wedge e^b,
 \qquad
 T^a=D_\omega e^a,
 \label{eq:MM-curvature-components}
\end{align}
where
\[
 R^{ab}:=\dd\omega^{ab}+\omega^a{}_c\wedge\omega^{cb},
 \qquad
 D_\omega e^a:=\dd e^a+\omega^a{}_b\wedge e^b.
\]
These formulas express the covariant Lorentz and coset projections
of the curvature in the chosen adapted frame. They retain their form
under the residual local Lorentz transformations that preserve
the chosen gauge.

The Lorentz-projected curvature and its complement are therefore
\begin{equation}
 K:=\cP_{\rm L}(V)\mathcal F,
 \qquad
 K^{ab}=\mathcal F^{ab},
 \qquad
 K^{a4}=0,
 \qquad
 \cP_\m(V)\mathcal F=\frac1\ell T^aP_a.
 \label{eq:MM-projected-and-complementary-curvature}
\end{equation}
Here $\cP_{\rm L}(V)$ acts on the internal Lie-algebra directions
of the curvature. In a fixed generator basis, the projection can be expressed either
through its action on the curvature components or, after the Clifford
lift of Eq.~\eqref{eq:MM-Clifford-embedding}, through its action on the
generator matrices. Identifying $J_{AB}=\frac12\Gamma_{AB}$ in the
Clifford representation, one has
\[
 K
 =\frac12
   \bigl(\cP_{\rm L}(V)\mathcal F\bigr)^{AB}J_{AB}
 =\frac12\mathcal F^{AB}\,
   \cP_{\rm L}(\slashed V)(J_{AB}).
\]
In the middle expression, the tensorial projector acts on the
antisymmetric internal indices of the curvature,
\[
 \bigl(\cP_{\rm L}(V)\mathcal F\bigr)^{AB}
 =h^A{}_C h^B{}_D\mathcal F^{CD}.
\]
In the right-hand expression, its Clifford realization acts on the
generator matrices,
\[
 \cP_{\rm L}(\slashed V)(J_{AB})
 =J_{AB}-\frac1{4V^2}
 [\slashed V,[\slashed V,J_{AB}]].
\]
These are equivalent realizations of the same Lie-algebra projection.
In the adapted frame, the projector retains $J_{ab}$ and annihilates
$P_a$. It acts only on the internal algebra indices, leaving the
spacetime two-form structure unchanged. In particular, it retains
the vierbein contribution
$\sigma e^a\wedge e^b/\ell^2$ contained in $\mathcal F^{ab}$.

The MacDowell-Mansouri action uses the internal Lorentz dual selected by $V$,
not the spacetime Hodge operator used in the Yang-Mills action:
\begin{equation}
 (\star_VX)^{AB}
 :=\frac1{2\ell}\epsilon^{AB}{}_{CDE}V^EX^{CD}.
 \label{eq:MM-internal-dual}
\end{equation}

The origin of $\star_V$ is the volume form induced on the four-dimensional
internal hyperplane $V^\perp$.  The invariant five-index tensor of $G_\sigma$
and the normalized compensator define
\begin{equation}
 \epsilon^{(V)}_{ABCD}
 :=
 \frac1\ell\epsilon_{ABCDE}V^E.\label{volumeVperp}
\end{equation}
Recall that $V^\perp$ is the four-dimensional vector space on which the stabilizer $SO(3,1)$ acts, and \eqref{volumeVperp} is tangent to $V^\perp$, because contraction with $V$ in any free index vanishes. Therefore \eqref{volumeVperp} is the volume tensor on $V^\perp$, and its action on internal bivectors
defines
\begin{equation}
 (\star_VX)^{AB}
 :=
 \frac12\epsilon^{(V)\,AB}{}_{CD}X^{CD}
 =
 \frac1{2\ell}\epsilon^{AB}{}_{CDE}V^EX^{CD}.
\end{equation}
In the gauge $V^A=\ell\delta^A{}_4$, it reduces to
\begin{equation}
 (\star_VX)^{ab}
 =
 \frac12\epsilon^{ab}{}_{cd}X^{cd},
\end{equation}
which is the ordinary internal Lorentz dual.  Moreover,
$\star_VX_{\m}=0$ for $X_{\m}^{AB}\propto V^{[A}w^{B]}$, and hence
\begin{equation}
 \star_VX=\star_V\cP_{\rm L}(V)X.
\end{equation}
The operator $\star_V$ acts only on internal gauge indices and should not be
confused with the spacetime Hodge operator $\ast$ acting on differential
forms.

In the gauge above,
$(\star_VX)^{ab}=\frac12\epsilon^{ab}{}_{cd}X^{cd}$, and
\begin{align}
 S_{\rm MM}
 &\propto\int_M\langle K,\star_VK\rangle
 \nonumber\\
 &\propto\int_M\epsilon_{abcd}
 \left(R^{ab}+\frac{\sigma}{\ell^2}e^a\wedge e^b\right)
 \wedge
 \left(R^{cd}+\frac{\sigma}{\ell^2}e^c\wedge e^d\right).
 \label{eq:MM-projected-action}
\end{align}
The three terms are the Euler density, the Einstein-Hilbert term, and the
cosmological term.  Projection alone is therefore not sufficient to turn the
gauge-theory action into the Einstein-Hilbert action with an algebraically
eliminable spin connection:
the internal $\epsilon$ tensor and the first-order decomposition
$\mathcal A=(\omega,e/\ell)$ are equally important.

There are three equivalent ways of displaying the projected structure of this
action.  First, using the tensorial projector
\eqref{eq:MM-Lorentz-projector-tensor}, one may write
\begin{equation}
 S_{\rm MM}[\mathcal A,V]
 =\alpha\int_M
 \epsilon_{ABCDE}\frac{V^E}{\ell}\,
 \bigl(\cP_{\rm L}(V)\mathcal F\bigr)^{AB}
 \wedge
 \bigl(\cP_{\rm L}(V)\mathcal F\bigr)^{CD}.
 \label{eq:MM-action-explicit-tensor-projector}
\end{equation}
Second, after embedding the compensator and curvature in the Clifford algebra
as in Eq.~\eqref{eq:MM-Clifford-embedding}, the same action is, up to the
representation-dependent normalization of the trace,
\begin{equation}
 S_{\rm MM}[\mathcal A,V]
 =\widetilde\alpha\int_M
 \Tr\left(
 \frac{\slashed V}{\ell}\,
 \cP_{\rm L}(\slashed V)\mathcal F
 \wedge
 \cP_{\rm L}(\slashed V)\mathcal F
 \right),
 \label{eq:MM-action-Clifford-projector}
\end{equation}
where the Clifford form of the projector is
\begin{equation}
 \cP_{\rm L}(\slashed V)\mathcal F
 =\mathcal F-\frac1{4V^2}
 [\slashed V,[\slashed V,\mathcal F]].
 \label{eq:MM-action-double-commutator-curvature}
\end{equation}
Equation~\eqref{eq:MM-action-explicit-tensor-projector} and
Eq.~\eqref{eq:MM-action-Clifford-projector} are therefore the same projector
written, respectively, on antisymmetric tensor indices and on the bivector
subspace of the Clifford algebra.  The latter form makes the analogy with the
Higgs-dependent Yang-Mills projector~\eqref{eq:dynamicprojector} manifest.

Third, the explicit Lorentz projectors may be omitted altogether:
\begin{align}
 &\epsilon_{ABCDE}V^E\,
 \mathcal F^{AB}\wedge\mathcal F^{CD}
 \nonumber\\
 &\qquad=
 \epsilon_{ABCDE}V^E\,
 \bigl(\cP_{\rm L}(V)\mathcal F\bigr)^{AB}
 \wedge
 \bigl(\cP_{\rm L}(V)\mathcal F\bigr)^{CD}.
 \label{eq:MM-projector-implicit-in-epsilon}
\end{align}
To see this, decompose
$\mathcal F=\cP_{\rm L}\mathcal F+\cP_{\m}\mathcal F$.  Every coset
bivector is proportional to $V^{[A}w^{B]}$, so any term containing at least
one $\cP_{\m}\mathcal F$ has two compensator directions contracted into the
totally antisymmetric tensor and vanishes.  In the gauge
$V^A=\ell\delta^A{}_4$, this simply states that a coset curvature already
carries an index $4$, while $V^E$ supplies another one.  Thus
\begin{equation}
 \epsilon V\mathcal F\mathcal F \,\sim\,
 \epsilon V(\cP_{\rm L}\mathcal F)(\cP_{\rm L}\mathcal F) \,\sim\,
 \Tr\left(\slashed V\,
 \cP_{\rm L}\mathcal F\wedge\cP_{\rm L}\mathcal F\right)
 \label{eq:MM-three-equivalent-representations}
\end{equation}
are equivalent representations of the same MacDowell-Mansouri invariant,
with the appropriate normalizations understood.  The first hides the
projection in the $\epsilon V$ contraction, the second displays its tensorial
geometry, and the third displays its double-commutator realization.

%%%%%%%%%%%%%%%%%%%%%%%%%%%%%%%%%%%%%%%%%%%%%%%%%%%%%%%%%%%%
\subsection{The projected equation and the torsion mechanism}
%%%%%%%%%%%%%%%%%%%%%%%%%%%%%%%%%%%%%%%%%%%%%%%%%%%%%%%%%%%%

We now turn to the field equations to recall how the Lorentz-projected action constrains the torsion. For the fixed Lorentz projector, variation of the full AdS connection gives
the compact equation
\begin{equation}
 D_{\mathcal A}\star_VK=0.
 \label{eq:MM-compact-field-equation}
\end{equation}
This is the analogue of $D_A\ast K=0$ in
Eq.~\eqref{eq:fixed-connection-equation}, with the crucial replacement of the
spacetime Hodge dual by the internal Lorentz dual.  Its Lorentz component is
\begin{equation}
 D_\omega\star_VK=0.
 \label{eq:MM-Lorentz-equation}
\end{equation}
Using $D_\omega R^{ab}=0$ gives
\begin{equation}
 D_\omega\mathcal F^{ab}
 =\frac{\sigma}{\ell^2}
 \left(T^a\wedge e^b-e^a\wedge T^b\right),
 \label{eq:MM-Bianchi-torsion}
\end{equation}
and hence
\begin{equation}
 \epsilon_{abcd}e^c\wedge T^d=0.
 \label{eq:MM-torsion-equation}
\end{equation}
For an invertible coframe this implies
\begin{equation}
 T^a=0.
 \label{eq:MM-zero-torsion}
\end{equation}
The spin connection is therefore auxiliary: its equation determines
$\omega=\omega(e)$ algebraically.

The coset component arises from the commutator of $e := e^a P_a$ with the Lorentz-valued
$K$,
\begin{equation}
 \cP_\m D_{\mathcal A}\star_VK
 =\frac1\ell[e,\star_VK]=0,
 \label{eq:MM-coset-equation}
\end{equation}
or
\begin{equation}
 \epsilon_{abcd}e^b\wedge
 \left(R^{cd}+\frac{\sigma}{\ell^2}e^c\wedge e^d\right)=0.
 \label{eq:MM-Einstein-equation}
\end{equation}
This is the Einstein equation with cosmological constant, up to the sign
convention encoded by $\sigma$.

The torsion equation also controls the variation under the
internal translations broken by the fixed projector.
For the parameter $\rho^aP_a/\ell$, the naive (A)dS connection
transformation gives
\[
 \delta_\rho e^a=D_\omega\rho^a,
 \qquad
 \delta_\rho\omega^{ab}
 =\frac{\sigma}{\ell^2}
 \left(e^a\rho^b-e^b\rho^a\right).
\]
The action varies as
\begin{equation}
 \delta_\rho S_{\rm MM}
 \propto\int_M\epsilon_{abcd}\,
 \rho^aT^b\wedge K^{cd}.
 \label{eq:MM-translation-variation}
\end{equation}
This variation vanishes when the spin-connection equation
imposes $T^a=0$. However, the naive transformation does not
generally preserve that equation, since
\[
 \delta_\rho T^a
 =D_\omega^2\rho^a
  +\delta_\rho\omega^a{}_b\wedge e^b
 =K^a{}_b\rho^b.
\]
On a torsion-free Einstein solution, $K^{ab}$ is the Weyl
curvature two-form and need not vanish. Thus the conditional
vanishing of Eq.~\eqref{eq:MM-translation-variation} does not
establish a symmetry of the solution space under arbitrary
naive (A)dS translations.

After eliminating $\omega=\omega(e)$ on the nondegenerate
coframe branch, its variation must instead be induced by
the variation of $e$. Define the spacetime vector field $\xi$
by $\rho^a=\iota_\xi e^a$. The transformation
$\delta_\rho^{\rm red}e^a=D_\omega\rho^a$ is then a
diffeomorphism combined with a Lorentz rotation of parameter
$-\iota_\xi\omega$. Its induced spin-connection variation is
\[
 \delta_\rho^{\rm red}\omega^{ab}(e)
 =\iota_\xi R^{ab}
 =\delta_\rho\omega^{ab}+\iota_\xi K^{ab}.
\]
These transformations preserve the reduced action.
The curvature correction distinguishes them from the naive
(A)dS connection transformations. Full internal (A)dS
covariance is realized by the transforming compensator in
the Stelle-West formulation discussed below.

Replacing $\star_V$ by the regular spacetime Hodge operator would instead
produce
\begin{equation}
 D_\omega\ast
 \left(R^{ab}+\frac{\sigma}{\ell^2}e^a\wedge e^b\right)=0.
 \label{eq:MM-regular-Hodge-equation}
\end{equation}
This is generically a differential Yang-Mills-type equation for $\omega$ and
does not reduce to the algebraic torsion constraint.  The relevant special
structure is therefore
\begin{equation}
 \text{Lorentz projection}
 \;+\text{internal dual}
 \;+\text{first-order variables }(e,\omega),
 \label{eq:MM-special-combination}
\end{equation}
not projection by itself.

%%%%%%%%%%%%%%%%%%%%%%%%%%%%%%%%%%%%%%%%%%%%%%%%%%%%%%%%%%%%%%%%
\subsection{Stelle-West gravity and the coset realization}
%%%%%%%%%%%%%%%%%%%%%%%%%%%%%%%%%%%%%%%%%%%%%%%%%%%%%%%%%%%%%%%%

In the Stelle-West formulation the choice of Lorentz subgroup is promoted to
a compensator field on the orbit
\begin{equation}
 V(x)\in SO(3,2)/SO(3,1),
 \qquad V^2=-\ell^2
 \quad(\sigma=+1).
 \label{eq:SW-compensator-orbit}
\end{equation}
The action can be written covariantly as
\begin{equation}
 S_{\rm SW}[\mathcal A,V]
 \propto\int_M\epsilon_{ABCDE}V^E
 \mathcal F^{AB}\wedge\mathcal F^{CD}
 \propto\int_M\langle\cP_{\rm L}(V)\mathcal F,
 \star_V\cP_{\rm L}(V)\mathcal F\rangle .
 \label{eq:SW-covariant-action}
\end{equation}
Because $V$ and $\mathcal F$ transform together,
Eq.~\eqref{eq:MM-projector-equivariance} makes the full $SO(3,2)$ covariance
exact off shell.  Choosing $V^A=\ell\delta^A{}_4$ is a local unitary gauge,
not the introduction of a fixed background vector, and recovers the usual
MacDowell-Mansouri expression.
If $V$ is parametrized only by coset coordinates subject to
Eq.~\eqref{eq:MM-compensator-norm}, those coordinates are pure compensators
rather than additional physical scalar modes.  Equivalently, when $V$ is
varied as a constrained field, its orbit equation is tied to the connection
equation by the $SO(3,2)$ Noether identity and does not constitute an
independent propagating equation.

This can also be formulated in terms of a nonlinear realization
of the AdS group. Locally, choose a coset representative
\begin{equation}
 u(\tau)=\exp(\tau^aP_a),
 \qquad
 V(\tau)=u(\tau)V_0,
 \label{eq:SW-coset-representative}
\end{equation}
where $V_0$ is stabilized by $SO(3,1)$.
The dressed connection decomposes as
\begin{equation}
 \mathcal A^u:=u^{-1}\mathcal A u+u^{-1}\dd u
 =\frac12W^{ab}J_{ab}+\frac1\ell E^aP_a.
 \label{eq:SW-dressed-connection}
\end{equation}
For $g\in SO(3,2)$, the local factorization
\begin{equation}
 g^{-1}u(\tau)=u(\tau')h^{-1}(g,\tau),
 \qquad h(g,\tau)\in SO(3,1),
 \label{eq:SW-nonlinear-factorization}
\end{equation}
gives $u'=g^{-1}uh$. Since $hV_0=V_0$, it follows that
\[
 V(\tau')=u'V_0=g^{-1}V(\tau),
\]
as required by the compensator transformation law.
Together with
$\mathcal A'=g^{-1}\mathcal A g+g^{-1}\dd g$,
the same factorization gives
\[
 (\mathcal A')^{u'}
 =h^{-1}\mathcal A^u h+h^{-1}\dd h.
\]
Thus $W$ transforms as a Lorentz connection and $E$ as a
Lorentz vector. The coordinates $\tau^a$ are compensating
Goldstone fields; the local gauge $\tau=0$, equivalently
$V=V_0$, reduces $W,E$ to the usual $\omega,e$ variables.

The orbit analogy with the exactly constrained Yang-Mills model of
Sec.~\ref{sec:constrained} is
\begin{equation}
 \Phi(x)\in SO(2n)/U(n)
 \quad\longleftrightarrow\quad
 V(x)\in SO(3,2)/SO(3,1).
 \label{eq:YM-SW-orbit-analogy}
\end{equation}
In each theory, a field on a coset selects a local stabilizer subalgebra and an
equivariant projector.  Freezing that field gives a fixed projector with only
the stabilizer manifest off shell; transforming it restores the full group.
The dynamical distinction is that the MacDowell-Mansouri internal-dual action
makes the Lorentz spin-connection equation a torsion constraint. The
ordinary-Hodge projected Yang-Mills action has no corresponding
torsion mechanism.

%%%%%%%%%%%%%%%%%%%%%%%%%%%%%%%%%%%%%%%%%%%%%%%%%%%%%%%%%%%%%%%%
\subsection{The gravitational Higgs sector and the induced metric}
\label{subsec:SW-Higgs-metric}
%%%%%%%%%%%%%%%%%%%%%%%%%%%%%%%%%%%%%%%%%%%%%%%%%%%%%%%%%%%%%%%%

The fixed-norm formulation has a dynamical Higgs extension in which the
norm of the compensator is allowed to vary. We briefly recall the
Stelle-West Higgs sector reviewed by Leclerc
\cite{Stelle:1979aj,Leclerc:2005qc}, before giving its projector
formulation. We retain the AdS convention $\sigma=+1$ and work on the
open domain
\begin{equation}
 s:=V^AV_A<0,
 \qquad r:=\sqrt{-s}>0,
 \qquad V^A=r n^A,
 \qquad n^An_A=-1.
 \label{eq:SW-Higgs-radial-decomposition}
\end{equation}
Here $\ell$ remains a constant reference length and the preferred Higgs
radius. Thus Eq.~\eqref{eq:MM-compensator-norm} is imposed only in the
exactly constrained theory, not throughout this extension.

Define the spacetime one-form valued in the internal hyperplane
$V^\perp$, tangent to the compensator orbit, and its normalized
coframe by
\begin{equation}
 q^A:=h^A{}_B(V)D_{\mathcal A}V^B=rD_{\mathcal A}n^A,
 \qquad
 e_V^A:=-\frac{\ell}{r}q^A=-\ell D_{\mathcal A}n^A.
 \label{eq:SW-Higgs-tangent-coframe}
\end{equation}
The minus sign matches our convention
$\mathcal A^{a4}=e^a/\ell$ and $\eta_{44}=-1$.
The gauge-invariant metric used in the Higgs action is
\begin{align}
 g_{\mu\nu}
 &:=\eta_{AB}e_{V\mu}^Ae_{V\nu}^B
 =-\frac{\ell^2}{s}\,h_{AB}(V)
 D_\mu V^A D_\nu V^B
 \nonumber\\
 &=\frac{\ell^2}{s}
 \left[
 \frac{(V_A D_\mu V^A)(V_B D_\nu V^B)}{s}
 -D_\mu V^A D_\nu V_A
 \right].
 \label{eq:SW-Higgs-induced-metric}
\end{align}
This is the normalized metric of the Stelle-West Higgs construction
\cite{Leclerc:2005qc}. At fixed norm it agrees with the usual
Cartan metric~\cite{Westman:2012zk}. Every internal contraction in
Eq.~\eqref{eq:SW-Higgs-induced-metric} is invariant when $V$ and
$\mathcal A$ transform together, so its gauge invariance holds before
choosing an adapted frame. We restrict to configurations where $e_V$
has rank four. The metric then has signature $(-,+,+,+)$, and defines
the spacetime Hodge operator $\ast_g$ and volume form
$\vol_g:=\ast_g1$. In the local gauge $V^A=r\delta^A{}_4$,
\begin{equation}
 D_{\mathcal A}V^a=-\frac r\ell e^a,
 \qquad D_{\mathcal A}V^4=\dd r,
 \qquad e_V^a=e^a,
 \qquad g_{\mu\nu}=\eta_{ab}e^a{}_\mu e^b{}_\nu.
 \label{eq:SW-Higgs-metric-unitary-gauge}
\end{equation}

With the kinetic sign appropriate to our spacetime signature, take
\begin{align}
 S_{\rm H}[\mathcal A,V]
 &=\frac{Z_V}{2}\int_M
 \frac{(V_A D_{\mathcal A}V^A)\wedge
 \ast_g(V_B D_{\mathcal A}V^B)}{s}
 -\int_M U(s)\vol_g,
 \label{eq:SW-Higgs-action}\\
 U(s)&=\frac{\lambda_V}{4}(s+\ell^2)^2+U_0,
 \qquad Z_V>0,\quad\lambda_V>0.
 \label{eq:SW-Higgs-potential}
\end{align}
This is the radial kinetic structure of Ref.~\cite{Leclerc:2005qc},
with an arbitrary positive normalization and an additive potential
constant. Since $V_A D_{\mathcal A}V^A=\frac12\dd s=-r\dd r$,
\begin{equation}
 S_{\rm H}
 =-\frac{Z_V}{2}\int_M\dd r\wedge\ast_g\dd r
 -\int_M U(-r^2)\vol_g.
 \label{eq:SW-Higgs-radial-action}
\end{equation}
The Higgs minimum fixes $r=\ell$ and $\dd r=0$, while the tangent
one-form in Eq.~\eqref{eq:SW-Higgs-tangent-coframe} supplies the
coframe. In particular, imposing $D_{\mathcal A}V=0$ would make the
metric degenerate. A minimum of $U$ must also satisfy the equations of
the complete gravitational action to define a vacuum.

Westman and Zlosnik develop a broader class of Cartan actions with
dynamical $V^2$ \cite{Westman:2013mf,Westman:2014yca}. Their commonly
used metric $\bar g_{\mu\nu}=h_{AB}D_\mu V^A D_\nu V^B$ is related
to ours by $g_{\mu\nu}=(\ell^2/r^2)\bar g_{\mu\nu}$ on the present
AdS domain. The conformal factor matters when $r$ varies. Those
polynomial actions can generate scalar dynamics through their
gravitational couplings, without adding
Eq.~\eqref{eq:SW-Higgs-action} separately. Depending on the action,
they yield scalar-tensor dynamics or include propagating torsion.
Here we use the explicit Higgs sector above and the
$\epsilon V\mathcal F\mathcal F$ gravitational invariant as a definite
model, rather than identifying it with the whole Cartan family.

%%%%%%%%%%%%%%%%%%%%%%%%%%%%%%%%%%%%%%%%%%%%%%%%%%%%%%%%%%%%%%%%
\subsection{A Clifford-projector formulation of the full Higgs-gravity action}
\label{subsec:SW-full-projector-action}
%%%%%%%%%%%%%%%%%%%%%%%%%%%%%%%%%%%%%%%%%%%%%%%%%%%%%%%%%%%%%%%%

The curvature projector, the Higgs kinetic term, and the induced
metric can all be expressed through the same Clifford involution.
This gives a common projector formulation of the gravitational and
Higgs sectors, with the radial and tangent modes distinguished
without choosing a gauge.
Introduce the Clifford lifts
\begin{equation}
 \Psi:=\slashed V=V^A\Gamma_A,
 \qquad
 \widehat{\mathcal A}:=\frac14\mathcal A^{AB}\Gamma_{AB},
 \qquad
 \widehat{\mathcal F}:=\frac14\mathcal F^{AB}\Gamma_{AB},
 \qquad \Psi^2=s\mathbf1,
 \label{eq:SW-full-Clifford-lifts}
\end{equation}
and the covariant derivative
\[
 \mathcal D\Psi:=\dd\Psi+[\widehat{\mathcal A},\Psi]
 =(D_{\mathcal A}V^A)\Gamma_A.
\]
Here $\Psi$ denotes the Clifford lift of the dynamical Higgs field,
whose invariant norm $s=V^AV_A$ is allowed to fluctuate away from
its vacuum value $-\ell^2$. The projectors retain the form of
Eq.~\eqref{eq:MM-double-commutator-projectors} and depend only on
the internal direction of $V^A$, independently of its radial
magnitude.

For any Clifford element $Y$, define
\begin{equation}
 \vartheta_\Psi(Y):=\Psi Y\Psi^{-1},
 \qquad
 \Pi_\pm(Y):=\frac12\bigl(Y\pm\vartheta_\Psi(Y)\bigr).
 \label{eq:SW-full-Clifford-projectors}
\end{equation}
Because $\Psi^2=s\mathbf1$ and $s\ne0$,
\begin{equation}
 \Pi_\pm^2=\Pi_\pm,
 \qquad \Pi_+\Pi_-=0,
 \qquad \Pi_++\Pi_-=\mathbf1.
 \label{eq:SW-full-projector-identities}
\end{equation}
These projector identities remain exact when $r(x)$ fluctuates. Indeed, the
projectors depend on $n=V/r$ alone. On Clifford bivectors they are
precisely $\Pi_+=\cP_{\rm L}(\slashed V)$ and
$\Pi_-=\cP_\m(\slashed V)$ from
Eq.~\eqref{eq:MM-double-commutator-projectors}.

Their action on Clifford vectors has a different geometric
interpretation. For $\slashed w=w^A\Gamma_A$, the Clifford relation
$\{\Psi,\slashed w\}=2(V\cdot w)\mathbf1$ gives
\begin{equation}
 \vartheta_\Psi(\slashed w)
 =\frac{2V\cdot w}{s}\Psi-\slashed w,
 \qquad
 \Pi_+(\slashed w)=\frac{V\cdot w}{s}\Psi,
 \qquad
 \Pi_-(\slashed w)=(h^A{}_B w^B)\Gamma_A.
 \label{eq:SW-vector-projector-action}
\end{equation}
Thus the same $+$ and $-$ projectors select Lorentz and coset
bivectors, but radial and tangent vectors, respectively. Applying
them to the Higgs derivative yields
\begin{align}
 Q_+&:=\Pi_+(\mathcal D\Psi)
 =\frac{\dd s}{2s}\Psi=\slashed n\,\dd r,
 \nonumber\\
 Q_-&:=\Pi_-(\mathcal D\Psi)
 =q^A\Gamma_A=r\mathcal D\slashed n,
 \qquad \slashed n:=n^A\Gamma_A.
 \label{eq:SW-projected-Higgs-derivatives}
\end{align}
For vector-valued forms, use the normalized Clifford trace pairing
$\langle\slashed a,\slashed b\rangle_{\rm Cl}
:=N_S^{-1}\Tr_S(\slashed a\slashed b)=\eta_{AB}a^Ab^B$,
where $N_S$ is the spinor dimension. In particular,
\begin{align}
 \langle Q_+\wedge\ast_g Q_+\rangle_{\rm Cl}
 &=\frac{(V_A D_{\mathcal A}V^A)\wedge
 \ast_g(V_B D_{\mathcal A}V^B)}{s}
 =-\dd r\wedge\ast_g\dd r,
 \label{eq:SW-radial-projector-kinetic-identity}\\
 g_{\mu\nu}
 &=-\frac{\ell^2}{s}
 \langle Q_{-\mu},Q_{-\nu}\rangle_{\rm Cl}.
 \label{eq:SW-tangent-projector-metric}
\end{align}
The radial projection therefore supplies the Higgs kinetic term,
while the tangent projection determines the metric used in that
term. Under a local spin representative $S(g)$, $\Psi$,
$\widehat{\mathcal F}$, and $\mathcal D\Psi$ transform by
conjugation and $\Pi_\pm$ transform equivariantly. The trace pairing
then makes Eq.~\eqref{eq:SW-tangent-projector-metric} gauge invariant
before gauge fixing. The Hodge operator acts only on spacetime form
indices, for example $\ast_g(q^A\Gamma_A)=(\ast_gq^A)\Gamma_A$.

Take the full action to be the Stelle-West curvature invariant
plus Eq.~\eqref{eq:SW-Higgs-action},
\begin{equation}
 S_{\rm full}[\mathcal A,V]
 :=\alpha\int_M\epsilon_{ABCDE}\frac{V^E}{\ell}
 \mathcal F^{AB}\wedge\mathcal F^{CD}+S_{\rm H}[\mathcal A,V].
 \label{eq:SW-full-tensor-action}
\end{equation}
This choice extends the fixed-norm invariant of
Eq.~\eqref{eq:SW-covariant-action} and is among the Higgs-gravity
couplings discussed in Refs.~\cite{Stelle:1979aj,Leclerc:2005qc}.
Its exact projector representation is
\begin{equation}
 \begin{aligned}
 S_{\rm full}[\widehat{\mathcal A},\Psi]
 &=\widetilde\alpha\int_M
 \langle \frac{\Psi}{\ell}\,\widehat K\wedge\widehat K \rangle_{\rm Cl}
 +\frac{Z_V}{2}\int_M
 \langle Q_+\wedge\ast_g Q_+\rangle_{\rm Cl}
 -\int_M U(s)\vol_g,\\
 \widehat K&:=\Pi_+(\widehat{\mathcal F}),
 \qquad Q_\pm:=\Pi_\pm(\mathcal D\Psi),
 \qquad
 g_{\mu\nu}=-\frac{\ell^2}{s}
 \langle Q_{-\mu},Q_{-\nu}\rangle_{\rm Cl}.
 \end{aligned}
 \label{eq:SW-full-projector-action}
\end{equation}
The constant $\widetilde\alpha$ absorbs the Clifford trace
normalization, as in Eq.~\eqref{eq:MM-action-Clifford-projector}.
The equality with Eq.~\eqref{eq:SW-full-tensor-action} follows from
the epsilon-trace identity, the implicit curvature projection in
Eq.~\eqref{eq:MM-projector-implicit-in-epsilon}, and
Eq.~\eqref{eq:SW-radial-projector-kinetic-identity}.
Unlike a Yang-Mills action on a prescribed metric, here $g$, $\ast_g$,
and $\vol_g$ depend on the fields and must be varied accordingly.
The tangent term $\langle Q_-\wedge\ast_gQ_-\rangle_{\rm Cl}$
is not an additional Higgs kinetic term in this model. In fact,
the metric definition implies
\begin{equation}
 \langle Q_-\wedge\ast_gQ_-\rangle_{\rm Cl}
 =-\frac{4s}{\ell^2}\vol_g,
 \label{eq:SW-tangent-kinetic-volume}
\end{equation}
so a constant coefficient multiplying it would shift the scalar
potential in this conformal frame.

The projectors also separate an arbitrary infinitesimal Higgs variation into its radial component and its component tangent to the gauge orbit, without requiring an expansion around the vacuum:
\begin{equation}
 \delta\Psi=\slashed n\,\delta r+r\delta\slashed n,
 \qquad n_A\delta n^A=0,
 \qquad
 \Pi_+(\delta\Psi)=\slashed n\,\delta r,
 \quad
 \Pi_-(\delta\Psi)=r\delta\slashed n.
 \label{eq:SW-radial-tangent-variations}
\end{equation}

Locally, the four tangent scalar coordinates parametrize the
$SO(3,2)/SO(3,1)$ gauge orbit and can be removed by choosing
$n^A=\delta^A{}_4$. By contrast, the radial field
$r=\sqrt{-V^AV_A}$ is invariant under the full local $SO(3,2)$
gauge group and therefore cannot be removed by a gauge
transformation. In this
gauge Eq.~\eqref{eq:SW-full-projector-action} becomes
\begin{align}
 S_{\rm full}
 &=\alpha\int_M\frac r\ell\epsilon_{abcd}
 \mathcal K^{ab}\wedge\mathcal K^{cd}
 -\frac{Z_V}{2}\int_M\dd r\wedge\ast_g\dd r
 -\int_M U(-r^2)\vol_g,
 \nonumber\\
 \mathcal K^{ab}&:=R^{ab}+\frac1{\ell^2}e^a\wedge e^b,
 \qquad g_{\mu\nu}=\eta_{ab}e^a{}_\mu e^b{}_\nu.
 \label{eq:SW-full-action-unitary-gauge}
\end{align}
Although the projectors are independent of $r$, the explicit
insertion $\Psi/\ell$ retains its coupling to the gravitational
sector. Replacing it by $\Psi/r$ would define a different
off-orbit action.

For example, varying $r$ at fixed $e,\omega$ gives
\begin{equation}
 Z_V\dd\ast_g\dd r
 -\frac{\dd U(-r^2)}{\dd r}\vol_g
 +\frac{\alpha}{\ell}\epsilon_{abcd}
 \mathcal K^{ab}\wedge\mathcal K^{cd}=0.
 \label{eq:SW-radial-equation}
\end{equation}
Thus $r=\ell$ at a minimum of the Higgs potential need not solve
the full radial equation on an arbitrary gravitational background.
For $U_0=0$, a torsion-free AdS background with
$\overline{\mathcal K}^{ab}=0$ and $\bar r=\ell$ is a solution.
Writing $r=\ell+\rho$ around that background gives the decoupled
quadratic radial sector
\begin{equation}
 S_\rho^{(2)}=-\frac12\int_M
 \left(Z_V\dd\rho\wedge\ast_{\bar g}\dd\rho
 +2\lambda_V\ell^2\rho^2\vol_{\bar g}\right),
 \qquad m_\rho^2=\frac{2\lambda_V\ell^2}{Z_V}.
 \label{eq:SW-radial-quadratic-sector}
\end{equation}
The term $\rho\mathcal K\wedge\mathcal K$ starts at cubic order
there. This establishes the radial scalar mode on that background,
without assigning independent physical Goldstone scalars to the
four tangent directions.

Away from constant $r$, the spin-connection equation is instead
\begin{equation}
 \epsilon_{abcd}\left[
 \dd r\wedge\mathcal K^{cd}
 +\frac r{\ell^2}
 \left(T^c\wedge e^d-e^c\wedge T^d\right)
 \right]=0.
 \label{eq:SW-variable-radius-torsion-equation}
\end{equation}
The term $\dd r\wedge R^{cd}$ prevents a general identification
with the fixed-radius algebraic torsion equation. In particular,
the scalar-dependent Euler term cannot be discarded as a boundary
term, in agreement with the dynamical-torsion possibilities discussed
in Refs.~\cite{Leclerc:2005qc,Westman:2013mf,Westman:2014yca}.
Imposing $r=\ell$ as an exact constraint removes the radial
variation and recovers the compensator theory, with the possible
volume term $-U_0\int\vol_g$. Merely setting $r=\ell$ in the
unrestricted equations must also respect
Eq.~\eqref{eq:SW-radial-equation}.

The underlying Higgs-gravity construction is established in the
literature. The contribution here is its explicit common projector
representation and its comparison with projected Yang-Mills theory.
The quadratic analysis around the torsion-free AdS background
establishes the radial scalar mode in
Eq.~\eqref{eq:SW-radial-quadratic-sector}; it does not establish a
generic reduction to a metric and one scalar, since the
variable-radius spin-connection equation
\eqref{eq:SW-variable-radius-torsion-equation} is not generally
algebraic in the connection.

%%%%%%%%%%%%%%%%%%%%%%%%%%%%%%%%%%%%%%%%%%%%%%%%%%%%%%%%%%%%%
\subsection{Contrast with the projected Yang-Mills models}
%%%%%%%%%%%%%%%%%%%%%%%%%%%%%%%%%%%%%%%%%%%%%%%%%%%%%%%%%%%%%

We now contrast the torsion mechanism in gravity with the projected
Yang-Mills models, focusing on the complementary field equation
and the variation of the action under broken gauge transformations. For the fixed Yang-Mills projector, the complementary equation is
\begin{equation}
 [X,\ast K]=0,
 \label{eq:MM-comparison-YM-X-equation}
\end{equation}
whereas a broken transformation $\xi\in\m$ gives
\begin{equation}
 \delta_\xi S_{\rm fix}
 =\frac1{g^2}\int_M\langle\xi,[F_\m,\ast K]\rangle.
 \label{eq:MM-comparison-YM-broken-variation}
\end{equation}
Because $F_\m$ contains $D_BX$, the broken variation involves
derivatives of $X$ and does not generally vanish upon imposing
the algebraic equation $[X,\ast K]=0$ alone.
The condition $F_\m=0$ defines a special sector preserved by
the residual $U(n)$ transformations, but it is not the general
equation of motion of the projected-out connection.
For $n\ge3$, imposing the algebraic $X$ equation alone therefore
does not generally make the broken Yang-Mills variation vanish.
The gravitational comparison concerns the more specific fact
that the torsion equation makes
Eq.~\eqref{eq:MM-translation-variation} vanish.
In either fixed-projector theory, conditional vanishing of
the action variation must be distinguished from preservation
of the solution space by the naive broken transformations.

For the Higgs-dependent Yang-Mills projector, no torsion-like on-shell
mechanism is required.  Equivariance gives
\begin{equation}
 (\delta_\eta\cP_{\su})F
 +\cP_{\su}(\Phi)[F,\eta]=[K,\eta],
 \label{eq:MM-comparison-dynamic-YM-equivariance}
\end{equation}
so the full $SO(2n)$ symmetry is exact off shell.  This is structurally
parallel to the Stelle-West identity
\eqref{eq:MM-projector-equivariance}.  Nevertheless, on the exact orbit its equation
\begin{equation}
 D_A\ast K+g^2Z_\Phi[\Phi,\ast D_A\Phi]=0
 \label{eq:MM-comparison-dynamic-YM-equation}
\end{equation}
contains the spacetime dual and does not become a first-order torsion
constraint.  Off orbit the curvature flux is instead
$\mathcal H_\Phi=\cP_{\su}(\Phi)^2F$, as in
Eq.~\eqref{eq:Aeomdynamic}. Equivariance and gauge invariance still hold
where this kinetic map is defined.  The compensator constructions share an equivariant geometry but
not the special dynamics of MacDowell-Mansouri gravity.

\begin{center}
\begin{tabular}{@{}llll@{}}
\toprule
Theory & Projector & Dual & Full internal group \\
\midrule
fixed projected YM & fixed $\cP_{\su}$ & spacetime $\ast$
 & explicitly reduced \\
Higgs-projected YM & $\cP_{\su}(\Phi)$ & spacetime $\ast$
 & exact off shell \\
fixed MacDowell-Mansouri & fixed $\cP_{\rm L}$ & internal $\star_V$
 & explicitly reduced \\
Stelle-West & $\cP_{\rm L}(V)$ & internal $\star_V$
 & exact off shell \\
\bottomrule
\end{tabular}
\end{center}

The table concerns internal gauge transformations for $n \ge 3$.
After eliminating the spin connection, the translation-like
transformations of MacDowell-Mansouri gravity are realized
by spacetime diffeomorphisms combined with field-dependent
Lorentz rotations. Their action on $\omega(e)$ is induced
by the coframe variation and includes the curvature correction
displayed above. They therefore differ from the naive
internal (A)dS connection transformations.

%%%%%%%%%%%%%%%%%%%%%%%%%%%%%%%%%%%%%%%%%%%%%%%%%%%%%
\subsection{Projected $SO(2n)$ Pontryagin theory}\label{sec:SO(2n)PPtheory}
%%%%%%%%%%%%%%%%%%%%%%%%%%%%%%%%%%%%%%%%%%%%%%%%%%%%%

The preceding comparison raises a natural intermediate question.  Suppose that
one keeps the fixed $SO(2n)\to U(n)$ projector but replaces the spacetime-Hodge
Yang-Mills term by a projected Pontryagin term.  Can the absence of the Hodge
operator turn some connection components into regular auxiliary fields, as in
MacDowell-Mansouri gravity?

For this question it is useful first to retain the full stabilizer algebra and
use the symmetric decomposition
\begin{equation}
 \so(2n)=\uu(n)\oplus\m,
 \qquad
 [\uu(n),\uu(n)]\subset\uu(n),
 \quad
 [\uu(n),\m]\subset\m,
 \quad
 [\m,\m]\subset\uu(n).
 \label{eq:projected-Pontryagin-symmetric-pair}
\end{equation}
Writing the connection as
\begin{equation}
 A=B+X,
 \qquad
 B\in\Omega^1(M,\uu(n)),
 \qquad
 X\in\Omega^1(M,\m),
 \label{eq:projected-Pontryagin-connection-split}
\end{equation}
gives
\begin{align}
 F(A)&=F_B+D_BX+\frac12[X,X],
 \nonumber\\
 K_{\uu}:=\cP_{\uu}F
 &=F_B+\frac12[X,X],
 \qquad
 F_{\m}=D_BX.
 \label{eq:projected-Pontryagin-curvature-split}
\end{align}
Here and below, the bracket denotes the graded Lie bracket,
including the exterior product. In four spacetime dimensions,
consider
\begin{equation}
 S_{\rm PP}[B,X]
 =\frac{\kappa}{2}\int_M
 \left\langle K_{\uu}\wedge K_{\uu}\right\rangle .
 \label{eq:projected-Pontryagin-action}
\end{equation}
Although the unprojected Pontryagin form
$\langle F\wedge F\rangle$ represents a characteristic class,
Eq.~\eqref{eq:projected-Pontryagin-action} is not generally a
characteristic integral, since
$K_{\uu}=F_B+\frac12[X,X]$ is not the curvature of $B$.
Moreover, the fixed projector is equivariant only under the
stabilizer $U(n)$.

The resulting field equations nevertheless have a useful
algebraic structure: the $X$ equation contains no derivatives
of $X$, while the $B$ equation contains no derivatives of $B$.
Each equation is therefore algebraic in the field being varied.
The coupled system remains differential, however, because
the $X$ equation contains $\dd B$ and the $B$ equation
contains $\dd X$.

The important kinematical fact is that $K_{\uu}$ contains no derivative of
$X$.  Since
\begin{equation}
 \delta_XK_{\uu}=[X,\delta X],
 \label{eq:projected-Pontryagin-X-variation}
\end{equation}
variation with respect to the complementary connection gives
\begin{equation}
 [K_{\uu},X]=0.
 \label{eq:projected-Pontryagin-X-equation}
\end{equation}

Thus the $X$ equation is algebraic, as in the spacetime-Hodge
projected Yang-Mills theory. In the present case, however, the
graded Jacobi identity gives $[[X,X],X]=0$, so the equation reduces to
\begin{equation}
 \left[F_B+\frac12[X,X],X\right]
 = [F_B,X]=0.
 \label{eq:projected-Pontryagin-X-equation-explicit}
\end{equation}
It is therefore exactly linear in $X$ at fixed $B$ and always
admits $X=0$. Whether this solution is regularly eliminable
depends on the pointwise linear map
\[
 \mathcal L_{F_B}:
 T^*M\otimes\m\longrightarrow\Lambda^3T^*M\otimes\m,
 \qquad
 \mathcal L_{F_B}(X):=[F_B,X].
\]
Where this map is invertible, the equation uniquely determines
$X=0$, and the quadratic Hessian in the $X$ sector is
nondegenerate. Where it has a nontrivial kernel, this equation
alone leaves the corresponding components of $X$ undetermined;
the remaining field equations must also be imposed.
In particular, for $F_B=0$ the $X$ equation vanishes identically.
Regularity of the $X=0$ branch therefore depends on the
background curvature.

Variation with respect to $B$ gives
\begin{equation}
 D_BK_{\uu}=0.
 \label{eq:projected-Pontryagin-B-equation}
\end{equation}
Using the $U(n)$ Bianchi identity, this becomes
\begin{equation}
 D_BK_{\uu}=[D_BX,X]=0.
 \label{eq:projected-Pontryagin-B-equation-explicit}
\end{equation}
Although this equation contains derivatives of $X$, the connection
$B$ enters algebraically through $D_BX=\dd X+[B,X]$.
This is the distinction from the spacetime-Hodge projected
Yang-Mills theory, whose $B$ equation is differential in $B$.

Define $Q:=\frac12[X,X]\in\Omega^2(M,\uu(n))$.
The graded Jacobi identity identifies the algebraic map as
\begin{equation}
 B\longmapsto [[B,X],X]=[B,Q].
 \label{eq:projected-Pontryagin-B-algebraic-map}
\end{equation}
Writing
\[
 B=C+aJ,\qquad Q=Q_{\su}+qJ,
 \qquad C\in\Omega^1(M,\su(n)),
 \qquad Q_{\su}:=\cP_{\su}Q,
\]
centrality of $J$ in $\uu(n)$ gives
\[
 [B,Q]=[C,Q_{\su}].
\]
Every central one-form $aJ$ therefore belongs to the kernel,
so the map is never invertible on the full $\uu(n)$ connection.

The equation $D_BQ=0$ separates into
\[
 \dd Q_{\su}+[C,Q_{\su}]=0,
 \qquad
 \dd q=0.
\]
Algebraic elimination of $C$ is consequently governed by the
reduced pointwise map
\[
 \mathcal N_X:
 T^*M\otimes\su(n)\longrightarrow
 \Lambda^3T^*M\otimes\su(n),
 \qquad
 \mathcal N_X(C):=[C,Q_{\su}].
\]
Where this map is invertible, it determines
$C=-\mathcal N_X^{-1}(\dd Q_{\su})$.
Its invertibility depends on $n$ and on the configuration of $X$
and requires a separate analysis. The central connection $a$
is not determined by the $B$ equation, although its curvature
$\dd a$ still enters the $X$ equation through $F_B$.

MacDowell-Mansouri gravity illustrates how a coframe supplies
a geometric criterion for algebraic elimination. The coset-valued
one-form is the coframe $e^a$, and the Lorentz connection equation
is Eq.~\eqref{eq:MM-torsion-equation}. The associated torsion map,
\[
 T^a\longmapsto\epsilon_{abcd}e^c\wedge T^d,
\]
has $4\binom42=6\binom43=24$ input and output components.
On the nondegenerate-coframe branch, this map is invertible:
the equation implies $T^a=0$ and determines
$\omega=\omega(e)$ algebraically.

The $\m$-valued one-form $X$ in the projected $SO(2n)$ theory
has no canonical coframe interpretation, so there is no
corresponding coframe criterion guaranteeing invertibility.
This does not, by itself, obstruct algebraic elimination of
the noncentral connection. Such elimination is governed by
the reduced map $\mathcal N_X$, whose rank must be assessed
for the chosen value of $n$ and configuration of $X$.
A soldering interpretation is therefore not a prerequisite
for investigating this algebraic problem.

If one further projects from $\uu(n)$ to $\su(n)$, as in the
main model, write $B=C+aJ$ and replace $K_{\uu}$ by
\begin{equation}
 K_{\su}=F_C+\cP_{\su}\!\left(\frac12[X,X]\right).
 \label{eq:projected-Pontryagin-su-curvature}
\end{equation}
The $X$ equation becomes
\begin{equation}
 [K_{\su},X]=0,
 \label{eq:projected-Pontryagin-su-X-equation}
\end{equation}
while the central connection $a$ drops out of the projected
Pontryagin action altogether. Its variation therefore yields
no independent equation.

The remaining field equations are algebraic in the field
being varied, but this property alone does not guarantee
unique or regular elimination. The relevant invertibility
conditions depend on the chosen projection, the value of $n$,
and the field configuration, and must therefore be analysed
case by case. A detailed study of these conditions and of
the branches on which algebraic elimination is possible
will be presented elsewhere.

A related four-dimensional construction was studied in
Ref.~\cite{Alvarez:2026zqy}, using the pair $(SU(3),U(2))$.
There the coset-valued one-form supplies a nondegenerate
coframe, and the algebraic equation for the $SU(2)$
connection identifies it with the anti-self-dual part of
the Levi-Civita connection. The remaining equations imply
an almost-K\"ahler geometry with constant scalar curvature
and a $J$-invariant Ricci tensor. On compact manifolds with
nonnegative scalar curvature, the solutions are K\"ahler.
They are further restricted to be K\"ahler-Einstein when
$2\pi c_1(TM,J)$ is proportional to $[\omega]$, where
$\omega$ is the fundamental two-form.
This provides an explicit example of algebraic connection
elimination supported by a nondegenerate coframe.
The corresponding invertibility conditions for the
present $SO(2n)/U(n)$ models require a separate analysis.

%%%%%%%%%%%%%%%%%%%%%%%%%%%%%%%%%%%%%%%%%%%%%%%%
\section{Full-rank equivariant kinetic operator}
\label{sec:fullrank}
%%%%%%%%%%%%%%%%%%%%%%%%%%%%%%%%%%%%%%%%%%%%%%%%

We now construct a full-rank equivariant completion of the
gauge kinetic operator. On the exact complex-structure orbit,
the first term of Eq.~\eqref{eq:dynamicalaction} is degenerate,
so this requires adding kinetic terms in the omitted directions.
The activation of coset kinetic terms by normal Higgs
displacements, discussed in Sec.~\ref{subsec:off-orbit-scope},
is a separate mechanism in which the rank changes away from
the orbit.

An adjoint complex structure intrinsically determines the
decomposition $\so(2n)=\uu(n)\oplus\m$, with
$\m=\uu(n)^\perp$. Thus the natural two-sector full-rank theory
assigns kinetic terms to the $\uu(n)$ and $\m$ curvatures.
Separating the central $\uu(1)$ direction from $\uu(n)$
is an additional refinement.

On the complex-structure orbit, define the complementary pair
\begin{equation}
 \cP_{\perp}(\Phi)
 :=\mathbf1-\cP_{\uu}(\Phi)
 =\cP_{\m}(\Phi),
 \qquad
 \cP_{\m}(\Phi)X
 :=-\frac1{4v^2}[\Phi,[\Phi,X]].
 \label{eq:dynamic-complementary-projector}
\end{equation}
Here $\cP_\m$ selects the tangent space to the orbit $SO(2n)/U(n)$.  Therefore,
in a vacuum-adapted gauge,
\begin{equation}
 \operatorname{Im}\cP_{\perp}
 =\m,
 \qquad
 \uu(n)^\perp=\m.
 \label{eq:u-orthogonal-complement}
\end{equation}
The two maps obey
\begin{equation}
 \cP_{\uu}^2=\cP_{\uu},
 \qquad
 \cP_{\perp}^2=\cP_{\perp},
 \qquad
 \cP_{\uu}\cP_{\perp}=0,
 \qquad
 \cP_{\uu}+\cP_{\perp}=\mathbf1,
 \label{eq:complementary-projector-identities}
\end{equation}
provided $\Phi^2=-v^2\mathbf1$.  Away from this orbit, the polynomial maps
in Eqs.~\eqref{eq:dynamicprojector} and
\eqref{eq:dynamic-complementary-projector} need not be idempotent; the same
off-shell qualification discussed after Eq.~\eqref{eq:equivariance} applies.

The simplest two-coupling action is
\begin{align}
 S_{\mathrm{full}}[A,\Phi]
 =&-\frac1{2g_{\uu}^2}\int_M
 \left\langle\cP_{\uu}(\Phi)F,
 \ast\cP_{\uu}(\Phi)F\right\rangle
 \nonumber\\
 &-\frac1{2g_{\perp}^2}\int_M
 \left\langle\cP_{\perp}(\Phi)F,
 \ast\cP_{\perp}(\Phi)F\right\rangle
 \nonumber\\
 &-\frac{Z_\Phi}{2}\int_M
 \langle D_A\Phi,\ast D_A\Phi\rangle
 -\int_M V(\Phi)\vol .
 \label{eq:fullrank-action}
\end{align}
For positive kinetic coefficients, the curvature kinetic
operator is
\begin{equation}
 \mathcal Q_{\rm full}(\Phi)
 =\frac1{g_{\uu}^2}\cP_{\uu}(\Phi)^2
 +\frac1{g_\perp^2}\cP_\perp(\Phi)^2.
 \label{eq:fullrank-kinetic-operator}
\end{equation}
It remains positive definite throughout the adjoint field
space. Indeed, the polynomial maps are self-adjoint and obey
$\cP_{\uu}+\cP_\perp=\mathbf1$ even away from the orbit.
Consequently, for every nonzero algebra element $Y$,
\[
 \langle Y,\mathcal Q_{\rm full}Y\rangle
 =\frac1{g_{\uu}^2}\|\cP_{\uu}Y\|^2
 +\frac1{g_\perp^2}\|\cP_\perp Y\|^2>0,
\]
where $\|Y\|^2:=\langle Y,Y\rangle$.
Both terms could vanish only if $Y=0$.
In particular,
\[
 \cP_{\uu}(0)=\mathbf1,\qquad
 \cP_\perp(0)=0,\qquad
 \mathcal Q_{\rm full}(0)=g_{\uu}^{-2}\mathbf1.
\]
The two-sector kinetic operator is therefore polynomial,
regular, and of full rank, including at $\Phi=0$.
Away from the orbit, the constituent maps generally cease
to be projectors, but this does not spoil positivity or
nondegeneracy.

On the complex-structure orbit, the projector pair assigns
independent kinetic weights to the two complementary sectors.
There, the limit $g_\perp^{-2}\to0$ leaves a degenerate
$\uu(n)$-projected theory, retaining the central kinetic term,
while $g_\perp=g_{\uu}$ gives the ordinary $SO(2n)$
Yang-Mills kinetic term.

To recover the $\su(n)$-projected theory as a limit, one must distinguish the
central and coset sectors further:
\begin{align}
 S_{\mathrm{kin}}=-\frac12\int_M\bigg[&
 \frac1{g_{\su}^2}\langle\cP_{\su}F,\ast\cP_{\su}F\rangle
 +\frac1{g_1^2}\langle\cP_{\Phi}F,\ast\cP_{\Phi}F\rangle
 \nonumber\\
 &+\frac1{g_{\m}^2}\langle\cP_{\m}F,\ast\cP_{\m}F\rangle
 \bigg],
 \label{eq:three-coupling-kinetic-action}
\end{align}
where every projector in Eq.~\eqref{eq:three-coupling-kinetic-action} is
understood to depend on $\Phi$.  Around $\vev{\Phi}=vJ$, the fields
$A_{\su}$, $aJ$, and $W_++W_-$ then have couplings $g_{\su}$, $g_1$, and
$g_\m$, respectively.  The Higgs kinetic term gives $W_\pm$ a mass because
$[J,W_\pm]\neq0$, while $A_{\su}$ and $aJ$ remain massless because they
commute with $J$.  A second Higgs is still required to Higgs the central
$U(1)$ if the desired unbroken group is exactly $SU(n)$.

All terms in Eqs.~\eqref{eq:fullrank-action} and
\eqref{eq:three-coupling-kinetic-action} are fully local $SO(2n)$ invariant,
subject to an important distinction.  Since
\begin{equation}
 \Phi\mapsto g^{-1}\Phi g,
 \qquad
 F\mapsto g^{-1}Fg,
 \qquad
 \cP_I(g^{-1}\Phi g)(g^{-1}Xg)
 =g^{-1}\cP_I(\Phi)Xg,
 \label{eq:all-projectors-equivariance}
\end{equation}
for $I=\uu,\su,\Phi,\m,\perp$, each projected curvature transforms by conjugation.
The invariant inner product then makes every curvature term gauge invariant.
Likewise, $D_A\Phi$ transforms by conjugation, and the trace potential
\eqref{eq:potential} is a class function.  Independent numerical couplings do
not break the symmetry because they multiply separately invariant terms.

For $n\ge3$, generic unequal sector weights with fixed $J$
preserve only the connected subgroup $U(n)$. Reduction to
$SU(n)$ additionally requires an $\Omega$-dependent structure
in the action. Special choices of weights can enhance the
symmetry since equal weights recover the ordinary Yang-Mills
kinetic term. For $n=2$ the three-sector curvature
action is $SO(4)$ invariant whenever $g_1=g_{\m}$, even if
this differs from $g_{\su}$. This is because
$$\so(4) \cong \su(2) \oplus \su(2) \cong \su(2) \oplus (\mathbb{R} J \oplus \m)$$
and each ideal can have its own coupling. Promoting the defining structure
to an adjoint-transforming compensator restores full gauge
covariance for arbitrary sector weights and a Higgs kinetic
term is not required for this covariance.

This freedom is not typical of the minimal renormalizable Yang-Mills sector
of a conventional grand unified theory.  For a simple unified group, the
gauge algebra admits, up to normalization, a single invariant quadratic form,
and the minimal action contains one unified coupling,
\begin{equation}
 S_{\rm YM}=-\frac1{2g_G^2}\int_M\langle F,\ast F\rangle.
\end{equation}
After spontaneous symmetry breaking, subgroup couplings are consequently
related at the matching scale.  The case $g_{\uu}\neq g_\perp$ is instead
analogous to a nonminimal Higgs-dependent gauge kinetic operator: the full
$SO(2n)$ symmetry is intact, but the vacuum gives different kinetic
normalizations, and hence different effective interaction strengths, to the
two sectors.  Indeed, on the complex-structure orbit,
\begin{align}
 S_{\rm kin}={}&-\frac1{2g_\perp^2}\int_M\langle F,\ast F\rangle
 \nonumber\\
 &-\frac12\left(\frac1{g_{\uu}^2}-\frac1{g_\perp^2}\right)
 \int_M\left\langle\cP_{\uu}(\Phi)F,
 \ast\cP_{\uu}(\Phi)F\right\rangle .
 \label{eq:fullrank-nonminimal-form}
\end{align}
Thus unequal couplings modify or abandon conventional coupling unification,
without explicitly breaking the unified gauge symmetry.  The two-sector completion in Eq.~\eqref{eq:fullrank-action}
uses polynomial maps and requires no exclusion of $\Phi=0$.
The three-sector refinement in
Eq.~\eqref{eq:three-coupling-kinetic-action}, however,
uses the normalized projector $\cP_\Phi$, which contains
$\langle\Phi,\Phi\rangle^{-1}$.
For unequal central and traceless-sector weights, the
resulting kinetic operator is generally nonpolynomial and
does not extend continuously to $\Phi=0$.
This refinement is therefore naturally defined on
$\Phi\neq0$, or on the exactly constrained orbit.

When $g_1=g_{\su}$, the central and traceless curvature terms
combine into the $\cP_{\uu}$ term, even away from the orbit.
The dependence on the normalized projector then cancels,
recovering the regular polynomial two-sector completion.
Possible applications of this full-rank equivariant
construction will be investigated elsewhere.

%%%%%%%%%%%%%%%%%%%%%%%%%%%%%%%%%%%%%%%%%%%%%%%%
\section{Summary and Discussion}
%%%%%%%%%%%%%%%%%%%%%%%%%%%%%%%%%%%%%%%%%%%%%%%%

We have studied gauge theories in which an internal projector defines a
degenerate curvature kinetic term. For $SO(2n)$, an orthogonal complex
structure selects $\uu(n)$, and subtraction of its central direction
gives the kinetic image $\su(n)$. The contribution is the use of this
splitting in the defining kinetic operator and the analysis of the
resulting constraints. In particular, the coset connection can be
auxiliary even when a Higgs kinetic term supplies an algebraic quadratic
term. The stabilizer of the adjoint vacuum remains $U(n)$, while its
central connection is null in the exactly projected orbit theory.
We have distinguished a fixed projector, an unrestricted adjoint Higgs,
and an exactly orbit-constrained Higgs, and compared their dynamics with
projected gravity and Pontryagin constructions.

The principal results of this paper are as follows:
\begin{enumerate}
 \item \emph{Equivariant projectors and the distinction between field spaces.}
 On the constrained orbit, Eqs.~\eqref{eq:constrained-projectors} and
 \eqref{eq:constrained-projector-identities} give exact orthogonal
 projectors. The intrinsic action~\eqref{eq:constrained-action} and the
 matrix-multiplier action~\eqref{eq:constrained-multiplier-action} have
 equivalent local classical equations, with the orbit Higgs equation
 related to the connection equation by
 Eq.~\eqref{eq:constrained-Noether}. For the unrestricted continuation,
 the kinetic operator~\eqref{eq:off-orbit-kinetic-operator} and the
 weights~\eqref{eq:off-orbit-root-weights} exhibit a change of rank away
 from the orbit. The quadratic vacuum spectrum therefore does not
 establish a nonlinear degree-of-freedom count for that theory.

 \item \emph{Regularity of the constrained model.}
 For $v\ne0$ and $Z_\Phi>0$, the Higgs kinetic term contributes the
 nondegenerate algebraic matrix in Eq.~\eqref{eq:SCmatrixdynamic}.
 This makes the coset primary and secondary constraints second class
 at the zero-field background and throughout a neighborhood in the
 constrained phase space. Their elimination gives the local
 weak-curvature branch $X=0$ and the reduced Yang-Mills
 action~\eqref{eq:constrained-reduced-YM}. The nonlinear count is
 $2(n^2-1)$ physical degrees of freedom in four spacetime dimensions,
 as in Eq.~\eqref{eq:dyncount}. This regularity uses $Z_\Phi>0$ and
 does not follow from restoring gauge covariance alone.

 \item \emph{Criteria for additional algebraic branches.}
 Equation~\eqref{eq:X-algebraic-branch-equation} admits a nonlinear
 eigenvalue formulation, Eq.~\eqref{eq:X-nonlinear-eigenvalue}, and a
 Cartan-root reduction,
 Eqs.~\eqref{eq:cartan-root-branch-equations1}-\eqref{eq:cartan-root-branch-equations2}.
 A branch meeting $X=0$ requires the singularity condition
 \eqref{eq:X-critical-background}. Under the stated assumptions on
 the critical modes, its leading amplitude obeys
 Eq.~\eqref{eq:X-amplitude-equation}. These are criteria for possible
 branches, not a general existence theorem: the remaining connection
 equation~\eqref{eq:nontrivial-branch-B-equation} must also hold, and
 its effect on the cubic coefficient is displayed in
 Eq.~\eqref{eq:X-coupled-cubic-coefficient}.

 \item \emph{A common Clifford-projector formulation of Higgs-gravity.}
 The involution in Eq.~\eqref{eq:SW-full-Clifford-projectors} selects
 Lorentz and coset bivectors, while its action on Clifford vectors
 separates radial and tangent components,
 Eq.~\eqref{eq:SW-vector-projector-action}. The same pair of projectors
 therefore expresses the gravitational curvature, radial Higgs
 kinetic term, and induced metric in the full
 action~\eqref{eq:SW-full-projector-action}. This is a reformulation
 of the established Stelle-West Higgs-gravity construction. Its
 contribution is the common projector description, valid for
 $s=V^AV_A<0$ and a nondegenerate induced metric, with exact projector
 identities even when the radial field fluctuates.

%  \item \emph{Projected Pontryagin theories.}
%  Subsection~\ref{sec:SO(2n)PPtheory} defines the $\uu(n)$-projected
%  action~\eqref{eq:projected-Pontryagin-action} and its $\su(n)$
%  refinement using Eq.~\eqref{eq:projected-Pontryagin-su-curvature}.
%  The equations~\eqref{eq:projected-Pontryagin-X-equation} and
%  \eqref{eq:projected-Pontryagin-B-equation-explicit} are algebraic
%  in the field being varied, although the coupled system remains
%  differential. The map~\eqref{eq:projected-Pontryagin-B-algebraic-map}
%  identifies the connection-elimination problem. Its invertibility
%  requires a separate analysis for each projection and configuration.
 
 \item \emph{Projected Pontryagin theories.}
Subsection~\ref{sec:SO(2n)PPtheory} defines the $\uu(n)$-projected
action~\eqref{eq:projected-Pontryagin-action} and its $\su(n)$
refinement using Eq.~\eqref{eq:projected-Pontryagin-su-curvature}.
The equations~\eqref{eq:projected-Pontryagin-X-equation} and
\eqref{eq:projected-Pontryagin-B-equation-explicit} are algebraic
in the field being varied, although the coupled system remains
differential. The full connection map in
Eq.~\eqref{eq:projected-Pontryagin-B-algebraic-map} is always
singular because its kernel contains every central one-form.
Algebraic elimination of the noncentral connection is instead
governed by the reduced map $\mathcal N_X(C)=[C,Q_{\su}]$.
Its invertibility depends on $n$ and on the configuration of
$X$, and requires a separate case-by-case analysis.

 \item \emph{A full-rank equivariant completion.}
 Equations~\eqref{eq:fullrank-action} and
 \eqref{eq:three-coupling-kinetic-action} assign independent kinetic
 weights to complementary sectors on the orbit. Their full
 $SO(2n)$ gauge invariance follows from projector equivariance,
 Eq.~\eqref{eq:all-projectors-equivariance}, and does not require
 equal sector couplings. Giving kinetic terms to the omitted
 directions changes the theory and its auxiliary-field content.
\end{enumerate}

These results leave several questions about the classical extensions
and quantum theory. A first issue is whether the degenerate kinetic
operator is stable under quantum corrections. The complementary term
\begin{equation}
 \int_M\left\langle\cP_\m(\Phi)F,
 \ast\cP_\m(\Phi)F\right\rangle
\end{equation}
is allowed by $SO(2n)$ covariance, but is not generally invariant under
the exact central shift of the constrained action described in
Sec.~\ref{subsec:constrained-unitary}. Its generation therefore cannot
be inferred from gauge covariance alone. A quantum analysis must account
for this additional symmetry and for any interactions that break it.
If a positive complementary kinetic term is consistently included,
the coset vectors can propagate and the auxiliary-field count no longer
applies. Quantum equivalence of the intrinsic and multiplier
formulations also requires a comparison of their measures, gauge fixing,
and redundant multiplier components.

For the unrestricted Higgs theory, the main unresolved issue is the
Hamiltonian analysis where the kinetic rank differs from its value on
the vacuum orbit. Equation~\eqref{eq:off-orbit-canonical-momentum}
shows why the constraints of the orbit theory cannot simply be carried
over to that field space. Establishing interaction scales and a
controlled low-energy relation to the constrained model requires
integrating out the normal Higgs fields, including their curvature
couplings, and accounting for the change in vector kinetic rank.
Covariant constancy of $\Phi$ does not replace this analysis: it gives
idempotence on compatible curvature components, not on the full algebra,
and leaves the Higgs equation to be checked. An alternative off-orbit
completion would require its own specification of the field domain
and constraint structure.

The classical branches also require further study. Solving the
pointwise algebraic equation for $X$ is only one part of constructing
a solution of the coupled system. The existence and stability of
nonzero branches, including those disconnected from $X=0$, remain open.
At the singular loci~\eqref{eq:singularlocus}, the constraint algorithm
must be reconsidered before assigning a degree-of-freedom count.
The corresponding invertibility problem for the projected Pontryagin
theories is likewise sensitive to the field configuration. A detailed
analysis of those conditions, together with possible applications of
the full-rank equivariant construction, will be presented elsewhere.

Matter couplings must account for the null central connection. Matter
charged under that direction can turn its field equation into a current
constraint, so ordinary $SU(n)$ matter dynamics does not follow merely
from choosing an $\su(n)$ kinetic image. Possible extensions include
matter neutral under the central direction or an additional
complex-volume-form Higgs that reduces the stabilizer to $SU(n)$.
Their constraint structure and quantum consistency require a separate
analysis. Finally, the local unitary-gauge description does not settle
the global classification of $SO(2n)/U(n)$ configurations and reduced
bundles, or the role of boundary charges and possible defects.

%%%%%%%%%%%%%%%%%%%%%%%%%%%%%%%%
\section*{Acknowledgements}
%%%%%%%%%%%%%%%%%%%%%%%%%%%%%%%%

The author acknowledges support from ANID-FONDECYT Regular grant No. 1230112.

\appendix

%%%%%%%%%%%%%%%%%%%%%%%%%%%%%%%%%%%%%%%%%%%%%%5
\section{Direct derivation of the projected \texorpdfstring{$X$}{X} equation}
\label{app:X-variation}
%%%%%%%%%%%%%%%%%%%%%%%%%%%%%%%%%%%%%%%%%%%%%%5

This appendix derives the algebraic equation obeyed by the connection
component in the projected-out sector.  We use the orthogonal decomposition
\begin{equation}
 \so(2n)=\su(n)\oplus\mathbb R J\oplus\m
\end{equation}
and write
\begin{equation}
 A=B+cJ+X,
 \qquad
 B\in\Omega^1(M,\su(n)),
 \qquad
 X\in\Omega^1(M,\m).
 \label{eq:app-A-decomposition}
\end{equation}
The fixed projector is self-adjoint with respect to the invariant bilinear
form and, using the symmetric-pair relations, the curvature entering the
action is
\begin{equation}
 K:=\cP_{\su}F(A)
   =F_B+\cP_{\su}(X\wedge X)
   \in\Omega^2(M,\su(n)).
 \label{eq:app-K-definition}
\end{equation}
In particular, $D_BX+c[J,X]$ is $\m$-valued and is annihilated by
$\cP_{\su}$.  The action therefore reads
\begin{equation}
 S_{\rm fix}
 =-\frac{1}{2g^2}\int_M\langle K,\ast K\rangle .
 \label{eq:app-fixed-action}
\end{equation}

We now vary $X$ while holding $B$ and $c$ fixed.  In the convention in which
$X\wedge X$ denotes the Lie-algebra product combined with the exterior
product, its variation is
\begin{equation}
 \delta_X(X\wedge X)
 =\delta X\wedge X+X\wedge\delta X.
 \label{eq:app-variation-XX}
\end{equation}
Equivalently, if the curvature is written as
$F=\dd A+\tfrac12[A,A]$, then
\begin{equation}
 \delta_X\!\left(\frac12[X,X]\right)=[\delta X,X],
 \label{eq:app-graded-bracket-variation}
\end{equation}
because the graded bracket of two Lie-algebra-valued one-forms is symmetric:
$[X,\delta X]=[\delta X,X]$.  Hence
\begin{equation}
 \delta_XK
 =\cP_{\su}[\delta X,X].
 \label{eq:app-delta-K}
\end{equation}
Self-adjointness of $\cP_{\su}$ and
$K=\cP_{\su}K$ give
\begin{align}
 \delta_XS_{\rm fix}
 &=-\frac1{g^2}\int_M
   \left\langle\cP_{\su}[\delta X,X],\ast K\right\rangle
 \nonumber\\
 &=-\frac1{g^2}\int_M
   \left\langle[\delta X,X],\ast K\right\rangle .
 \label{eq:app-first-variation}
\end{align}
Using invariance of the bilinear form, with the exterior-product signs
included, one may move the commutator from the first slot to the second:
\begin{equation}
 \left\langle[\delta X,X],\ast K\right\rangle
 =\left\langle\delta X,[X,\ast K]\right\rangle .
 \label{eq:app-invariance-identity}
\end{equation}
Therefore
\begin{equation}
 \delta_XS_{\rm fix}
 =-\frac1{g^2}\int_M
   \left\langle\delta X,[X,\ast K]\right\rangle .
 \label{eq:app-X-variation-final}
\end{equation}

The variation $\delta X$ is arbitrary only within
$\Omega^1(M,\m)$; it is not an arbitrary $\so(2n)$-valued one-form.
Orthogonality of the decomposition consequently implies the Euler-Lagrange
equation
\begin{equation}
 \cP_{\m}[X,\ast K]=0.
 \label{eq:app-projected-X-equation}
\end{equation}
This projection is the variational origin of the equation: stationarity tests
only the component paired with the allowed $\m$-valued variations.

For the present symmetric pair,
\begin{equation}
 [\su(n),\m]\subset\m,
 \qquad
 X\in\m,
 \qquad
 K\in\su(n),
 \label{eq:app-symmetric-pair-use}
\end{equation}
and therefore $[X,\ast K]$ is already $\m$-valued.  The projector in
Eq.~\eqref{eq:app-projected-X-equation} is consequently redundant, and the
equation reduces to
\begin{equation}
 [X,\ast K]=0,
 \qquad
 K=F_B+\cP_{\su}(X\wedge X),
 \label{eq:app-unprojected-X-equation}
\end{equation}
which is Eq.~\eqref{eq:Xeomfixed}.  Since $K$ contains no $D_BX$ term,
Eq.~\eqref{eq:app-unprojected-X-equation} is algebraic and nonlinear in $X$.

%%%%%%%%%%%%%%%%%%%%%%%%%%%%%%%%%%%%%%%%%%%%%%5
\section{Goldstone fields and the auxiliary coset connection near the vacuum}
\label{app:goldstone-auxiliary}
%%%%%%%%%%%%%%%%%%%%%%%%%%%%%%%%%%%%%%%%%%%%%%5

This appendix makes explicit the fate of the tangent Higgs fluctuations near
the vacuum orbit.  Decompose the connection as
\begin{equation}
 A=A_H+X,
 \qquad
 A_H:=B+cJ\in\Omega^1(M,\uu(n)),
 \qquad
 X\in\Omega^1(M,\m),
 \label{eq:app-goldstone-A-decomposition}
\end{equation}
and parametrize a neighborhood of the orbit by
\begin{equation}
 \Phi=e^{-\tau}(vJ+\rho)e^\tau,
 \qquad
 \tau\in\m,
 \qquad
 \rho\in\uu(n).
 \label{eq:app-goldstone-parametrization}
\end{equation}
Here $\tau$ is dimensionless and describes the tangent directions, whereas
$\rho$ describes fluctuations normal to the orbit.  To first order,
\begin{equation}
 \Phi=vJ+\rho+[vJ,\tau]+O(\tau^2,\tau\rho),
 \qquad
 \varphi_{\m}=[vJ,\tau].
 \label{eq:app-tangent-fluctuation}
\end{equation}

Because $[A_H,J]=0$, the $\m$-valued part of the covariant derivative is
\begin{align}
 (D_A\Phi)_{\m}
 &=D_{A_H}[vJ,\tau]+[X,vJ]+O(2)
 \nonumber\\
 &=[vJ,D_{A_H}\tau-X]+O(2).
 \label{eq:app-DPhi-tangent}
\end{align}
The tangent field and the coset connection thus enter through the
Stueckelberg combination
\begin{equation}
 \mathcal X:=X-D_{A_H}\tau.
 \label{eq:app-stueckelberg-combination}
\end{equation}
Indeed, for an infinitesimal broken transformation with parameter
$\alpha\in\m$, the convention
$A\mapsto g^{-1}Ag+g^{-1}\dd g$ and
$\Phi\mapsto g^{-1}\Phi g$ gives, to leading order,
\begin{equation}
 \delta\tau=\alpha,
 \qquad
 \delta X=D_{A_H}\alpha,
 \qquad
 \delta\mathcal X=0.
 \label{eq:app-stueckelberg-transformations}
\end{equation}

The tangent contribution to the quadratic Higgs kinetic action is therefore
\begin{align}
 S_{\Phi,\m}^{(2)}
 &=-\frac{Z_\Phi v^2}{2}
 \int_M\left\langle[J,\mathcal X],
 \ast[J,\mathcal X]\right\rangle
 \nonumber\\
 &=-2Z_\Phi v^2
 \int_M\langle\mathcal X,\ast\mathcal X\rangle.
 \label{eq:app-tangent-quadratic-action}
\end{align}
In the second line we used
$[J,[J,Y]]=-4Y$ for $Y\in\m$ and the invariance of the inner product.
Equation~\eqref{eq:app-tangent-quadratic-action} is Stueckelberg-like, but it
does not describe an ordinary massive vector: the projected Yang-Mills term
contains no quadratic kinetic operator
$\langle D_{A_H}X,\ast D_{A_H}X\rangle$.

Locally one may choose the vacuum-adapted gauge $\tau=0$.  Then
\begin{equation}
 S_{\Phi,\m}^{(2)}
 =-2Z_\Phi v^2\int_M\langle X,\ast X\rangle,
 \label{eq:app-unitary-gauge-X-term}
\end{equation}
which is the algebraic term encoded by the matrix
\eqref{eq:Xmassmatrix}.  Thus $X$ is auxiliary rather than Proca: its free
equation is simply $X=0$.

The same conclusion follows without choosing this local gauge.  Around the
free vacuum, variation of Eq.~\eqref{eq:app-tangent-quadratic-action} with
respect to $X$ gives
\begin{equation}
 X=D_{A_H}\tau.
 \label{eq:app-X-eliminates-goldstone}
\end{equation}
Substitution back into the quadratic action sets $\mathcal X=0$, so the
apparent derivative term for $\tau$ disappears.  Consequently $\tau$ is a gauge
orbit coordinate and $X$ is an auxiliary connection component; neither one
provides a propagating degree of freedom.  In particular, the Goldstone field
does not supply the longitudinal polarization of a massive vector.

This differs from the ordinary Higgs mechanism, in which the same
Stueckelberg term is accompanied by a nondegenerate kinetic term for $X$.
If a complementary curvature kinetic term is added, $X$ becomes a conventional
massive vector and $\tau$ is then eaten in the usual sense.  In the purely
projected model, by contrast, the Goldstone-connection pair is removed by
gauge symmetry and the algebraic $X$ equation.
This conclusion is a quadratic statement about the zero-curvature vacuum
of the finite-$\lambda_H$ theory, and a nonlinear statement only on the
regular exactly constrained branch analyzed in the main text.  Normal
displacements can activate the derivative term in
Eq.~\eqref{eq:off-orbit-induced-X-kinetic}; this appendix does not
eliminate those off-orbit vector modes.

\bibliographystyle{unsrt}
\bibliography{projected_yang_mills_draft_v5.bib}

\end{document}